\def\preprint{1}                
\def\comment#1{}

\if\preprint1
        \documentclass[usenatbib]{mn2e}
         \usepackage{natbib}  
          \usepackage{times}
        \usepackage{graphicx}
        \usepackage{calc}
        \usepackage{rotating}
         \usepackage{lscape}
         \bibpunct{(}{)}{;}{a}{}{,} 
         \usepackage{longtable}
         \usepackage{tgcursor}
         \usepackage[toc,page]{appendix}
         \usepackage{enumitem}
\else
        \documentstyle[astrop-bib,referee,times]{mn2e}
        \newcommand{\includegraphics}[1]{}
\fi

\def\oversim#1#2{\lower0.5pt\vbox{\baselineskip0pt \lineskip-0.5pt
     \ialign{$\mathsurround0pt #1\hfil##\hfil$\crcr#2\crcr\sim\crcr}}}

\def\apj {{ApJ}}

\def\aap {{A\&A}}
\def\mnras {{MNRAS}}

\def\pasp  {{PASP}}
\def\apjl {{ApJ}}
\def\aj {{ApJ}}

\def\aaps {{A\&A}}
\def\apss {{Astrophysics and Space Science}}
\def\PASA {{PASA}}

\def\rmxaa {{Revista Mexicana de Astronom\'ia y Astrof\'isica}}
\def\nar {{NAR}}
\def\physrep {{Physics Reports}}

\title[Catalogue of IPHAS PNe]{First release of the IPHAS Catalogue of New Extended Planetary Nebulae}
\author[L. Sabin et al.]{\thanks{E-mail:laurence.sabin@gmail.com(LS)}
L. Sabin$^{1}$, Q.A. Parker$^{2,3,4}$, R.L.M Corradi$^{5,6}$, L. Guzman-Ramirez$^{7}$, R.A.H. Morris$^{8}$, \newauthor A.A. Zijlstra$^{9}$, I.S. Boji\v{c}i\'c$^{2,3,4}$,  D.J. Frew$^{2,3}$, M. Guerrero$^{10}$, M. Stupar$^{2,3}$,  M.J. Barlow$^{11}$,\newauthor F. Cort\'es Mora$^{1}$, J.E. Drew$^{12}$,  R. Greimel$^{13}$, P. Groot$^{14}$, J.M. Irwin$^{15}$, M.J. Irwin$^{16}$, \newauthor  A. Mampaso$^{5,6}$, B. Miszalski$^{17,18}$, L. Olgu\'in$^{19}$, S. Phillipps$^{8}$, M. Santander Garc\'ia$^{20,21}$,\newauthor K. Viironen$^{22}$ and N.J. Wright$^{12}$ \\
$^{1}$Instituto de Astonom{\'i}a y Meteorolog{\'i}a, Departamento de F{\'i}sica, CUCEI, Universidad de Guadalajara, Av. Vallarta 2602, C.P. 44130, Guadalajara, Jal., M\'exico\\
$^{2}$Macquarie University Research Centre in Astronomy, Astrophysics \& Astrophotonics, Sydney, NSW 2109, Australia\\
$^{3}$Department of Physics and Astronomy, Macquarie University, Sydney, NSW 2109, Australia\\
$^{4}$Australian Astronomical Observatory, PO Box 296, Epping, NSW 1710, Australia\\
$^{5}$Instituto de Astrof{\'{\i}}sica de Canarias, E-38200 La Laguna, Tenerife, Spain\\
$^{6}$Departamento de Astrof{\'{\i}}sica, Universidad de La Laguna, E-38206 La Laguna, Tenerife, Spain\\
$^{7}$European Southern Observatory, Alonso de C\'ordova 3107, Casilla 19001, Santiago, Chile \\
$^{8}$School of Physics, Bristol University, Tyndall Avenue, Bristol, BS8 1TL, UK\\
$^{9}$Jodrell Bank Centre for Astrophysics, Alan Turing Building, Manchester, M13 9PL, UK\\
$^{10}$Instituto de Astrof\'isica de Andaluc\'ia, IAA-CSIC, Glorieta de la Astronom\'ia s/n, 18008 Granada, Spain\\
$^{11}$Department of Physics and Astronomy, University College London, Gower Street, London WC1E 6BT, UK\\
$^{12}$School of Physics, Astronomy \& Mathematics, University of Hertfordshire, College Lane, Hatfield, AL10 9AB, UK\\
$^{13}$IGAM, Institute of Physics, NAWI Graz, University of Graz, Universit\"{a}tsplatz 5/II, 8010 Graz, Austria\\
$^{14}$Department of Astrophysics/IMAPP, Radboud University Nijmegen, P.O. Box 9010, 6500 GL Nijmegen, The Netherlands\\
$^{15}$Harvard-Smithsonian Center for Astrophysics, 60 Garden St., Cambridge, MA, 02138, US \\
$^{16}$Institute of Astronomy, University of Cambridge, Madingley Road, Cambridge CB3 0HA, UK  \\
$^{17}$South African Astronomical Observatory, PO Box 9, Observatory, 7935, South Africa\\
$^{18}$Southern African Large Telescope Foundation, PO Box 9, Observatory, 7935, South Africa\\
$^{19}$Departamento de Investigaci\'on en F\'isica, Universidad de Sonora, M\'exico \\
$^{20}$Observatorio Astron\'omico Nacional, Ap 112, 28803 Alcal\'a de Henares, Spain\\
$^{21}$CAB, INTA-CSIC, Ctra de Torrej\'on a Ajalvir, km 4, 28850 Torrej\'on de Ardoz, Madrid, Spain\\
$^{22}$Centro de Estudios de F\'isica del Cosmos de Arag\'on, Plaza San Juan 1, Planta 2, Teruel, 44001, Spain\\
}

\begin{document}

\date{Accepted . Received}

\pagerange{\pageref{firstpage}--\pageref{lastpage}} \pubyear{2014}

\maketitle

\label{firstpage}

\begin{abstract}

We present the first results of our search for new, extended Planetary Nebulae (PNe) based on careful, systematic, visual scrutiny of the imaging data from the INT Photometric H$\alpha$ Survey of the Northern Galactic Plane (IPHAS). The newly uncovered PNe will help to improve the census of this important population of Galactic objects that serve as key windows into the late stage evolution of low to intermediate mass stars. They will also facilitate study of the faint end of the ensemble Galactic PN luminosity function. The sensitivity and coverage of IPHAS allows PNe to be found in regions of greater extinction in the Galactic Plane and/or those PNe in a more advanced evolutionary state and at larger distances compared to the general Galactic PN population. Using a set of newly revised optical diagnostic diagrams in combination with access to a powerful, new, multi-wavelength imaging database, we have identified 159 true, likely and possible PNe for this first catalogue release. The ability of IPHAS to unveil PNe at low Galactic latitudes and towards the Galactic Anticenter, compared to previous surveys, makes this survey an ideal tool to contribute to the improvement of our knowledge of the whole Galactic PN population.

\end{abstract}


\section{Introduction}

Planetary nebulae (PNe) are strong astrophysical tools allowing us to understand the late stage stellar evolution and the chemical evolution of our entire Galaxy. The ionised shell exhibits strong and numerous emission lines that are excellent laboratories for plasma physics. PNe are also visible to great distances where their strong lines permit determination of the sizes, expansion velocities and ages of the PNe, so probing the physics and timescales of stellar mass loss (e.g. \citealt{Iben1995}). We can also use them to derive luminosity, temperature and mass of their central stars, and the chemical composition of the ejected gas. Finally PNe can be used to directly probe Galactic stellar and chemical evolution (\citealt{Dopita1997,Maciel2003}).  Adding to their number, particularly at their more evolved extremes, can help inform general models describing the physical and chemical processes occurring during this crucial late stage of stellar evolution.

The  general knowledge of the Galactic PN population has traditionally been based on the $\sim$1500 objects listed in the Strasbourg-ESO Catalogue and its supplement \citep{Acker92,Acker96} and the largely overlapping compendium of \citet{Kohoutek2001}. More recently the Macquarie-AAO-Strasbourg H$\alpha$ Survey (MASH) catalogues (\citealt{parker2006a}, \citealt{Miszalski2008}) uncovered an additional $\sim$1500 spectroscopically confirmed PNe.
These discoveries were based on careful scrutiny of the SuperCOSMOS AAO/UKST H$\alpha$ Survey (SHS) of 4000 square degrees of the Southern Galactic plane  which is described in full in \citet{Parker2005} and \citet[a]{Frew2014a}. The detection rate of new Galactic PNe has been relatively low since the release of the MASH catalogue. Nevertheless, a further 200 or so confirmed PNe have been uncovered subsequently by a medley of other researchers including \citet{Boumis2006} and \citet{Gorny2006} who found 44 and 24 new PNe in the Galactic Bulge region respectively. Significant numbers ($\sim$70) have also come from the Deep Sky Hunters (DSH) 'amateur' consortium via the analysis of the on-line Digital Sky Survey plates e.g. \citet{Jacoby2010}, \citet{Kronberger2006,Kronberger2012,Kronberger2014} and also from a group of French amateurs,  \citet{Acker2012}. Finally, several hundred unconfirmed compact candidates from the Isaac Newton Telescope Photometric H$\alpha$ Survey IPHAS survey \citep{Drew2005} have also been found by \citet{Viironen2009b}. Their proper investigation and veracity has now been assessed during the construction of a new, comprehensive, multi-wavelength Galactic PN database by Bojicic et al. (in preparation and see later). 

All the recent, confirmed discoveries takes the total current Galactic PN population to $\sim3300$, double what it was a decade ago. However, even  this number falls a factor of $\sim$1.5 short of even the most conservative Galactic PN number estimates. Population synthesis yields 6,600-46,000 PNe depending on whether the binary hypothesis for PN formation is invoked. For example,  \citet{Frew2006} predict a global PN population of 28\,000$\pm$5000, \citet{Moe2006} derived 46\,000$\pm$13\,000 Galactic PNe with a radius r$<$0.9 pc, \citet{Zijlstra1991} gave an estimation of the total number of PNe in the Galactic disk of $\sim$ 23\,000 $\pm$ 6000 while Moe \& De Marco (2005) predicted only $\sim$6600 if close binaries (e.g. a common envelope phase) is required to form PNe. In this last case at least there are prospects to rule out the PN binary hypothesis as known PN numbers are now within less than a factor or two of this prediction. This is especially true given that a significant population of Galactic PNe must still be lurking behind the extensive clouds of gas and dust that obscure large regions of our view across the optical regime (e.g. Parker et al. 2012). Indeed, it is the extension of previous PN discovery techniques away from the optically dominant  [OIII] emission line in un-reddened PN spectra towards the longer wavelength 
H$\alpha$ emission line (that can peer at least partially through the dust), that has led to the major discoveries of the previous decade.

The most studied PNe (particularly those used for abundance studies) currently belong to the bright end of the luminosity function \citep{Ciardullo2010}. This means that they are nearby and/or relatively young and as such may not be representative of the true, underlying PNe population. Until the advent of MASH the faintest and more evolved PNe were not well represented and this remains the case for the Northern Galactic plane. PN studies have also mainly concentrated on the solar neighbourhood and on the inner Galaxy, with fewer objects investigated towards the Galactic Anticentre. This imbalance becomes very important when considering the existence of an abundance gradient in the Galaxy and the behaviour of that gradient in the outer regions of the Galactic plane. 
The IPHAS H$\alpha$ survey of the inner regions of the Northern Galactic Plane \citep{Drew2005} allows us scope to tackle these issues.

We present here the first significant discoveries of spectroscopically confirmed, extended PNe from candidates selected via careful visual scrutiny of  the IPHAS data. 
This paper represents the outcome of nearly eight years of candidate detections and spectroscopic follow-up including many evolved and (very) faint nebulae located in the Northern Galactic Plane. 
In this, the first of several papers on new IPHAS PNe,  we present the basic information for 159 newly confirmed Galactic PNe including positions, sizes and morphologies concentrated primarily in a two hour Right Ascension (RA) zone between 18 and 19 hours.  There still remain hundreds of IPHAS resolved PN candidates still waiting final confirmation that will be the subject of additional papers in the series. This paper is structured as follows.
First the IPHAS survey itself is briefly described  ($\S$2), then the PNe candidate detection method ($\S$3) and then the subsequent spectroscopic follow-up ($\S$4). The catalogue is presented in $\S$5 with some statistics based on the PN parameters given in $\S$6. The online version of the catalogue which is included as a subset of the new Macquarie-AAO-Strasbourg multi-wavelength and spectroscopic PN database (MASPN; Bojicic et al, in preparation) is briefly presented in $\S$7. Our concluding remarks are discussed in $\S$8.

\section{The INT Photometric H$\alpha$ Survey: IPHAS}

IPHAS is a fully photometric CCD survey of the Northern Galactic Plane that began in 2003 and is now essentially complete (see \citealt{Drew2005}, \citealt{Solares2008}). A careful photometric calibration of the survey has now been undertaken and is presented in Barentsen et al. 2014 (submitted). IPHAS targeted the inner regions of the Northern  plane over the latitude range of -5$^\circ$ $<$ b $<$ 5$^\circ$ and a longitude range of  29$^\circ$ $<$ l $<$ 215$^\circ$ covering a total of 1800 square degrees  or about 45\% of the coverage of the SHS in the south due to the more restricted range in b. The survey used the 2.5 m Isaac Newton Telescope (INT)  at La Palma in the  Canary Islands, Spain equiped  with the Wide Field Camera (WFC). The WFC offered a field of view of 34$\times$34 arcmin$^2$ thanks to its four EEV 2k$\times$4 CCDs\footnote{http://www.ing.iac.es/Astronomy/telescopes/int/}. In addition to the 120~second  H$\alpha$ filter exposures (95 \AA~ FWHM, central wavelength at 6568 \AA), IPHAS was also conducted with two broadband filters: Sloan {\it r'} (central wavelength at 6240\AA, 30s exposure) and Sloan {\it i'} (central
wavelength at 7743 \AA, 10s exposure). The {\it r'} filter is a continuum 'off-band' filter which can be used with the narrow-band  H$\alpha$ filter to detect emission line stars and nebulae.  The IPHAS survey offered two main advantages over previous surveys of this kind
in the north. First, the WFC has a small pixel scale (0.33~arcsec/pixel) and the observing site has generally good seeing (with a median value of 1.1 arcseconds) that resulted in better resolution than other existing wide-field, narrow band surveys.  Secondly, IPHAS is generally deeper, covering point sources with r' magnitudes down to 19.5-20 \citep{Solares2008}, and also extended emission with a H$\alpha$ detection limit down to
2.5$\times$10$^{-16}$ erg cm$^{-2}$ s$^{-1}$ arcsec$^{-2}$ at full spatial resolution and $\sim$10$^{-17}$ erg cm$^{-2}$ s$^{-1}$ arcsec$^{-2}$ with a 5" binning \citep{Corradi2005}. The resolution and the sensitivity offered by IPHAS make it an ideal tool for the detection of emission nebulae of all kinds.  IPHAS is a survey sensitive to very low surface brightness nebulae and our group has been able to detect individual, morphologically exceptional PNe \citep[and we can also cite the external work by \citealt{Hsia2014}]{Mampaso2006,Wesson2008,Corradi2011,Viironen2011} , PNe interacting with the ISM  \citep{Wareing2006, Sabin2010,Sabin2012}, symbiotic stars \citep{Corradi2008,Corradi2010}, proplyd-like objects \citep{Wright2012} and new Galactic supernova remnants \citep{Sabin2013}. Crucially though IPHAS has the sensitivity  to reveal many new PNe belonging to the faint end of the PNe luminosity function and  sample more of the evolved PN population previously unavailable for study in the Northern Galactic plane. Furthermore, the high resolution and sensitivity of IPHAS allows the discovery of new morphological structures including extremely faint  Asymptotic Giant Branch (AGB) haloes around some known PNe. Finally, IPHAS is able to detect PNe through the more extinguished regions of Northern plane due to the longer narrow-band H$\alpha$ wavelength. Such PNe would not have been optically detected in [OIII] or broad-band optical filters.

\section{Detection process}

\subsection{Mosaicking}

As described by \citet{Sabin2008}, the search for new ionised nebulae is performed on IPHAS image mosaics based on two scales of binned data. This pragmatic approach was adopted as careful visual scrutiny of the 0.3~arcsecond/pixel full resolution data would have been too time consuming given the scale of the survey. Pixel binning has the significant advantage of making coherent, low surface brightness features easier to detect. Of course once any candidate nebulae has been found it can be subsequently examined at full resolution for further confirmation and examination.\\
Mosaicked $H\alpha - r$ difference maps are generated for a set of pre-defined $2^\circ \times 2^\circ$ regions on the sky, following a two-step process. First, a pair of $H\alpha$ and $r$ band exposures is selected for each IPHAS survey field together with a corresponding separate pair for the ``offset'' fields used in IPHAS to fill the inter-chip CCD gaps on the Wide Field Camera. We use the data quality control parameters as stored in a PostgreSQL database generated automatically from the FITS headers of the existing IPHAS pipeline object catalogues (e.g. \citealt{Irwin2001}; \citealt{Drew2005}) to select survey field exposures  taken in good observing conditions. These constraints are currently: sky brightness $< 2400\
{\rm ADU}$; median image ellipticity $< 0.3$; seeing $< 2.0$~arcseconds; $5 \sigma$ magnitude limit $> 18.0$; astrometric fit rms $< 0.75$~arcseconds).
Where multiple images are available, the one with the best calculated limiting magnitude is used.
All pairs of $H\alpha$ and $r$ images comprising the $2^\circ \times2^\circ$ tile are then subtracted, storing one subtracted image per
CCD (four per telescope pointing, and eight per IPHAS field including the ``offset'' fields).

We use a simple pixel-by-pixel subtraction method based on using an accurately assigned image world coordinate system (WCS) to re-bin the two images onto the same pixel coordinate system using bilinear interpolation.  The object catalogues are used to refine the frame-to-frame transformation by fitting for a standard $6$-coefficient linear plate solution. An accurate astrometric registration is critical for difference imaging to avoid introducing additional, unwanted artefacts into the images.  The confidence maps (e.g. \citealt{Irwin2001}) are used to flag bad pixels and other low-confidence regions of the image.  We then derive the median sky background on the subtracted images, and subtract this (constant) offset from the pixel values to remove the effects of any varying difference in sky background between the $H\alpha$ and $r$ images in the mosaic.

This subtraction method frequently introduces image artefacts when the PSF match between the image pair is poor (e.g. if the seeing was substantially different for the two images despite then being taken consecutively).  Although we have implemented an adaptive kernel method based on that of \citet{Alard1998} and \citet{Alard2000},  the computational cost of this technique was prohibitively expensive given the size of the IPHAS data-set and the number of mosaics which must be processed ($\sim 1500$), so it has not been used in practice for this project. However, it is made available via the Wide Field Survey interface\footnote{http://www.ast.cam.ac.uk/~wfcsur/data/dqc/} to allow individual
fields to be processed in this fashion if required for other projects/purposes. The presence of image artefacts, although ubiquitous, does not prevent the recognition and discovery of resolved nebulae even if they are of exceptionally low surface brightness as they have the same general form and character.

Finally, the subtracted images of all the CCDs in each survey field are combined into a full mosaic for that field.
The resulting images are initially binned by $5\times 5$ pixels in each dimension that corresponds to $\sim$ 1.7 arcseconds/pixel. This is only a little poorer than the median site seeing and is similar to the native seeing of the equivalent SHS survey in the south. The human visual system is quite immune to the effects of undersampling the PSF.  This binning is done as before, by interpolating the input images onto the output map by using the WCS information stored in the FITS headers, and flagging bad pixels using the confidence maps.  Since each pixel of the output image is typically covered by $\gg 1$ input pixel, this effectively removes a large fraction of the CCD artefacts present in the full resolution data quotient images. A second coarser binning factor of 15 pixels $\times$ 15 pixels (equivalent resolution of $\sim$5 arcseconds/pixel) was also employed to capture larger-scale diffuse nebulosity more easily (see below).

\begin{figure*}
\begin{center}
{\includegraphics[height=7cm]{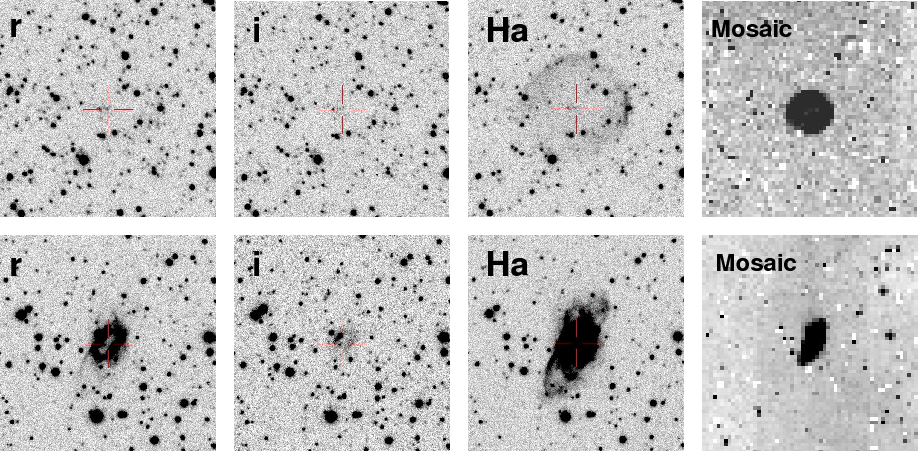}}
\caption{\label{IP1905} The top row shows the faint, new PN IPHASX J190512.4+161347 in the r', i' and H$\alpha$ filter (FoV=2 arcmin and resolution 0.33 arceconds/pixel) and the region of the mosaic where it was first clearly spotted (FoV= 5 arcmin and resolution 4.95 arcseconds/pixel). For comparison purposes we show in the second row the IPHAS images of the known and brighter PN M1-75 (FoV=2 arcmin and resolution 0.33 arcseconds/pixel) and also the region where it lies in the 15 pixels binned mosaic (FoV= 5 arcmin and resolution 4.95 arcseconds/pixel). The advantage in using the binned mosaicking technique to unveil low surface brightness nebulae such as those found with IPHAS is clear. In both cases North is up and East to the left.}
\end{center}
\end{figure*}

\subsection{Visual search}

Searches for extended but discrete ionised nebulae within the binned mosaics were done via careful and painstaking visual scrutiny of each processed IPHAS survey field. 
The most interesting tool, for our purposes, is the setting of two mosaic image scale factors or binning levels at 15 pixels $\times$ 15 pixels (resolution of $\sim$5 arcseconds/pixel) and at  the original 5 pixels $\times$ 5 pixels (resolution of $\sim$1.7 arcseconds/pixel) as mentioned above. The cruder binning level helps to resolve low surface brightness objects (Fig.\ref{IP1905}-Top row) down to the IPHAS limit \citep{Sabin2010} and to
accentuate the contours/shape of such extended nebulae (this is particularly useful to see the full extent of an outflow  for example). The second, finer binning is used to detect intermediate size nebulae, i.e those smaller than $\sim$15-20 arcseconds in diameter which constitutes a decent fraction of the total discoveries. These objects are too small to be seen with the crudest binning level (they are not resolved, and thus not distinguishable). They are also generally too faint to be detected via point source $H\alpha - r$ colour photometry as used by \citep{Viironen2009b} to uncover unresolved PN candidates as they drop out of the r-band completely. Some of the smallest new IPHAS PN candidates from $\sim$3-4 arcseconds up to $\sim$10-15 arcseconds were discovered using a semi-automated detection method. The technique, also used in the southern MASH survey \citep{Miszalski2008}, relies upon quotient imaging and combining H$\alpha$, r'  and i' images into an RGB composite.  Some smaller extended nebulae can be detected and we
can also discriminate them from ``late-type star'' contaminants which can mimic emitters in H$\alpha$ - r'  due to their increasingly strong TiO bands whereas in the i' band such stars are brighter.


Careful examination of the mosaics was done according to the following procedure. First, all the detected ionised nebulae were noted regardless of whether they are already known in the literature. This helped to ensure the completeness of the search. The large overlap between two adjacent mosaics is also a guarantee that no real object is easily missed. Candidate selection was then refined based on the objects' morphology (which is particularly valid for PNe) and on whether the nebulae are detached or isolated,  i.e. they are not merely part of a larger nebular conglomerate such as a large supernova remnant, HII region or other large-scale meandering, diffuse nebulae. Environmental considerations were also assessed such that objects in areas of high extinction, low stellar number density and general HII regions were considered to be likely of young provenance (but the candidates were not discarded). Finally, objects unknown in Simbad and Vizier \footnote{http://cdsportal.u-strasbg.fr/} were separated from  known sources.
Despite the great care taken while performing the extensive search  and in creating the mosaics we were still confronted with three main problems which complicated the selection process:
\begin{itemize}[noitemsep]

\item
The presence of image artefacts. These can be very easily mistaken for real objects under certain binning conditions. Such artefacts generally result from instrumental effects (e.g. CCD edge reflections) or are generated during the H$\alpha$-r image subtraction due to psf mismatch. A simple way to check the veracity of a candidate is to look at the un-binned images (0.33 arseconds/pixel) and also in the native H$\alpha$ band as each IPHAS field is observed at least twice. An artefact is highly unlikely to be repeated in all frames and so can be easily eliminated.
\item
Based on some independent comparisons from the DSH team it is clear that our selection process is not sensitive to a group of true candidate objects  i.e. those extending over a couple of pixels in the 15$\times$15 mosaics (which were also not picked up during the semi-automated search of 5$\times$5 pixel mosaics). Some of these are likely to be genuine PNe. The only solution is to perform a scan of the full resolution mosaics. This massive work will be undertaken in the future. 
\item
The bad quality of some mosaics early in the search process due to the need to initially use frames taken in non-optimal weather conditions. Many of these frames have now been replaced with higher quality equivalents as the survey nears completion and it would be worthwhile revisiting the affected fields in the future.
\end{itemize}
 
As this is the first in a series of papers of new IPHAS PNe we mainly present the visual detection and spectroscopic confirmation from the hundreds of PN candidates in the two hour right ascension range RA=18  to 19~hours that has been examined most thoroughly. However,  we have also included a limited number of additional confirmed PNe in the RA range from 20~hours through to 06h30m acquired and confirmed due to the vagaries of telescope time allocations for the spectroscopic follow-up. All the newly discovered objects with IPHAS, including the new PNe presented here, are named according the International Astronomical Union convention: IPHASX JHHMMSS.s+DDMMSS (exclusively used for extended sources).

\section{Follow-up Spectroscopy}

\begin{table*}
\begin{center}
\addtolength{\tabcolsep}{-4pt}
\caption[Observing logs of follow-up spectroscopic program]{\label{telescopes} Details of the spectroscopic follow-up performed on a variety of telescopes.}
\begin{tabular}{|l|l|c|c|c|c|l|c|}
\hline
Telescope     &  Instrument   &  Grating(s)   &    Wavelength Coverage   &  Dispersion  & Resolution  &  Run dates     & Observers \\
              &               &               &            \AA             &   \AA /pix    &   \AA      &   yyyy-mm-dd &  \\
\hline 
\hline

     WHT-4.2m & ISIS  &   R300B        &   3500-6100    &  0.86         &    3 	 &    2004-09-23:27 & -$^{\dagger}$  \\
       -     &   -    &   R158R        &   6000-10500    &  1.82         &    6 	 &   & \\
	SPM-2.1m & B\&Ch &  600 l/mm      &   3590-5650    &    2          &    3.2  &	     2005-12-03:04 & KV\\
        -    &  -     &                &   5310-7450    &    2          &    3.2  &    &\\
	INT-2.5m & IDS	 &   R300V        &   3030-8960    &   1.8         &    5 	 &    2006-06-14:15 & RC,LS\\
	INT-2.5m & IDS	 &   R300V        &   3030-8960    &   1.8         &    5 	 &    2006-08-01 & RC\\
	WHT-4.2m & ISIS  &   R300B        &   3040-5460    &  0.86         &    3 	 &    2006-08-23:24  & MB,LS\\
        -    &   -    &   R158R        &   5140-9700    &  1.82         &    6 	 &     &\\
	INT-2.5m & IDS 	 &   R300V        &   3030-8960    &   1.8         &    5    &     2006-08-28:29  & MB,LS\\
	INT-2.5m & IDS 	 &   R300V        &   3030-8960    &   1.8         &    5    &     2006-09-08 & RG,RC \\
	SPM-2.1m & B\&Ch & 400 l/mm       &   4330-7530    &    5          &    4.8  &	     2007-01-23:25 & KV\\
	WHT-4.2m & ISIS  &   R300B        &   3040-5460    &  0.86         &    3 	 &    2007-07-16:17 & KV,LS\\
       -     &   -    &   R158R        &   5140-9700    &  1.82         &    6 	 &   & \\
          MSSO-2.3m & DBS$^{(2)}$	 & 300B           &   3630-7390    &   1.9         &    6    &     2007-09-05 & BM\\
	INT-2.5m & IDS	 &   R300V        &   3030-8960    &   1.8         &    5 	 &    2007-08-02 & -$^{\dagger}$ \\
	INT-2.5m & IDS	 &   R300V        &   3030-8960    &   1.8         &    5 	 &    2008-06-27:29 & RC \\
	KPNO & GoldCam	 & 240            &   3980-7020    & 1.52          &   4.1   &     2009-08-12:18 & LG,KV\\
    GTC-10.2m & Osiris &   R1000B     &    3600-7760   &   2.12        &   2.15  & 2009-11-09 & -$^{\dagger}$ \\
	SPM-2.1m & B\&Ch & 400 l/mm       &   4330-7530    &    5          &    4.8  & 		 2010-06-02-03 & LS\\
	SPM-2.1m & B\&Ch & 400 l/mm       &   4330-7530    &    5          &    4.8  &		 2010-07-09:12 & LS\\
	SPM-2.1m & B\&Ch & 400 l/mm       &   4330-7530    &    5          &    4.8  &		 2010-09-17 &LS \\
	OSN-1.5m $^{(3)}$ & Albireo 	&   R600      &  3650-7180     &    3.49       &   6.5   &     2010-09-18  & MG\\
	SPM-2.1m & B\&Ch & 400 l/mm       &   4330-7530    &    5          &    4.8  &		 2010-09-20 & LS\\
	GTC-10.2m & Osiris 	&   R1000B    &    3600-7760   &   2.12        &   2.15  &	 2011-04-13 & -$^{\dagger}$\\
	SPM-2.1m & B\&Ch & 400 l/mm       &   4330-7530    &    5          &    4.8  &		 2011-05-04:06 & LS \\ 
	GTC-10.2m & Osiris &   R1000B     &    3600-7760   &   2.12        &   2.15  &		 2011-06-07:28 & -$^{\dagger}$\\
	SAAO-1.9m & CCDSPEC	&  Grating 7  &   3000-7200    & 210 (\AA /mm) &   5	 & 2011-07-01:04  & QP,MS\\
	SAAO-1.9m & CCDSPEC	&  Grating 7  &   3000-7200    & 210 (\AA /mm) &   5	 & 2011-07-05:11  & AZ\\
	MSSSO-2.3m & WiFeS	&  B7000      &   4180-5580    &               &         &	 2011-07-01:05 & LS,LG \\
	     -      &   -     &  R7000      &   5290-7060    &               &         &	 &  \\
	SPM-2.1m & B\&Ch & 400 l/mm       &   4330-7530    &    5          &    4.8  &		 2011-09-22:26 & LS\\
	OSN-1.5m & Albireo  &   R600      &  3650-7180     &    3.49       &   6.5   &	     2011-10-04:10 & MG\\
	OSN-1.5m & Albireo 	 &   R600     &  3650-7180     &    3.49       &   6.5   &     2011-11-23:28 & MG\\
	SPM-2.1m & B\&Ch & 400 l/mm       &   4330-7530    &    5          &    4.8  &        2012-04-12:15 & IB\\
	OSN-1.5m & Albireo 	 &   R600     &  3650-7180     &    3.49       &   6.5   &     2012-05-17:25  & MG\\
	SPM-2.1m & B\&Ch & 400 l/mm       &   4330-7530    &    5          &    4.8  &		 2013-02-10:11  & LS\\
	SPM-2.1m & B\&Ch & 400 l/mm       &   4330-7530    &    5          &    4.8  &		 2013-05-07:09  & LS\\
 \hline
 \hline
 \end{tabular}
\begin{minipage}[!t]{8cm}
\hspace{-2cm}$^{\dagger}$ Observations performed in service mode.
\end{minipage}
\end{center}
\end{table*}

\subsection{Observations}
The large number of PN candidates discovered  required the use of several telescopes worldwide for an efficient spectroscopic follow-up program as given in Table \ref{telescopes}. On the whole 2-m class telescopes were sufficient to provide the necessary spectroscopic confirmation which is all that is required at this stage. Detailed abundance, kinematics and photo-ioinisation studies for selected high-interest candidates will be performed later but has already been done for a few selected objects including the so-called 'Necklace' PN (e.g \citealt{Mampaso2006,Corradi2011,Viironen2011}).

\begin{figure*}
\begin{center}
{\includegraphics[height=12cm]{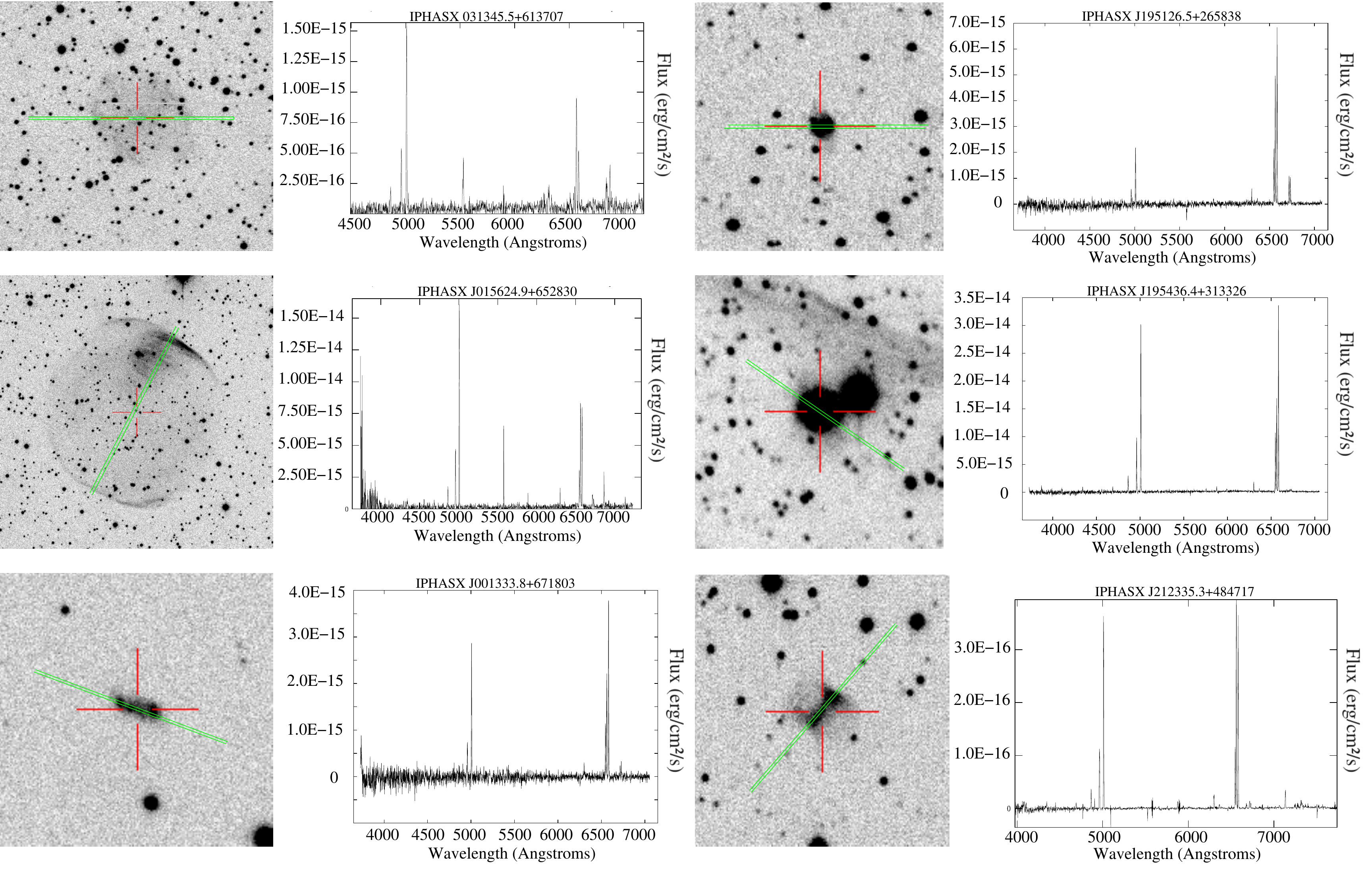}}
\caption{\label{spectra} Example images and spectra of selected, newly discovered IPHAS PNe. For each object we indicate its official name followed by the field of view (in arcminutes), the binning of the image (in pixels) and the telescope where the confirmatory spectroscopy was performed such as (FoV,Bin,Telescope).  North is up and East is left. We show a wide variety of morphologies and peak fluxes ranging from 10$^{-14}$ to 10$^{-16}$ erg/cm$^{2}$/s. From top to bottom in the left column we present successively: IPHASX J031345.5+613707 (5,3,SPM) which is relatively faint and round, IPHASX J015624.9+652830 (5,2,SPM) a large (radius $\sim$100 arcseconds) and well defined transparent bubble and IPHASX J001333.8+671803 (1,1,KPNO) which has an irregular and knotty structure. In the right column we present: IPHASX J195126.5+265838 (1,1,KPNO) is a small (3.8$\times$3.4 arcseconds semi axis) new bipolar PN, IPHASX J195436.4+313326 (1,1,KPNO) is a ring PN and IPHASX J212335.3+484717 (1,1,GTC) is a faint bipolar (butterfly type) PN. All the PNe shown here are classified as True PNe (see text). The green lines indicate the position of the slit. }
\end{center}
\end{figure*}

During the initial stages of our spectroscopic follow-up we used the Intermediate Dispersion Spectrograph (IDS) on the 2.5m INT and the spectrograph ISIS on the 4.2 m William Herschel Telescope (WHT) located at the Observatorio del Roque de los Muchachos on La Palma in the Canary Islands. Many subsequent observations were performed with the 2.1m San Pedro Martir Telescope (SPM) in Mexico with its Boller $\&$ Chivens spectrograph and  with the 1.5m telescope associated to the ALBIREO spectrograph at the Observatory of Sierra Nevada (OSN) in Spain. One valuable run was performed  with the 2.1m telescope at the Kitt Peak National Observatory (KPNO) and a few observations were made with the 10.4m Gran Telescopio Canarias (GTC) and OSIRIS spectrograph also on La Palma. Other observations for those IPHAS PNe candidates also accessible from the south were observed with the ANU 2.3 m Telescope at Siding Spring Observatory in Australia, initially with its Dual Beam Spectrograph (DBS) and later with the Wide Field Spectrograph (WiFeS; \citealt{Dopita2007}) IFU and also with the 1.9m  Radcliffe Telescope at  the South African Astronomical Observatory (SAAO) at Sutherland. Details of  each telescope run together with the configurations of the associated spectrographs are listed in Table \ref{telescopes}. In total more than 500 spectroscopic observations were carried out mostly between 2006 and 2014 (see some examples in Fig.\ref{spectra}). The general faintness of the IPHAS targets also necessitated  long exposure times typically ranging from 900~seconds to 2$\times$1800~seconds, some of them repeated once or twice depending on the weather conditions or result obtained. Due to some modest overlapping with the coverage of the MASH survey in the southern Galactic plane as well as with the area searched by the DSH community some of our targets were common to both of these independent studies. In cases where the basic information on a common source has already been published by one of the two aforementioned groups, we only present the IPHAS detected PNe for which a clear classification and confirmation has not previously been clearly determined although such PNe are indicated in the main catalogue table.

The majority of the new IPHAS PNe uncovered have low surface brightness which is either intrinsic (evolved PNe),  due to heavy intervening extinction in the plane and/or is because the candidates are located at large distance. Consequently, for many nebulae only the strongest few emission lines could be detected. Fortunately,  these are also generally the minimum necessary to allow object identification following the usual diagnostic diagrams (i.e. H$\alpha$, [NII]$\lambda\lambda$6548,6583\AA; [SII]$\lambda\lambda$6717,6731\AA~ and [OIII]$\lambda$4959,5007\AA). \citet{Parker2012} showed that the most obscured PNe discovered from optical images have a $V$-band extinction in excess of 10 magnitudes. For such objects the H$\beta$ line can be very difficult to detect spectroscopically. For some of the more obscured IPHAS PNe, this line was undetected so we were not able to derive an extinction value for them based on the Balmer decrement technique, although with an estimation of the upper limit of the H$\beta$ flux we can derive a lower limit of cH$\beta$. Other basic physical properties such as  the electron density (n$_{e}$) can come from the observed [SII] line ratio available for many of our existing spectra. However, other parameters such as the electron temperature (T$_{e}$) and abundances for individual PNe require far higher S/N spectra to detect the faint diagnostic lines required. Such determinations will be reported in forthcoming papers and in the MASPN database as available spectra allow.

The spectral data reduction including bias removal, flat-fielding, cosmic ray removal, wavelength and flux calibration and 1-D spectrum extraction was generally performed with  standard long-slit IRAF routines \citep{Valdes1986} with some specific, minor differences according to the various data formats, calibration lamps used etc. In the case of the WiFeS IFU data specially developed reduction pipelines were used (e.g. \citealt{Dopita2010}). Given the large number of objects, for some suitable data sets the automatic analysis software {\sc ANNEB}, developed by \citet{Olguin2011}, which includes the {\it{Nebular}} packages of IRAF/STSDAS \citep{Shaw1995}, could be employed on the reduced 1-D spectra. It conveniently provides a set of useful information such as the emission line identifications, the extinction c(H$\beta$) according to the chosen extinction law and for any Balmer line, the dereddened fluxes, physical parameters (T$_{e}$,n$_{e}$) where the spectra S/N and detections allow and crucially, the most important line ratios for use in the diagnostic diagrams. The extinction correction applied to our spectra was performed using the Fitzpatrick and Massa extinction curve \citep{Fitzpatrick2007} for $R_{V}$ = 3.1. \\

\subsection{Spectroscopic object identification and confirmation}

An accurate determination of the PN nature of many of the IPHAS candidate nebulae is challenging for two reasons. First, there are a large number of other morphologically similar nebular mimics such as Wolf-Rayet (WR) shells, symmetric HII regions, symbiotic stars, Herbig Haro objects, CVs with highly collimated bipolar outflows, reflection nebulae and shell supernova remnants (e.g. \citealt{Sabin2013}) which can all be confused with true PNe. Secondly, the low surface brightness of the sources generally leads to  low S/N emission-line spectra  and  small numbers of diagnostic identified lines.  In several cases similar emission-line ratios are found for totally different types of equally faint astrophysical objects. In these cases the use of supplementary multi-wavelength imaging data, now increasingly available, has proven extremely valuable (see later). The removal of contaminants  (i.e. non-PNe) is a critical and important step in delivering a catalogue of IPHAS PNe of high integrity. The issue of PN mimics has been thoroughly discussed by \citet{parker2006a} and \citet{Frew10} who also provide robust tests and other indicators of PN veracity that were adopted here.

The new IPHAS PNe were confirmed through using a combination of new, multi-wavelength imagery from the UV through the optical, NIR, MIR and radio and improved sets of optical diagnostic diagrams by \citet{Frew10} and more recently \citet{Sabin2013} and \citet[b]{Frew2014b}. These diagrams are based on improved determinations of the traditional \citet{Sabbadin77} and \citet{Baldwin1981} emission line ratio plots. Contrary to older versions of such diagrams (from \citealt{Riesgo06} for example) these new sets have the advantage of including more robustly determined line measurements from the very best data and several more types of well identified ionised nebulae which occupy distinct zones in these diagrams. These newly established plots offer more reliable constraints for an accurate determination of the nature of the investigated source. In order to keep some coherency between the diverse PNe catalogues we chose to adopt the flags used by the MASH survey of \citet{parker2006Catb} to estimate the quality of the identification. Those flags are:\\
 {\it True ``T"}: To indicate a spectroscopically and morphologically well defined PN across perhaps several multi-wavelength images.\\
 {\it Likely ``L"}: To indicate a not completely conclusive spectroscopic and/or morphological identification though a PN ID is likely.\\
 {\it Possible ``P"}: To indicate a non conclusive identification due to the insufficient quality or ambiguous nature of spectroscopic and/or morphological data. Such objects cannot yet be ruled out as PNe but the current data  could also support identification as  several other possible astrophysical objects  such as HII regions.

\section{IPHAS Catalogue of newly discovered PNe}
As a result of our spectroscopic, multi-wavelength and morphological investigation we were able to identify 159 candidate PNe. Following our adopted criteria we have classified 113 True, 26 Likely and 20 Possible PNe. The complete list of PNe can be found at the end of this paper (Table \ref{catalogue}) as well as the catalogue of images (from Fig. A1 to Fig. A40). The different column entries for each object in the catalogue are described below.\\
\noindent{\it Flag}: True ``T", Likely ``L" and Possible ``P" as mentioned above.\\
{\it IAU designation}: Based on the Galactic coordinates and set to fit the general galactic PN nomenclature i.e. PN Glll.l+bb.b \\
{\it IPHAS designation}: IPHASX JHHMMSS.s+DDMMSS as described following the adopted IAU convention and based on the measured RA/DEC of the PNe.\\
{\it RA, DEC J2000 equatorial coordinates}: The coordinates were defined based on the best estimate of the geometric centre of each object from the IPHAS H$\alpha$ or quotient image,  or in the case of highly asymmetric or one sided nebulae, based on the middle of this arc or zone. When a candidate central star (CSPN) was identified its location was not adopted as the position of the PN. This is because in many cases of evolved and large angular size PNe the likely CSPN is clearly not centrally located. A separate list of unequivocally identified CSPN with positions, including those which exhibit [WR] or WELS or PG1159 spectral signatures, will be provided in a subsequent publication.\\
{\it Galactic coordinates l,b}: They are based on the RA,DEC defined previously.\\
{\it Major and where relevant,  minor axis dimensions in arcseconds}: The measurement of PNe size was done from the 120~second exposure H$\alpha$+[NII] images so we are limited in description of the exact extent of the nebulae.\\
{\it Morphological classification} Assigned following the identical scheme used for the sister MASH survey (see below).\\
{\it Telescope and date for first spectroscopic confirmation}: A two letter code is used to identify each telescope used for spectroscopic confirmations as follows: WH - WHT 4.2m; IN - INT 2.5m ; SM - San Pedro Martir 2m;  KP - KPNO 2m ; GC - Grantecan 10m, OS - OSN 1.5m, MS - ANU 2.3m with DBS, WI - ANU 2.3m with WiFeS; SA - SAAO 1.9m.

\subsection{Comments on table column values}
Although our search focused on extended objects, i.e. with a lower limit on the total size of 3~arcseconds in the 5$\times$5 binned pixel data, we also found  quasi-stellar (essentially unresolved) PNe which were still picked up  through the  binned images. Some objects are not well defined, e.g. when only a rim is seen. In such cases the exact size will be slightly inaccurate as we rely on the best determined geometric centre of the structure encompassing the detected nebulae to establish the major axis. A comment to this effect is made in the notes accompanying such cases in the MASPN database (e.g. IPHASX J194240.5+275109). This of course can introduce a (large) bias in the assumption of the optical size of the nebulae, but with no other high quality narrow-band optical data available,  we adopted this scheme for the time being.  Hence, the positional accuracy of some large, irregular nebulae (such as IPHASX J192534.9+200334), of certain examples  of bipolar nebulae (such as IPHASX J193718.6+202102) and some PNe with a possible ISM/interaction (e.g. IPHASX J195358.2+312120) could be off by tens-of-arcseconds compared to their actual centroids. Recall each observed nebulae is merely the 2-D projection of a 3-D source so we do not pretend to present a position for the projected physical object's centre. Deeper, high resolution images may be needed to establish the best coordinates to use for some of the IPHAS PNe listed here. Nevertheless, we have endeavoured to provide the best coordinates that the current IPHAS images can provide and also a best estimate of the angular extent to assist any observer in  locating these PNe for further studies. Of course these may still not be sufficiently deep to pick up any faint, external AGB haloes or the extremely faint lobes of highly evolved bipolar PNe.

In order to facilitate easy inter-comparison with both the earlier MASH survey in the south and the new, full-scale catalogue of all known PNe recently put together by \citet{Parker2014} and Bojicic et al. (in preparation) we adopted the ``ERBIAS"  morphological classifiers to indicate  Elliptical , Round, Bipolar, Irregular, Asymmetric or quasi-Stellar (unresolved or barely resolved) PNe. The additional sub-classifiers of ``amprs"  were also used where evident where a  one sided enhancement/asymmetries denoted with ``a", multiple shells or external structure  as ``m", point symmetry ``p", well defined ring structure or annulus ``r" and resolved, internal structure as ``s". We emphasize that this initial morphological classification is based on the short H$\alpha$+[NII] exposure IPHAS image. Further investigations based on IFU data coupled with morpho-kinematical modelling will be needed to assert the ``true" shape and geometry of many of these newly discovered PNe. Such detailed kinematical study would help, for example to disentangle cases of apparently round annular PNe that are actually face-on bipolar PNe (e.g. \citealt{Jones2012}), while longer exposure times will likely reduce the number of apparently irregular PNe by revealing an overall more coherent structure.

\begin{figure*}
\vspace{-5cm}
\hspace{1cm}
{\includegraphics[height=4.5in]{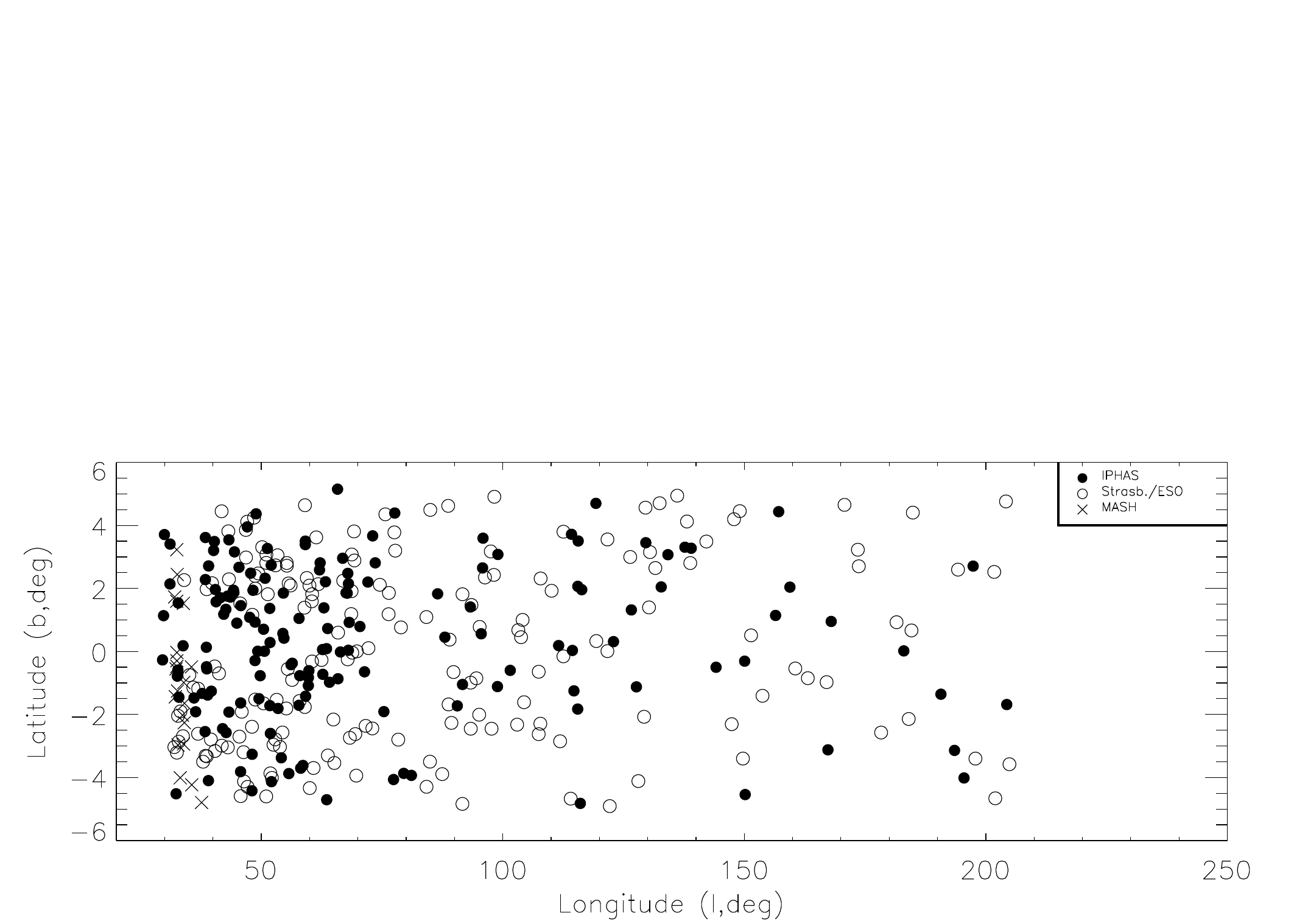}} 
{\includegraphics[height=4.5in]{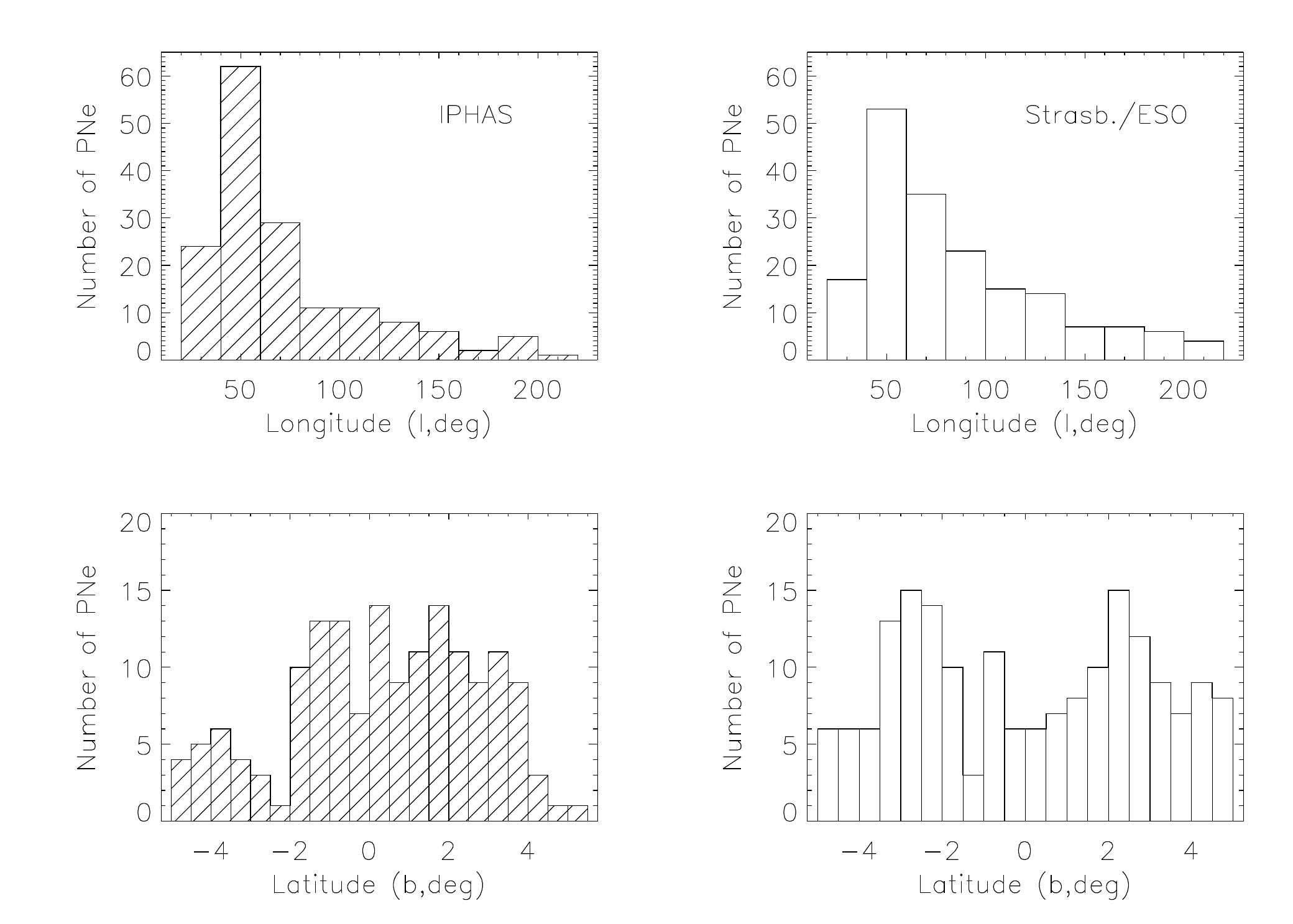}} 
\caption{\label{Distribution} Top panel: Galactic distribution of the new IPHAS PNe (filled circles) shown against the sample from the Strasbourg/ESO catalogue (open circles) and the MASH catalogues (crosses) in the same area. Bottom panels: same as before with a histogram representation of the longitude and latitude distribution of the IPHAS PNe on the left and the Strasbourg/ESO data on the right. These plots underline the gain in terms of coverage and gap filling obtained with our new IPHAS PNe catalogue.}
\end{figure*}

\begin{figure*}
{\includegraphics[height=15cm]{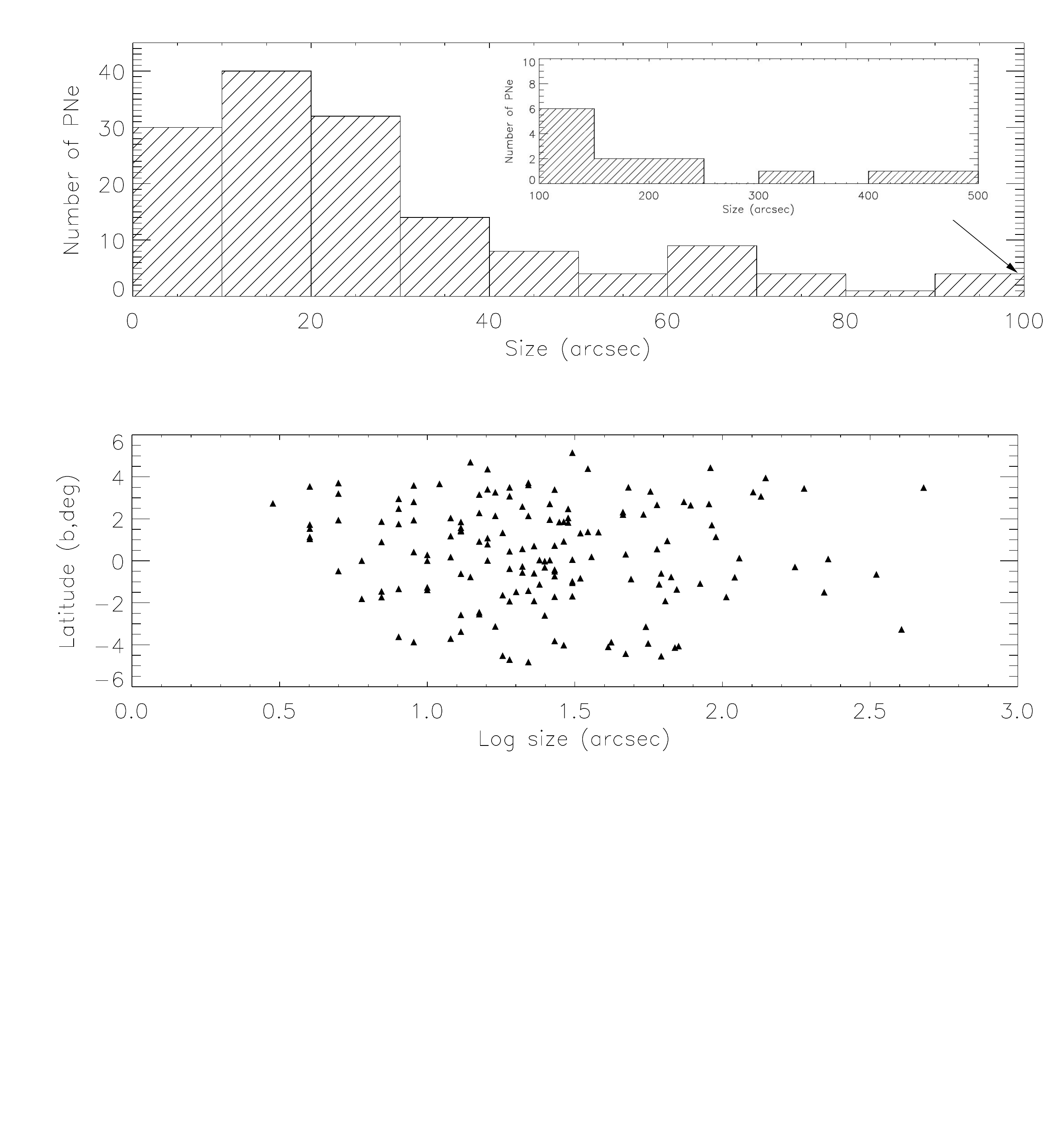}} 
\vspace{-5cm}\caption{\label{Size} Estimated angular size distribution of the IPHAS PNe sample in the Northern Galactic plane. We observe a large scatter from compact barely resolved objects  to those large objects 5 or more arcminutes in diameter even at low Galactic latitudes .}
\end{figure*}

\section{Preliminary Statistical analysis}
\subsection{Galactic distribution of new IPHAS PNe}

Fig.\ref{Distribution}-Top shows the general distribution profile of new IPHAS PNe found in the longitude range $l^{\circ}$=29--204 degrees. We compared our data with earlier surveys such as the Strasbourg/ESO survey \citep{Acker92,Acker96} and to a lesser extent with the overlapping zone of the southern MASH-I\&II surveys \citep{parker2006a,Miszalski2008} both restricted to the same area. Our new IPHAS sample increases the number of known PNe in this region of the Northern Galactic plane by nearly doubling it. Indeed, 181 PNe were previously found with the Strasbourg/ESO survey and 21 with the MASH survey in the region overlapping with the IPHAS survey area. As expected the detection rate declines when we move to larger Galactic longitudes but it is important to note that we have uncovered  additional new PNe towards the Galactic Anticenter region i.e. $l^{\circ}$ $>$ 115$^{\circ}$ (Fig.\ref{Distribution}-Middle). These new objects will be of high importance for the estimation of the metallicity gradient which is one of the major issues that PN studies can help address as they provide an easily detectable target population that can be traced to great distance and whose emission lines can provide decent abundance estimates given decent S/N \citep{Viironen2011,Henry2010,Costa2004}. We also increase the number of detections close to the Galactic Plane (Fig.\ref{Distribution}-Bottom) including in more heavily obscured zones as can be seen by comparing our new IPHAS PNe distribution to the extinction map by \citet{Solares2008}. This trend is particularly well seen at -0.5$^{\circ}$ $\leq$  b $\leq$ 0.5$^{\circ}$ where the detection rate is multiplied by a factor $\sim$2. \\ 

\begin{figure*}
{\includegraphics[height=4cm]{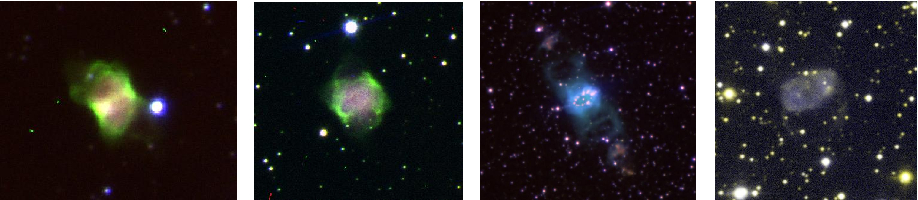}} 
\caption{\label{Deep_image} IPHAS image gallery of bipolar PNe with from left to right: IPHASX J194940.9+261521, IPHASX J205527.2+390359, IPHASX J194359.5+170901 (the Necklace) and IPHASX J221118.0+552841 (with its two faint lobes). All the images were taken at the Nordic Optical Telescope with ALFOSC and an average 20 min per filter (H$\alpha$,[OIII]~5007 and [NII]~6583\AA).}
\end{figure*}

\subsection{Estimated angular sizes}
\begin{figure}
\begin{center}
{\includegraphics[height=5cm]{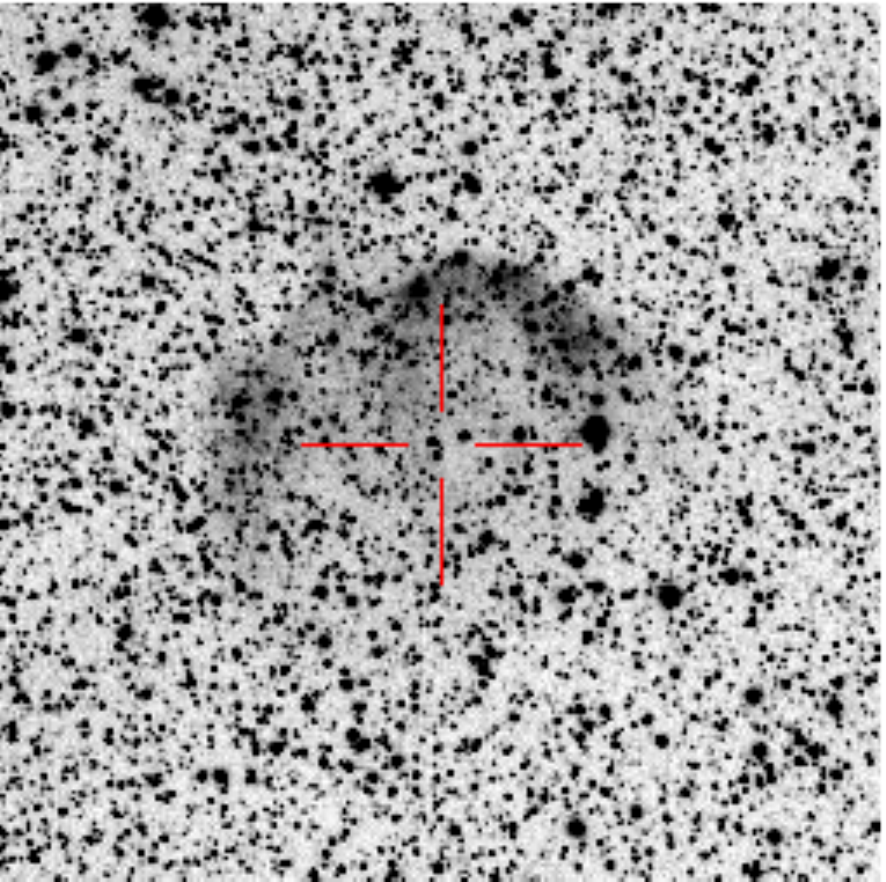}} 
\caption{\label{ISM_interac} H$\alpha$+[NII] image of IPHASX J195358.2+312120 (North is up, East on the left). This new and large PN ($\sim$3.8 arcmin size) is an example of the many objects found during the survey which only show a bright rim inferred to be coincident with an interaction with the surrounding interstellar medium. The accurate estimation of the angular sizes of such objects is not currently straightforward  due to the relatively short H$\alpha$ exposure time of 120~seconds. }
\end{center}
\end{figure}

Our investigation, which is strongly biased towards extended (nearby and/or evolved) PNe, shows a large scatter in observed angular size. We detected PNe with an average major-axis dimension of 42 arcseconds and a median of 22 arcseconds (Fig.\ref{Size}) which is comparable to the results found in the MASH survey by \citet{parker2006a} with had an average size of 51 arcseconds and also a median of 22 arcseconds as for IPHAS. Among our sample we identified a set of 13 large PNe with diameters ranging between 100 and 480 arcseconds.  The largest PNe in our catalogue has a diameter of $\simeq$8 arcmin (IPHASX J185225.8+005250). Our ability to detect such extended structures at low latitudes (Fig.\ref{Size}-bottom) underlines  both the depth reached with IPHAS and its ability to at least partially peer through dust.  Apart from their size, this group of larger PNe is spread over a wide range of different morphologies. As an example, while IPHASX J015624.9+652830 appears as a well defined circular structure (Fig.\ref{spectra}-left-2nd row), IPHASX J195358.2+312120 shows a bright rim structure probably indicating its interaction with the surrounding interstellar medium (Fig.\ref{ISM_interac}). 
The geometry of the latter PN does not allow an accurate measurement of its size and in those cases the coordinates and dimensions are derived based on the best estimate of the geometric centre from the best fitting circle or ellipse.

\subsection{Morphological classifications}
As previously mentioned the morphological classifications were made using the ``ERBIAS"  and ``aprms" classification scheme. The current IPHAS sample consists of 50 elliptical, 45 round, and 45 bipolar PNe, while 6 display an irregular morphology, 6 are asymmetric and 7 are classified as quasi-stellar (i.e. essentially unresolved) sources (Table \ref{morpho}). Even with the short exposure time of 120~seconds in H$\alpha$, we were able to observe additional or secondary morphological structures in many objects. Deeper imaging coupled in some cases with 3-D morpho-kinematic analysis will be required for a  more robust classification. We have already started an observational campaign with this aim  for some of the more interesting IPHAS PNe so far uncovered  (Fig.\ref{Deep_image}).

A large, relative fraction of PNe classified essentially as round have been found in the latitude range targeted by IPHAS. This is a potentially interesting result as difficulties in detection of such round objects is linked to their generally lower surface brightness compared to non-spherical and bipolar PNe and the likelihood of being detected in more restricted surveys in terms of depth \citep{Soker2005}. Our new data will contribute to better investigate this interesting ``group". Nevertheless the majority of these new IPHAS PNe are non-circular. 

\begin{table}
\begin{center}
\footnotesize\addtolength{\tabcolsep}{-4pt}
\caption[Morphological distribution of the new IPHAS PNe]{\label{morpho} Morphological distribution of the new IPHAS PNe.}
\begin{tabular}{|c|c|c|c|c|c|c|}
\hline 
Morphology & Elliptical & Round  & Bipolar& Irregular & Asymmetric & Quasi- \\
           &            &       &          &          &            & Stellar\\
\hline 
Total Number&  50 & 45 & 45 & 6 & 6 &7\\
\hline 
Fraction(\%) & 32 & 28 & 28 & 4& 4 & 4\\
\hline
$<|$b$|>$ $^{\circ}$& 1.89 & 2.36 & 1.96 & 0.84& 2.46 & 1.66 \\
\hline 
$<$Major axis$>$"&53 & 36 &30 &72 & 96 &--\\
\hline 
\end{tabular}
\end{center}
\end{table}

\subsection{Estimating IPHAS PNe distances} 
We are now in the process of determining distances to most of these new IPHAS PNe using the newly developed Surface-Brightness radius relation Frew et al. 2014 (submitted). These new distance estimates will help us to ascertain the evolutionary status of our objects. Meanwhile we can still infer that the large sizes associated with the generally low surface brightness PNe suggest (highly) evolved more local PNe and hence provide a new sample with which to study the end stages of the PN evolution (i.e. several PNe with ISM interactions have also been found). The extinction method described by \citet[]{Giammanco2011} and \citet[]{Sale2009}, which uses highly reliable photometric data from IPHAS, will be an additional, very useful tool to derive alternative distance estimates with which to determine the age and evolutionary stage of these IPHAS PNe.

\section{IPHAS PNe in the new online Macquarie-Strasbourg PN Database}

\begin{figure*}
{\includegraphics[width=\textwidth]{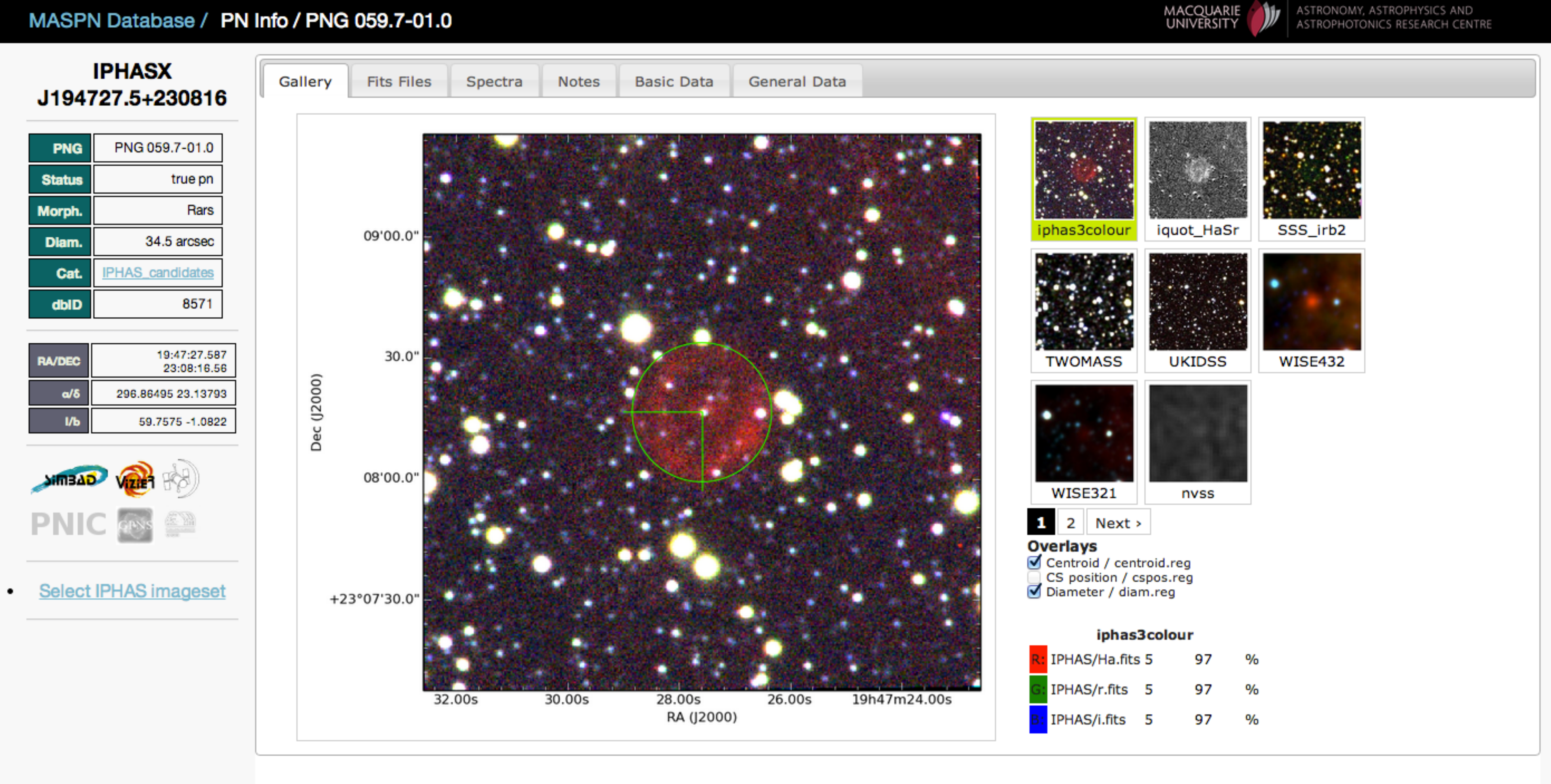}} 
{\includegraphics[width=\textwidth]{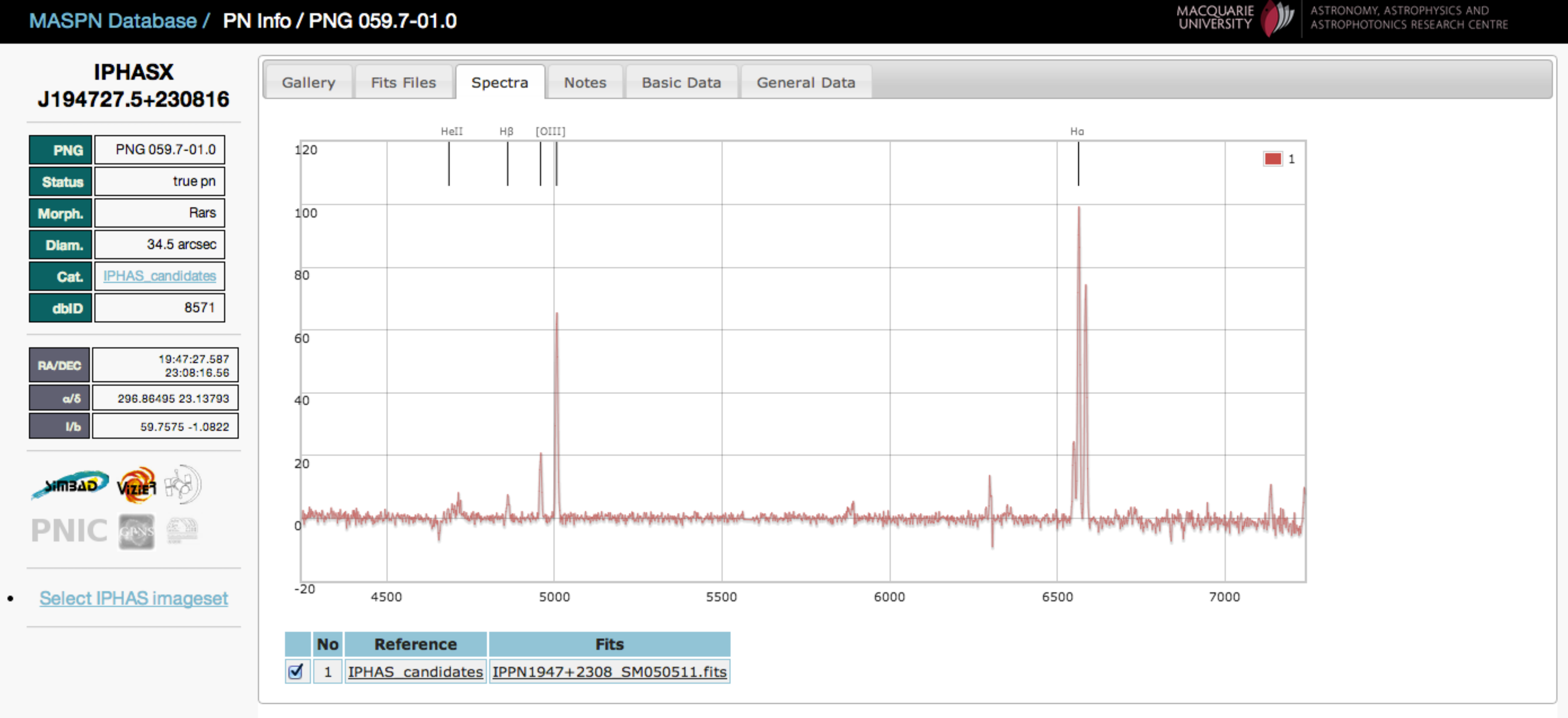}}   
\caption{\label{MASPN2} Multi-wavelength images of IPHASX J194727.5+230816 (Top) with its associated optical spectrum (Bottom).}
\end{figure*}

\begin{figure*}
{\includegraphics[height=12cm]{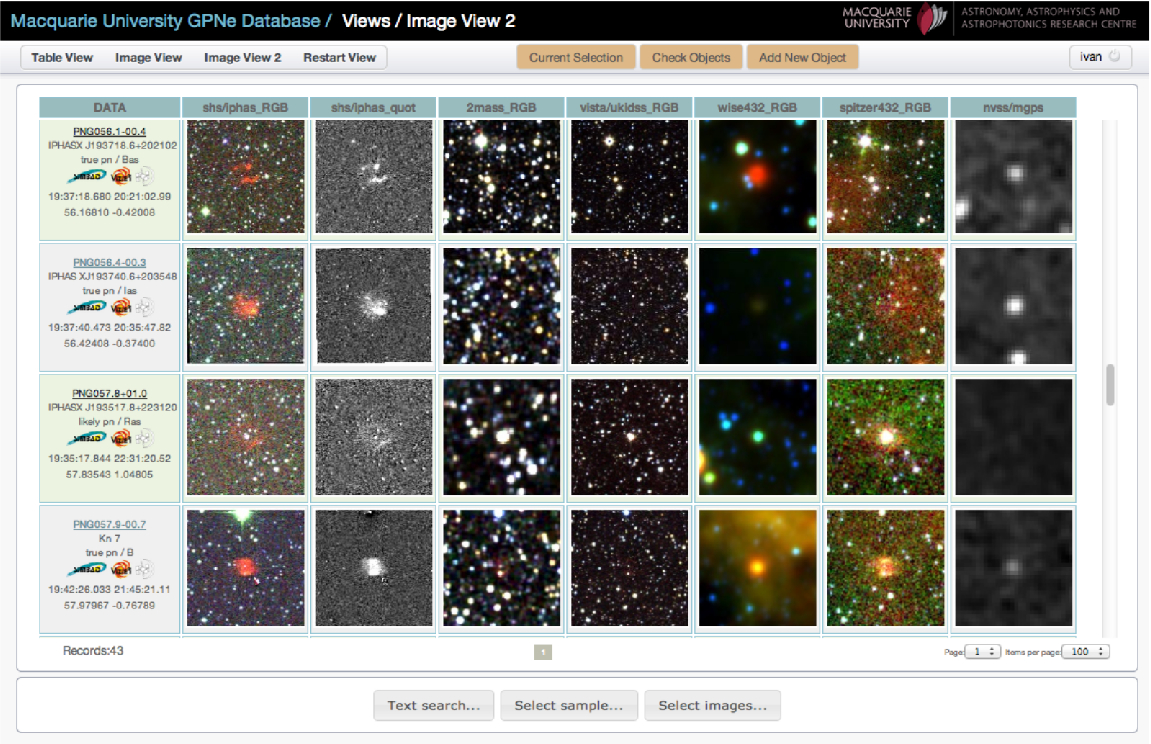}} 
\caption{\label{MASPN1} Selection of IPHAS PNe from the new MASPN database that gives an idea of how this new utility can be used.}
\end{figure*}

All the newly discovered IPHAS PNe described and listed in this paper will also be accessible through the new Macquarie-Strasbourg PNe database MASPN (Parker et al. 2014, Bojicic et al in preparation). This powerful, new database and research tool gathers all known Galactic  PNe in a single place. It provides, for the first time, an accessible, reliable, on-line ``one-stop'' shop for essential, up-to date information for all known Galactic PN and provides the community with the most complete data with which to undertake new science. It  provides  quick and easy access to information such as a multi-wavelength image service, spectroscopy, morphologies and other useful data. Fig. \ref{MASPN2} shows the specific dedicated MASPN entries avaialable for each PNe in the catalogue for the PN IPHASX J194727.5+230816. MASPN allows for the retrieval of detailed information such as the identification spectra and direct links to the SIMBAD and VIZIER entries for the selected object. As a preliminary demonstration of the MASPN database, the catalogue entries for the complete list of a set of our newly discovered IPHAS PNe  are presented  in Fig. \ref{MASPN1}  in the form of selected multi-wavelength images.

\newpage
\section{Conclusions and future work}
We present the first major release of the preliminary catalogue of new, extended PNe discovered in the framework of the IPHAS Survey. We have detected 159 new PNe by visually scanning binned IPHAS image mosaics predominately over a two hour RA region from 18-19 hours. These newly discovered objects nearly double the number of known Northern PNe close to the Galactic mid-plane in the regions covered. Most of the newly found objects are relatively faint, attesting to their generally more advanced evolutionary stage and/or location in more obscured regions. The survey now provides access to a class of evolved and distant Galactic PNe previously under-represented in the Northern plane and vital for a proper evaluation of the global PN population across the whole Galaxy. Our sample also includes new PNe towards the Galactic Anticentre, a crucial region for abundance gradient studies. The IPHAS PNe are generally of low surface brightnesses and also of low excitation which makes their spectroscopic confirmation more difficult due to similar lines ratios sometimes existing for totally different classes of objects (which are equally faint and also show a few number of emission lines). However,  the use of newly implemented diagnostic diagrams  and associated environmental, morphological and multi-wavelength analysis allows us to largely overcome this problem. The first statistical studies indicate that our catalogue shows a large scatter in PNe sizes (though we are biased towards non point source PNe) and morphological structures.  Also, within our sample we unveiled an important  group of, what appeared to be  genuine, round PNe at low Galactic latitudes. Their study would allow us to have a new look at this class of objects which are usually found at higher latitudes. By extension their progenitor characteristics could also be derived.

The work on extended, new IPHAS PNe presented here is the first in a series of associated papers in terms of discoveries as more IPHAS PNe have yet to be found and spectroscopically investigated. Additional work is planned on detailed  kinematical analysis using IFUs for velocity and morphological determination of selected sub-samples combined with deep spectroscopy for accurate chemical analysis and abundance determinations. The study of identified CSPN and derivation of the distances (which could both benefit from the future GAIA mission) are also part of a non-exhaustive list of works still to be performed by our team. We intend to measure the integrated H$\alpha$ fluxes of all these new nebulae directly from the IPHAS imaging data now that it has been properly calibrated (e.g. Barentsen et al. 2014, submitted), to compliment the recent catalogue of \citet[a]{Frew2014a}. The flux data is necessary to determine the distances of these new PNe independently using the H$\alpha$ surface brightness -- radius relation of Frew et al. 2014 (submitted), and to be reported in a forthcoming paper.
This investigation on the discovery and preliminary analysis of these new extended IPHAS PNe will be supplemented  with data from the UVEX survey (``blue'' counterpart of IPHAS in the North; \citealt{Groot2009}) and the VPHAS+ survey \citet{Drew2014} in the South.

\section*{Acknowledgements}
We would like to thank Margaret Meixner for her careful review and highly appreciate the comments which contributed to improving the quality of the publication.
LS is supported by the CONACYT grant CB-2011-01-0168078, MS was partially supported by Spanish MICINN within the program CONSOLIDER INGENIO 2010, under grant ``Molecular Astrophysics: The Herschel and ALMA Era, ASTROMOL'' (ref: CSD2009-00038),LO acknowledges support by project PROMEP/103.5/12/3590.
We also thank the Bristol University students Greg Mould, William Howie, Luke Davies, Heidi Naumann, Will Summers, Alex Townshend, Paul May, Matina Mitchell, Finn Hoolahan, Tom Burgess, Ashley Akerman, James Jordan, Simon Palmer, Anna Kovacevic, Jai Tailor, Olivia Smedley and Daniel Huggins for their participation in the search in the framework of their undergraduate thesis.\\
This paper makes use of data obtained as part of the INT Photometric H$\alpha$ Survey of the Northern Galactic Plane (IPHAS) carried out at the Isaac Newton Telescope (INT). All IPHAS data are processed by the Cambridge Astronomical Survey Unit, at the Institute of Astronomy in Cambridge.
The INT and WHT telescopes are operated on the island of La Palma by the Isaac Newton Group in the Spanish Observatorio del Roque de los Muchachos of the Instituto de Astrof\'isica de Canarias. This research has been partially carried out with telescope time awarded by the CCI International Time Programme. The Observatorio Astron\'omico Nacional at San Pedro M\'artir, is a facility operated by Instituto de Astronom\'ia of the Universidad Nacional Aut\'onoma de Mexico. We acknowledge the staff of the San Pedro M\'artir Observatory for their support. This research also made used of data from Kitt Peak National Observatory, National Optical Astronomy Observatory, which is operated by the Association of Universities for Research in Astronomy (AURA) under cooperative agreement with the National Science Foundation. This paper uses observations made at the South African Astronomical Observatory (SAAO) as well as at the Siding Spring Observatory (SSO-Australia), which is part of the Research School of Astronomy \& Astrophysics (RSAA) at the Australian National University (ANU). The 1.5m and ALBIREO spectrograph are operated by the Instituto de Astrof\'isica de Andaluc\'ia at the Sierra Nevada Observatory. Some observations were also made with the Gran Telescopio Canarias (GTC), instaled in the Spanish Observatorio del Roque de los Muchachos of the Instituto de Astrofísica de Canarias, in the island of La Palma.
This research has made use of the SIMBAD and Vizier databases, operated at CDS, Strasbourg, France. This research made use of APLpy, an open-source plotting package for Python hosted at http://aplpy.github.com. This research made use of Montage, funded by the National Aeronautics and Space Administration's Earth Science Technology Office, Computation Technologies Project, under Cooperative Agreement Number NCC5-626 between NASA and the California Institute of Technology. Montage is maintained by the NASA/IPAC Infrared Science Archive. This publication makes use of data products from the Wide-field Infrared Survey Explorer, which is a joint project of the University of California, Los Angeles, and the Jet Propulsion Laboratory/California Institute of Technology, funded by the National Aeronautics and Space Administration.

\clearpage
\clearpage
\bibliographystyle{mn2e}



\newpage
\newpage
\begin{table*}
\footnotesize\addtolength{\tabcolsep}{-4pt}
\caption{\label{catalogue} IPHAS PNe catalogue. See the text for the description of the database. There are 20 objects that were suggested as possible PN prior to IPHAS spectroscopic confirmation. These are identified by one of the following symbols appended to the PN status flag with the associated reference indicated: $\ast$ \citet{Ack1992},$\dagger$ \citet{Jacoby2010},$\ddagger$ \citet{Kohoutek2001}, $\oplus$ \citet{Urquhart2009}, $\star$ \citet{Preite1988}, $\times$ \citet{Kronberger2006} }
\begin{tabular}{l|l|l|c|l|l|l|r|c|c|c|c|c|}
\hline
Number & PN status	& IAU PNG	& IPHAS ID:	& RA	 & DEC	 & Galactic & Galactic & Major Diam & Morphology & No Spectra & Telescope  & Obs.date\\
& (T,L,P)	&  designation	& IPHASX	& (J2000)	& (J2000)	&	Longitude	& Latitude	& axis (arcsec)	& 		& 		&      & yyyy-mm-dd \\
\hline
\hline
1	&	P	&	G116.0$-$04.8	&	J000021.4$+$572207	&	00:00:21.4	&	57:22:07	&	116.062	&	-4.8202	&	22	&	B	&	1	&	KP	&	2009-08-15	\\ 
2	&	T$^{\dagger}$	&	G119.2$+$04.6	&	J001333.8$+$671803	&	00:13:33.8	&	67:18:03	&	119.28	&	4.6983	&	14	&	B	&	2	&	KP,OS	&	2009-08-17	\\ 
3	&	P	&	G122.9$+$00.3	&	J005123.2$+$631059	&	00:51:23.2	&	63:10:59	&	122.926	&	0.3113	&	47	&	A	&	1	&	KP	&	2009-08-16	\\ 
4	&	T	&	G126.6$+$01.3	&	J012507.9$+$635652	&	01:25:07.9	&	63:56:52	&	126.622	&	1.3177	&	33	&	Bapm	&	1	&	WH	&	2004-09-27	\\ 
5	&	T	&	G127.6$-$01.1	&	J013108.9$+$612258	&	01:31:08.9	&	61:22:58	&	127.67	&	-1.1229	&	24	&	Ra	&	1	&	KP	&	2009-08-16	\\ 
6	&	T	&	G129.6$+$03.4	&	J015624.9$+$652830	&	01:56:24.9	&	65:28:30	&	129.612	&	3.4496	&	189	&	Rar	&	1	&	SM	&	2011-09-23	\\ 
7	&	T	&	G132.8$+$02.0	&	J022045.0$+$631134	&	02:20:45.0	&	63:11:34	&	132.804	&	2.046	&	30	&	Eas	&	1	&	OS	&	2011-10-04	\\ 
8	&	P	&	G134.1$+$03.0	&	J023539.3$+$633823	&	02:35:39.3	&	63:38:23	&	134.194	&	3.0706	&	135	&	E	&	1	&	SM	&	2013-02-11	\\ 
9	&	T	&	G137.7$+$03.3	&	J030421.3$+$621802	&	03:04:21.3	&	62:18:02	&	137.714	&	3.3066	&	57	&	Rr	&	2	&	OS,SM	&	2010-09-19	\\ 
10	&	T	&	G139.0$+$03.2	&	J031345.5$+$613707	&	03:13:45.5	&	61:37:07	&	139.015	&	3.2741	&	127	&	Bs	&	1	&	SM	&	2011-09-24	\\ 
11	&	T	&	G144.1$-$00.5	&	J033105.3$+$553851	&	03:31:05.3	&	55:38:51	&	144.159	&	-0.5012	&	27	&	Rar	&	1	&	SM	&	2010-09-18	\\ 
12	&	L	&	G150.1$-$04.5	&	J034659.8$+$484900	&	03:46:59.8	&	48:49:00	&	150.186	&	-4.5403	&	62	&	Ra	&	1	&	OS	&	2012-09-20	\\ 
13	&	T$^{\dagger}$	&	G150.0$-$00.3	&	J040329.5$+$520825	&	04:03:29.5	&	52:08:25	&	150.074	&	-0.3084	&	25	&	Bp	&	1	&	SM	&	2010-09-18	\\ 
14	&	L	&	G156.4$+$01.1	&	J043826.1$+$483907	&	04:38:26.1	&	48:39:07	&	156.476	&	1.1425	&	95	&	Eas	&	2	&	OS,SM	&	2010-09-20	\\ 
15	&	P	&	G159.4$+$02.0	&	J045358.6$+$465842	&	04:53:58.6	&	46:58:42	&	159.444	&	2.0407	&	12	&	Am	&	1	&	OS	&	2011-10-08	\\ 
16	&	T	&	G157.1$+$04.4	&	J045627.6$+$501720	&	04:56:27.6	&	50:17:20	&	157.117	&	4.4347	&	91	&	Ra	&	1	&	SM	&	2011-09-24	\\ 
17	&	T	&	G167.3$-$03.1	&	J045847.7$+$373640	&	04:58:47.7	&	37:36:40	&	167.314	&	-3.1209	&	17	&	Rs	&	1	&	SM	&	2010-09-18	\\ 
18	&	P	&	G167.9$+$00.9	&	J051733.3$+$393027	&	05:17:33.3	&	39:30:27	&	167.988	&	0.9515	&	65	&	A	&	1	&	OS	&	2012-09-22	\\ 
19	&	L	&	G183.0$+$00.0	&	J055242.8$+$262116	&	05:52:42.8	&	26:21:16	&	183.022	&	0.0176	&	16	&	Rar	&	1	&	GC	&	2009-11-09	\\ 
20	&	T	&	G193.5$-$03.1	&	J060328.1$+$154108	&	06:03:28.1	&	15:41:08	&	193.522	&	-3.1388	&	55	&	Ba	&	1	&	SM	&	2011-09-24	\\ 
21	&	T	&	G190.7$-$01.3	&	J060412.2$+$190031	&	06:04:12.2	&	19:00:31	&	190.709	&	-1.3567	&	70	&	Ba	&	1	&	SM	&	2010-09-18	\\ 
22	&	T	&	G195.4$-$04.0	&	J060416.2$+$133250	&	06:04:16.2	&	13:32:50	&	195.486	&	-4.014	&	29	&	Ba	&	1	&	SM	&	2007-01-23	\\ 
23	&	T	&	G204.3$-$01.6	&	J062937.8$+$065220	&	06:29:37.8	&	06:52:20	&	204.33	&	-1.6839	&	31	&	Ba	&	1	&	SM	&	2011-09-25	\\ 
24	&	T	&	G197.3$+$02.7	&	J063223.9$+$150410	&	06:32:23.9	&	15:04:10	&	197.373	&	2.7078	&	90	&	B	&	1	&	SM	&	2011-09-26	\\ 
25	&	T	&	G029.9$+$03.7	&	J183249.6$-$005638	&	18:32:49.6	&	-00:56:38	&	29.9655	&	3.7077	&	5	&	S	&	1	&	SA	&	2011-07-10	\\ 
26	&	L$^{\star}$	&	G031.1$+$03.4	&	J183602.3$-$000227	&	18:36:02.3	&	-00:02:27	&	31.1381	&	3.4069	&	16	&	E	&	2	&	WI	&	2011-07-04	\\ 
27	&	L	&	G031.1$+$02.1	&	J184030.1$-$003822	&	18:40:30.1	&	-00:38:22	&	31.1151	&	2.1411	&	22	&	B	&	1	&	WI	&	2011-07-04	\\ 
28	&	L	&	G029.7$+$01.1	&	J184139.1$-$021649	&	18:41:39.1	&	-02:16:49	&	29.7865	&	1.1354	&	--	&	S	&	1	&	SA	&	2011-07-06	\\ 
29	&	T	&	G032.8$+$01.5	&	J184546.7$+$003631	&	18:45:46.7	&	00:36:31	&	32.8287	&	1.5367	&	4	&	R	&	1	&	SA	&	2011-07-12	\\ 
30	&	T	&	G029.5$-$00.2	&	J184616.3$-$030625	&	18:46:16.3	&	-03:06:25	&	29.578	&	-0.2684	&	4	&	Sm	&	1	&	SA	&	2011-07-10	\\ 
31	&	P	&	G038.4$+$03.6	&	J184834.6$+$063302	&	18:48:34.6	&	06:33:02	&	38.4449	&	3.6152	&	21	&	Ba	&	2	&	SA	&	2011-07-09	\\ 
32	&	T	&	G040.3$+$03.4	&	J185225.0$+$080843	&	18:52:25.0	&	08:08:43	&	40.3013	&	3.4883	&	22	&	R	&	3	&	WI,SM	&	2010-06-02	\\ 
33	&	P	&	G033.8$+$00.1	&	J185225.8$+$005250	&	18:52:25.8	&	00:52:50	&	33.8287	&	0.1807	&	480	&	Ears	&	2	&	SM	&	2010-06-01	\\ 
34	&	L$^{\star}$	&	G039.1$+$02.7	&	J185301.9$+$064415	&	18:53:01.9	&	06:44:15	&	39.114	&	2.7138	&	--	&	S	&	1	&	SA	&	2011-07-08	\\ 
35	&	T	&	G040.1$+$03.2	&	J185309.4$+$075241	&	18:53:09.4	&	07:52:41	&	40.1456	&	3.2038	&	12	&	Eam	&	3	&	SA,IN	&	2008-06-30	\\ 
36	&	L	&	G032.7$-$00.5	&	J185312.9$-$002529	&	18:53:12.9	&	-00:25:29	&	32.7563	&	-0.5893	&	26	&	E	&	1	&	WI	&	2011-07-03	\\ 
37	&	T	&	G038.4$+$02.2	&	J185321.7$+$055641	&	18:53:21.7	&	05:56:41	&	38.4445	&	2.2806	&	5	&	R	&	2	&	SA,KP	&	2009-08-18	\\ 
38	&	P	&	G032.6$-$00.7	&	J185342.6$-$003628	&	18:53:42.6	&	-00:36:28	&	32.6498	&	-0.7828	&	23	&	E	&	2	&	WI	&	2011-07-03	\\ 
39	&	T	&	G032.9$-$01.4	&	J185640.0$-$003804	&	18:56:40.0	&	-00:38:04	&	32.9632	&	-1.4528	&	15	&	Bp	&	1	&	SM	&	2010-07-09	\\ 
40	&	P	&	G043.3$+$03.5	&	J185744.4$+$105053	&	18:57:44.4	&	10:50:53	&	43.3109	&	3.5424	&	110	&	Eams	&	2	&	SM	&	2010-07-10	\\ 
41	&	T	&	G040.5$+$01.9	&	J185815.8$+$073753	&	18:58:15.8	&	07:37:53	&	40.5014	&	1.9648	&	7	&	Rrs	&	1	&	KP	&	2009-08-15	\\ 
42	&	T	&	G040.6$+$01.5	&	J185957.0$+$073544	&	18:59:57.0	&	07:35:44	&	40.6602	&	1.5766	&	4	&	B	&	2	&	SA	&	2011-07-10	\\ 
43	&	T$^{\ddagger}$	&	G041.5$+$01.7	&	J190105.7$+$082536	&	19:01:05.7	&	08:25:36	&	41.5293	&	1.7038	&	--	&	S	&	1	&	SA	&	2011-07-08	\\ 
44	&	T	&	G044.4$+$03.1	&	J190115.5$+$114150	&	19:01:15.5	&	11:41:50	&	44.4613	&	3.1594	&	26	&	Ear	&	1	&	KP	&	2009-08-12	\\ 
45	&	P	&	G038.6$+$00.1	&	J190125.4$+$050858	&	19:01:25.4	&	05:08:58	&	38.6517	&	0.1329	&	13	&	B	&	1	&	SM	&	2011-09-25	\\ 
46	&	T	&	G036.0$-$01.4	&	J190227.3$+$020804	&	19:02:27.3	&	02:08:04	&	36.0885	&	-1.4757	&	92	&	Ias	&	1	&	SM	&	2011-05-05	\\ 
47	&	T	&	G047.1$+$03.9	&	J190319.0$+$142524	&	19:03:19.0	&	14:25:24	&	47.1224	&	3.9521	&	15	&	E	&	2	&	WI	&	2011-07-05	\\ 
48	&	P	&	G038.6$-$00.4	&	J190340.4$+$045311	&	19:03:40.4	&	04:53:11	&	38.6745	&	-0.4857	&	114	&	I	&	1	&	WI	&	2011-07-01	\\ 
49	&	T	&	G043.0$+$01.7	&	J190340.7$+$094639	&	19:03:40.7	&	09:46:39	&	43.023	&	1.7542	&	20	&	Em	&	2	&	SA	&	2011-07-01	\\ 
50	&	P	&	G038.7$-$00.5	&	J190401.5$+$045433	&	19:04:01.5	&	04:54:33	&	38.7351	&	-0.5533	&	140	&	Ims	&	1	&	WI	&	2011-07-01	\\ 
51	&	T	&	G042.2$+$01.1	&	J190417.9$+$084916	&	19:04:17.9	&	08:49:16	&	42.2425	&	1.1804	&	--	&	S	&	1	&	SM	&	2010-07-10	\\ 
52	&	T	&	G042.6$+$01.3	&	J190432.9$+$091656	&	19:04:32.9	&	09:16:56	&	42.6809	&	1.3368	&	8	&	E	&	1	&	IN	&	2008-06-29	\\ 
53	&	P$^{\ddagger}$	&	G036.4$-$01.9	&	J190438.6$+$021424	&	19:04:38.6	&	02:14:24	&	36.4326	&	-1.9136	&	21	&	Bmp	&	3	&	SA,IN	&	2008-06-28	\\ 
54	&	T	&	G045.4$+$02.6	&	J190447.9$+$121844	&	19:04:47.9	&	12:18:44	&	45.4047	&	2.6692	&	12	&	Er	&	2	&	SM	&	2011-09-25	\\ 
55	&	T	&	G043.6$+$01.7	&	J190454.0$+$101801	&	19:04:54.0	&	10:18:01	&	43.6257	&	1.7262	&	18	&	Ears	&	1	&	SM	&	2011-05-04	\\ 
56	&	T	&	G037.7$-$01.3	&	J190503.1$+$034225	&	19:05:03.1	&	03:42:25	&	37.7839	&	-1.3321	&	64	&	Ear	&	3	&	OS,WI	&	2011-07-02	\\ 
57	&	T	&	G048.9$+$04.3	&	J190512.4$+$161347	&	19:05:12.4	&	16:13:47	&	48.9435	&	4.3662	&	60	&	Rar	&	2	&	KP,SM	&	2009-08-12	\\ 
58	&	T	&	G044.2$+$01.9	&	J190518.3$+$105750	&	19:05:18.3	&	10:57:50	&	44.2615	&	1.942	&	4	&	R	&	1	&	SA	&	2011-07-06	\\ 
59	&	T	&	G044.3$+$01.8	&	J190543.8$+$110018	&	19:05:43.8	&	11:00:18	&	44.346	&	1.8681	&	8	&	Rr	&	1	&	KP	&	2009-08-12	\\ 
\hline
\end{tabular}
\end{table*}

\begin{table*}
\footnotesize\addtolength{\tabcolsep}{-4pt}
\begin{flushleft}
{\bf Table~\ref{catalogue}.} (continued)
\end{flushleft}
\begin{tabular}{l|l|l|c|l|l|l|r|c|c|c|c|c|}
\hline
Number &PN status	& IAU PNG	& IPHAS ID:	& RA	& DEC  & Galactic & Galactic & Major Diam & Morphology & No Spectra & Telescope  & Obs.date\\
& (T,L,P)	&  designation	& IPHASX	& (J2000)	& (J2000)	&	Longitude	& Latitude	& axis (arcsec)	& 		& 		&      & yyyy-mm-dd \\
\hline
\hline
60	&	L	&	G032.3$-$04.5	&	J190631.5$-$023236	&	19:06:31.5	&	-02:32:36	&	32.3845	&	-4.5154	&	16	&	E	&	2	&	WI	&	2011-07-05	\\ 
61	&	L	&	G038.9$-$01.3	&	J190718.1$+$044056	&	19:07:18.1	&	04:40:56	&	38.908	&	-1.3827	&	9	&	I	&	1	&	SA	&	2011-07-02	\\ 
62	&	P	&	G039.6$-$01.2	&	J190816.8$+$052506	&	19:08:16.8	&	05:25:06	&	39.674	&	-1.2603	&	7	&	Es	&	2	&	SA	&	2011-07-08	\\ 
63	&	T$^{\times}$	&	G045.7$+$01.4	&	J190954.7$+$120455	&	19:09:54.7	&	12:04:55	&	45.7737	&	1.4535	&	18	&	B	&	1	&	SA	&	2011-07-08	\\ 
64	&	T	&	G047.8$+$02.4	&	J191001.1$+$142202	&	19:10:01.1	&	14:22:02	&	47.8155	&	2.4831	&	10	&	E	&	1	&	KP	&	2009-08-15	\\ 
65	&	T	&	G044.9$+$00.8	&	J191022.1$+$110538	&	19:10:22.1	&	11:05:38	&	44.9485	&	0.8986	&	10	&	Ear	&	1	&	KP	&	2009-08-15	\\ 
66	&	T	&	G038.3$-$02.5	&	J191027.4$+$034046	&	19:10:27.4	&	03:40:46	&	38.3792	&	-2.5422	&	13	&	Ra	&	1	&	KP	&	2009-08-16	\\ 
67	&	T	&	G048.2$+$01.9	&	J191255.4$+$143248	&	19:12:55.4	&	14:32:48	&	48.2994	&	1.9417	&	8	&	Rar	&	1	&	KP	&	2009-08-17	\\ 
68	&	T	&	G051.2$+$03.2	&	J191345.5$+$174752	&	19:13:45.5	&	17:47:52	&	51.2785	&	3.2661	&	7	&	Ras	&	1	&	WH	&	2007-07-17	\\ 
69	&	T	&	G047.6$+$01.0	&	J191445.1$+$133219	&	19:14:45.1	&	13:32:19	&	47.6116	&	1.0818	&	15	&	Ears	&	2	&	SA,WH	&	2006-08-24	\\ 
70	&	T	&	G050.8$+$02.3	&	J191621.4$+$165638	&	19:16:21.4	&	16:56:38	&	50.8082	&	2.3214	&	5	&	R	&	1	&	IN	&	2008-06-28	\\ 
71	&	L	&	G042.0$-$02.4	&	J191651.6$+$065608	&	19:16:51.6	&	06:56:08	&	42.0067	&	-2.4494	&	17	&	A	&	2	&	WI	&	2011-07-05	\\ 
72	&	L	&	G039.0$-$04.0	&	J191716.4$+$033447	&	19:17:16.4	&	03:34:47	&	39.0765	&	-4.0972	&	16	&	Emr	&	1	&	KP	&	2009-08-16	\\ 
73	&	T	&	G052.0$+$02.7	&	J191716.5$+$181518	&	19:17:16.5	&	18:15:18	&	52.0719	&	2.7369	&	46	&	Rrs	&	1	&	SM	&	2010-07-12	\\ 
74	&	T	&	G043.3$-$01.9	&	J191727.0$+$082036	&	19:17:27.0	&	08:20:36	&	43.3218	&	-1.9235	&	15	&	Ras	&	2	&	SA	&	2011-07-02	\\ 
75	&	L	&	G048.7$+$00.9	&	J191727.2$+$142735	&	19:17:27.2	&	14:27:35	&	48.732	&	0.9305	&	--	&	S	&	1	&	SA	&	2011-07-10	\\ 
76	&	T	&	G042.7$-$02.5	&	J191840.4$+$073131	&	19:18:40.4	&	07:31:31	&	42.7386	&	-2.5729	&	41	&	Ea	&	2	&	WI	&	2011-07-01	\\ 
77	&	L	&	G045.7$-$01.6	&	J192101.1$+$103734	&	19:21:01.1	&	10:37:34	&	45.7521	&	-1.6335	&	3	&	E	&	1	&	SA	&	2011-07-10	\\ 
78	&	L$^{\oplus}$	&	G050.4$+$00.7	&	J192140.4$+$155354	&	19:21:40.4	&	15:53:54	&	50.4801	&	0.7055	&	19	&	B	&	1	&	SM	&	2010-07-10	\\ 
79	&	T	&	G051.7$+$01.3	&	J192146.7$+$172055	&	19:21:46.7	&	17:20:55	&	51.772	&	1.3658	&	29	&	Ears	&	1	&	SM	&	2010-07-12	\\ 
80	&	P	&	G048.7$-$00.2	&	J192152.0$+$135223	&	19:21:52.0	&	13:52:23	&	48.7153	&	-0.2894	&	13	&	B	&	1	&	KP	&	2009-08-17	\\ 
81	&	T	&	G049.2$+$00.0	&	J192153.9$+$143056	&	19:21:53.9	&	14:30:56	&	49.2858	&	0.0065	&	18	&	Ear	&	1	&	KP	&	2009-08-16	\\ 
82	&	T	&	G050.6$+$00.0	&	J192436.3$+$154402	&	19:24:36.3	&	15:44:02	&	50.6681	&	0.0061	&	23	&	Ea	&	1	&	SM	&	2011-09-26	\\ 
83	&	L	&	G054.5$+$01.8	&	J192534.9$+$200334	&	19:25:34.9	&	20:03:34	&	54.5879	&	1.8515	&	38	&	E	&	1	&	OS	&	2012-09-19	\\ 
84	&	L	&	G049.7$-$00.7	&	J192543.2$+$143546	&	19:25:43.2	&	14:35:46	&	49.794	&	-0.7702	&	176	&	Rars	&	1	&	SP	&	2012-04-13	\\ 
85	&	T	&	G051.8$+$00.2	&	J192553.5$+$165331	&	19:25:53.5	&	16:53:31	&	51.8341	&	0.2838	&	6	&	E	&	1	&	WH	&	2007-07-18	\\ 
86	&	T$^{\star}$	&	G049.5$-$01.4	&	J192751.3$+$140127	&	19:27:51.3	&	14:01:27	&	49.5371	&	-1.4973	&	10	&	Rars	&	1	&	SA	&	2011-07-09	\\ 
87	&	T	&	G059.1$+$03.5	&	J192837.7$+$245024	&	19:28:37.7	&	24:50:24	&	59.135	&	3.5009	&	13	&	Eas	&	1	&	KP	&	2009-08-17	\\ 
88	&	T	&	G045.7$-$03.8	&	J192847.2$+$093436	&	19:28:47.2	&	09:34:36	&	45.7296	&	-3.8144	&	67	&	Br	&	1	&	WH	&	2006-08-25	\\ 
89	&	T	&	G059.1$+$03.3	&	J192902.5$+$244646	&	19:29:02.5	&	24:46:46	&	59.1263	&	3.3897	&	10	&	Rms	&	1	&	WH	&	2007-07-17	\\ 
90	&	T	&	G054.4$+$00.5	&	J193009.3$+$192129	&	19:30:09.3	&	19:21:29	&	54.4845	&	0.5712	&	221	&	Ears	&	2	&	SA,SM	&	2011-07-09	\\ 
91	&	T	&	G054.7$+$00.4	&	J193110.7$+$192905	&	19:31:10.7	&	19:29:05	&	54.7118	&	0.4203	&	19	&	Ba	&	1	&	SM	&	2011-09-26	\\ 
92	&	T	&	G048.1$-$03.2	&	J193127.0$+$115622	&	19:31:27.0	&	11:56:22	&	48.123	&	-3.2626	&	27	&	R	&	1	&	SM	&	2011-09-26	\\ 
93	&	T	&	G051.7$-$01.7	&	J193308.9$+$155354	&	19:33:08.9	&	15:53:54	&	51.7956	&	-1.7197	&	27	&	Ears	&	2	&	WI	&	2011-07-03	\\ 
94	&	L	&	G057.8$+$01.0	&	J193517.8$+$223120	&	19:35:17.8	&	22:31:20	&	57.8354	&	1.0481	&	21	&	Ras	&	1	&	OS	&	2012-09-22	\\ 
95	&	T	&	G048.0$-$04.4	&	J193532.1$+$112115	&	19:35:32.1	&	11:21:15	&	48.0901	&	-4.421	&	9	&	R	&	2	&	KP,SM	&	2009-08-14	\\ 
96	&	P	&	G065.8$+$05.1	&	J193630.2$+$312810	&	19:36:30.2	&	31:28:10	&	65.8057	&	5.1485	&	404	&	Aa	&	1	&	SP	&	2012-04-12	\\ 
97	&	T	&	G051.9$-$02.5	&	J193633.5$+$153345	&	19:36:33.5	&	15:33:45	&	51.8995	&	-2.6003	&	103	&	R	&	2	&	WI	&	2011-07-02	\\ 
98	&	T	&	G053.4$-$01.8	&	J193652.9$+$171940	&	19:36:52.9	&	17:19:40	&	53.4803	&	-1.8082	&	4	&	R	&	1	&	WH	&	2007-07-18	\\ 
99	&	T	&	G056.1$-$00.4	&	J193718.6$+$202102	&	19:37:18.6	&	20:21:02	&	56.1681	&	-0.4201	&	47	&	Bas	&	1	&	IN	&	2006-06-15	\\ 
100	&	T	&	G056.4$-$00.3	&	J193740.4$+$203547	&	19:37:40.4	&	20:35:47	&	56.4241	&	-0.374	&	31	&	Ias	&	1	&	SA	&	2011-07-10	\\ 
101	&	T	&	G062.1$+$02.8	&	J193752.2$+$271119	&	19:37:52.2	&	27:11:19	&	62.1977	&	2.8117	&	25	&	E	&	1	&	SM	&	2010-07-10	\\ 
102	&	T	&	G062.0$+$02.5	&	J193827.8$+$265752	&	19:38:27.8	&	26:57:52	&	62.0663	&	2.5868	&	6	&	R	&	1	&	WH	&	2007-07-18	\\ 
103	&	T$^{\times}$	&	G057.9$-$00.7	&	J194226.0$+$214521	&	19:42:26.0	&	21:45:21	&	57.9797	&	-0.7679	&	27	&	B	&	1	&	KP	&	2009-08-16	\\ 
104	&	T	&	G052.1$-$04.1	&	J194232.8$+$150034	&	19:42:32.8	&	15:00:34	&	52.1249	&	-4.1306	&	19	&	Rr	&	1	&	SM	&	2010-09-20	\\ 
105	&	T	&	G063.3$+$02.2	&	J194240.5$+$275109	&	19:42:40.5	&	27:51:09	&	63.3005	&	2.2106	&	74	&	Rar	&	1	&	OS	&	2012-09-22	\\ 
106	&	T	&	G054.2$-$03.4	&	J194359.5$+$170901	&	19:43:59.5	&	17:09:01	&	54.1615	&	-3.3735	&	21	&	Bamps	&	1	&	WH	&	2007-07-17	\\ 
107	&	T	&	G062.9$+$01.3	&	J194510.6$+$270930	&	19:45:10.6	&	27:09:30	&	62.9747	&	1.3839	&	14	&	B	&	1	&	KP	&	2009-08-18	\\ 
108	&	L	&	G057.8$-$01.7	&	J194533.6$+$210808	&	19:45:33.6	&	21:08:08	&	57.8047	&	-1.7077	&	69	&	Ear	&	1	&	SA	&	2011-07-10	\\ 
109	&	T	&	G059.8$-$00.6	&	J194556.2$+$232833	&	19:45:56.2	&	23:28:33	&	59.8739	&	-0.6101	&	54	&	Bas	&	1	&	SM	&	2010-07-10	\\ 
110	&	L	&	G059.7$-$00.8	&	J194633.0$+$231659	&	19:46:33.0	&	23:16:59	&	59.7778	&	-0.8285	&	13	&	Ea	&	1	&	IN	&	2007-08-03	\\ 
111	&	T	&	G059.7$-$01.0	&	J194727.5$+$230816	&	19:47:27.5	&	23:08:16	&	59.7575	&	-1.0822	&	35	&	Rars	&	1	&	SM	&	2011-05-05	\\ 
112	&	L	&	G059.1$-$01.4	&	J194728.8$+$222823	&	19:47:28.8	&	22:28:23	&	59.1857	&	-1.4214	&	27	&	Ras	&	1	&	SP	&	2012-04-12	\\ 
113	&	T	&	G066.8$+$02.9	&	J194751.9$+$311818	&	19:47:51.9	&	31:18:18	&	66.8587	&	2.9572	&	13	&	B	&	1	&	WH	&	2007-07-17	\\ 
114	&	P	&	G055.7$-$03.8	&	J194905.2$+$181503	&	19:49:05.2	&	18:15:03	&	55.7245	&	-3.8742	&	33	&	A	&	2	&	WI	&	2011-07-01	\\ 
115	&	T	&	G063.7$+$00.7	&	J194930.9$+$273028	&	19:49:30.9	&	27:30:28	&	63.7622	&	0.7277	&	84	&	R	&	1	&	OS	&	2012-09-24	\\ 
116	&	T	&	G062.7$+$00.0	&	J194940.9$+$261521	&	19:49:40.9	&	26:15:21	&	62.7024	&	0.0602	&	22	&	Bps	&	1	&	WH	&	2007-07-17	\\ 
117	&	T	&	G063.5$+$00.0	&	J195126.5$+$265838	&	19:51:26.5	&	26:58:38	&	63.5236	&	0.0893	&	8	&	B	&	2	&	KP,SM	&	2009-08-14	\\ 
118	&	T	&	G067.9$+$02.4	&	J195221.6$+$315859	&	19:52:21.6	&	31:58:59	&	67.9297	&	2.4762	&	42	&	Ear	&	1	&	SM	&	2011-09-25	\\ 
\hline
\end{tabular}
\end{table*}

\begin{table*}
\footnotesize\addtolength{\tabcolsep}{-4pt}
\begin{flushleft}
{\bf Table~\ref{catalogue}.} (continued)
\end{flushleft}
\begin{tabular}{l|l|l|c|l|l|l|r|c|c|c|c|c|}
\hline
Number &PN status	& IAU PNG	& IPHAS ID:	& RA	& DEC  & Galactic & Galactic & Major Diam & Morphology & No Spectra & Telescope  & Obs.date\\
& (T,L,P)	&  designation	& IPHASX	& (J2000)	& (J2000)	&	Longitude	& Latitude	& axis (arcsec)	& 		& 		&      & yyyy-mm-dd \\
\hline
\hline
119	&	T	&	G062.7$-$00.7	&	J195248.8$+$255359	&	19:52:48.8	&	25:53:59	&	62.7552	&	-0.7263	&	27	&	B	&	1	&	WH	&	2007-07-17	\\ 
120	&	T	&	G058.1$-$03.7	&	J195343.7$+$202635	&	19:53:43.7	&	20:26:35	&	58.1744	&	-3.7047	&	31	&	B	&	1	&	SA	&	2011-07-10	\\ 
121	&	T	&	G067.5$+$01.8	&	J195358.2$+$312120	&	19:53:58.2	&	31:21:20	&	67.5661	&	1.8605	&	228	&	Ems	&	1	&	SM	&	2011-05-06	\\ 
122	&	T	&	G068.0$+$02.1	&	J195400.8$+$315554	&	19:54:00.8	&	31:55:54	&	68.0655	&	2.1488	&	30	&	Rar	&	1	&	SM	&	2010-07-10	\\ 
123	&	T	&	G058.6$-$03.6	&	J195424.6$+$205252	&	19:54:24.6	&	20:52:52	&	58.6331	&	-3.6172	&	27	&	Bps	&	0	&	--	&	--	\\ 
124	&	T	&	G067.8$+$01.8	&	J195436.4$+$313326	&	19:54:36.4	&	31:33:26	&	67.8091	&	1.8479	&	12	&	Emr	&	2	&	KP,IN	&	2008-06-28	\\ 
125	&	T	&	G064.1$-$00.9	&	J195657.6$+$265713	&	19:56:57.6	&	26:57:13	&	64.1372	&	-0.9765	&	29	&	Bars	&	1	&	WH	&	2006-08-24	\\ 
126	&	L	&	G066.4$-$00.0	&	J195836.4$+$292314	&	19:58:36.4	&	29:23:14	&	66.4028	&	-0.016	&	17	&	Ea	&	1	&	SM	&	2011-09-26	\\ 
127	&	L	&	G068.2$+$00.9	&	J195919.0$+$312534	&	19:59:19.0	&	31:25:34	&	68.2195	&	0.9221	&	8	&	E	&	2	&	KP,SM	&	2009-08-14	\\ 
128	&	T$^{\ast}$	&	G073.0$+$03.6	&	J200018.6$+$365934	&	20:00:18.6	&	36:59:34	&	73.0711	&	3.6669	&	28	&	Ems	&	2	&	KP,IN	&	2006-09-08	\\ 
129	&	T	&	G065.8$-$00.8	&	J200041.5$+$283023	&	20:00:41.5	&	28:30:23	&	65.8928	&	-0.867	&	31	&	Ra	&	1	&	SM	&	2011-09-26	\\ 
130	&	T	&	G068.0$+$00.0	&	J200224.3$+$304845	&	20:02:24.3	&	30:48:45	&	68.0469	&	0.037	&	25	&	Bar	&	1	&	SM	&	2010-09-20	\\ 
131	&	T	&	G072.0$+$02.2	&	J200353.5$+$352250	&	20:03:53.5	&	35:22:50	&	72.0824	&	2.2017	&	15	&	B	&	1	&	SM	&	2011-09-26	\\ 
132	&	T	&	G073.6$+$02.8	&	J200522.0$+$36594	&	20:05:22.0	&	+36:59:4	&	73.6083	&	2.8113	&	11	&	Eas	&	1	&	KP	&	2009-08-17	\\ 
133	&	P	&	G070.4$+$00.7	&	J200525.3$+$331424	&	20:05:25.3	&	33:14:24	&	70.4413	&	0.791	&	49	&	I	&	1	&	OS	&	2011-10-08	\\ 
134	&	T	&	G063.5$-$04.7	&	J200937.3$+$242903	&	20:09:37.3	&	24:29:03	&	63.556	&	-4.7037	&	24	&	Ba	&	1	&	SM	&	2013-05-09	\\ 
135	&	T$^{\dagger}$	&	G077.6$+$04.3	&	J200940.9$+$411442	&	20:09:40.9	&	41:14:42	&	77.6494	&	4.3915	&	46	&	Rars	&	1	&	SM	&	2011-09-25	\\ 
136	&	T	&	G071.3$-$00.6	&	J201339.0$+$331507	&	20:13:39.0	&	33:15:07	&	71.3879	&	-0.6456	&	9	&	B	&	2	&	KP,GC	&	2009-08-15	\\ 
137	&	T	&	G075.3$-$01.9	&	J202946.0$+$354926	&	20:29:46.0	&	35:49:26	&	75.3928	&	-1.9088	&	16	&	Rr	&	1	&	IN	&	2006-08-29	\\ 
138	&	T	&	G077.4$-$04.0	&	J204414.1$+$360737	&	20:44:14.1	&	36:07:37	&	77.4005	&	-4.063	&	19	&	Es	&	1	&	SM	&	2013-05-07	\\ 
139	&	T$^{\times}$	&	G079.5$-$03.8	&	J205002.8$+$375315	&	20:50:02.8	&	37:53:15	&	79.5049	&	-3.8688	&	35	&	Ears	&	1	&	SM	&	2013-05-09	\\ 
140	&	L	&	G086.5$+$01.8	&	J205013.6$+$465515	&	20:50:13.6	&	46:55:15	&	86.5191	&	1.8283	&	332	&	Ea	&	2	&	WH	&	2007-07-17	\\ 
141	&	T$^{\dagger}$	&	G081.0$-$03.9	&	J205527.2$+$390359	&	20:55:27.2	&	39:03:59	&	81.0914	&	-3.9315	&	23	&	B	&	2	&	SM,IN	&	2006-08-30	\\ 
142	&	L$^{\times}$	&	G088.0$+$00.4	&	J210204.7$+$471015	&	21:02:04.7	&	47:10:15	&	88.0182	&	0.4528	&	71	&	Eamrs	&	3	&	OS,SM,IN	&	2006-08-29	\\ 
143	&	T$^{\dagger}$	&	G093.3$+$01.4	&	J212000.0$+$514105	&	21:20:00.0	&	51:41:05	&	93.3035	&	1.4107	&	9	&	Bps	&	1	&	GC	&	2010-06-07	\\ 
144	&	T	&	G090.5$-$01.7	&	J212151.8$+$473301	&	21:21:51.8	&	47:33:01	&	90.5918	&	-1.7251	&	30	&	Bs	&	1	&	IN	&	2006-08-29	\\ 
145	&	T$^{\dagger}$	&	G095.9$+$03.5	&	J212200.9$+$550430	&	21:22:00.9	&	55:04:30	&	95.9178	&	3.5929	&	56	&	Bams	&	1	&	WH	&	2006-08-25	\\ 
146	&	T	&	G091.6$-$01.0	&	J212335.3$+$484717	&	21:23:35.3	&	48:47:17	&	91.667	&	-1.0472	&	19	&	B	&	1	&	GC	&	2011-07-06	\\ 
147	&	T	&	G095.8$+$02.6	&	J212608.3$+$542015	&	21:26:08.3	&	54:20:15	&	95.8257	&	2.6495	&	13	&	Bas	&	1	&	GC	&	2011-06-28	\\ 
148	&	T$^{\star}$	&	G095.5$+$00.5	&	J213423.2$+$523727	&	21:34:23.2	&	52:37:27	&	95.5461	&	0.561	&	7	&	E	&	1	&	GC	&	2011-04-13	\\ 
149	&	T	&	G098.9$+$03.0	&	J214032.5$+$564751	&	21:40:32.5	&	56:47:51	&	98.9933	&	3.0779	&	9	&	E	&	1	&	IN	&	2006-08-01	\\ 
150	&	T	&	G098.9$-$01.1	&	J215842.3$+$533003	&	21:58:42.3	&	53:30:03	&	98.9076	&	-1.1146	&	31	&	Rar	&	1	&	SM	&	2011-09-25	\\ 
151	&	T	&	G101.5$-$00.6	&	J221118.0$+$552841	&	22:11:18.0	&	55:28:41	&	101.551	&	-0.6008	&	78	&	Bas	&	1	&	IN	&	2006-08-29	\\ 
152	&	L	&	G111.5$+$00.1	&	J231546.1$+$605551	&	23:15:46.1	&	60:55:51	&	111.572	&	0.187	&	60	&	Er	&	1	&	OS	&	2011-10-10	\\ 
153	&	T$^{\dagger}$	&	G114.2$+$03.7	&	J232713.2$+$650923	&	23:27:13.2	&	65:09:23	&	114.232	&	3.717	&	19	&	Rs	&	1	&	KP	&	2009-08-16	\\ 
154	&	T	&	G114.4$+$00.0	&	J233841.2$+$614146	&	23:38:41.2	&	61:41:46	&	114.421	&	0.0324	&	61	&	Ears	&	2	&	OS,IN	&	2006-08-30	\\ 
155	&	P	&	G115.6$+$03.5	&	J234025.5$+$652147	&	23:40:25.5	&	65:21:47	&	115.616	&	3.5051	&	62	&	Ra	&	1	&	OS	&	2011-10-08	\\ 
156	&	T	&	G115.5$+$02.0	&	J234318.0$+$635717	&	23:43:18.0	&	63:57:17	&	115.537	&	2.0656	&	36	&	R	&	1	&	OS	&	2011-10-07	\\ 
157	&	T	&	G114.7$-$01.2	&	J234403.8$+$603242	&	23:44:03.8	&	60:32:42	&	114.738	&	-1.2504	&	22	&	Ras	&	3	&	KP,IN	&	2006-08-29	\\ 
158	&	T	&	G116.3$+$01.9	&	J235044.2$+$640311	&	23:50:44.2	&	64:03:11	&	116.351	&	1.9598	&	26	&	Ba	&	1	&	SM	&	2011-09-23	\\ 
159	&	P	&	G115.5$-$01.8	&	J235114.7$+$601026	&	23:51:14.7	&	60:10:26	&	115.509	&	-1.8274	&	48	&	Ra	&	1	&	OS	&	2011-10-08	\\ 
\hline
\end{tabular}
\end{table*}
\clearpage

\appendix
\setcounter{figure}{0} \renewcommand{\thefigure}{A.\arabic{figure}}

\begin{figure*}
\includegraphics[height=5.1cm]{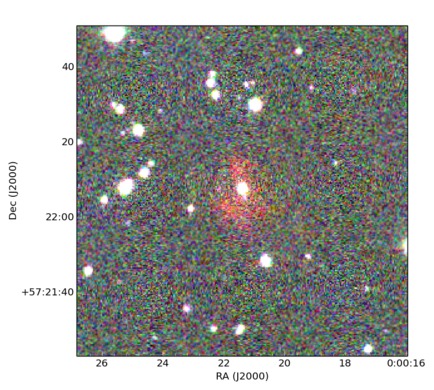}
\includegraphics[height=5.1cm]{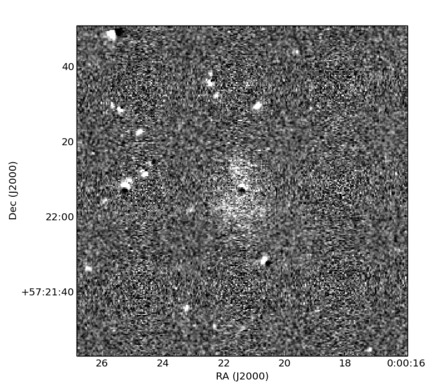}
\includegraphics[height=5.1cm]{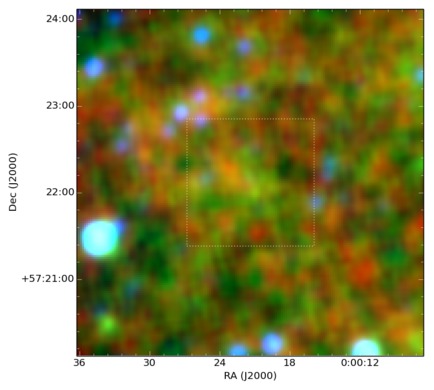}
\includegraphics[height=5.1cm]{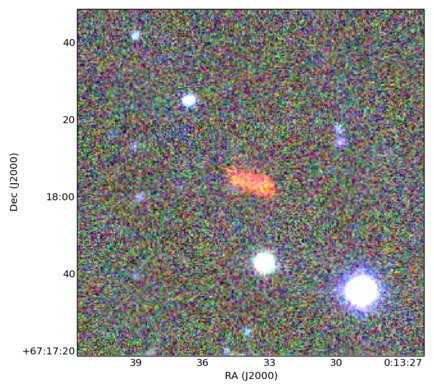}
\includegraphics[height=5.1cm]{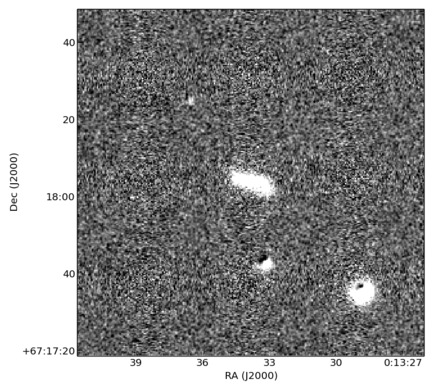}
\includegraphics[height=5.1cm]{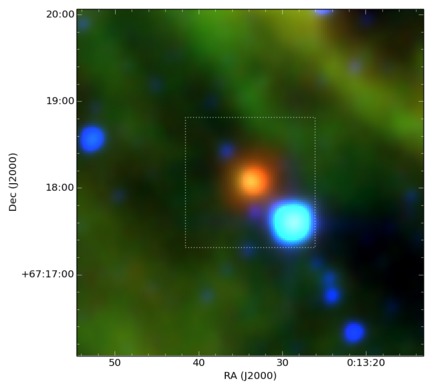}
\includegraphics[height=5.1cm]{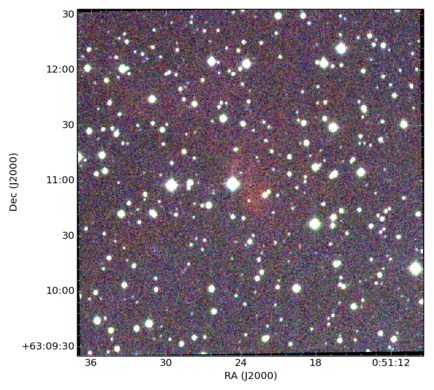}
\includegraphics[height=5.1cm]{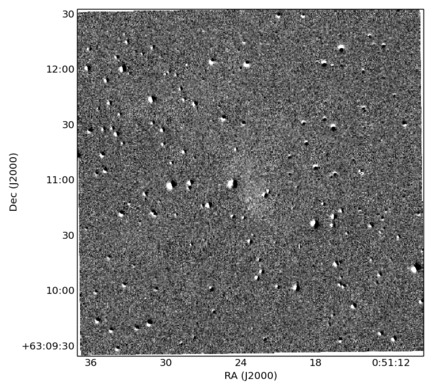}
\includegraphics[height=5.1cm]{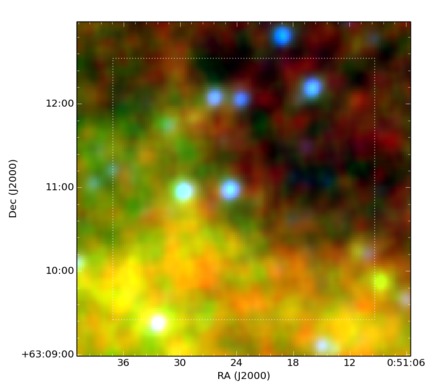}
\includegraphics[height=5.1cm]{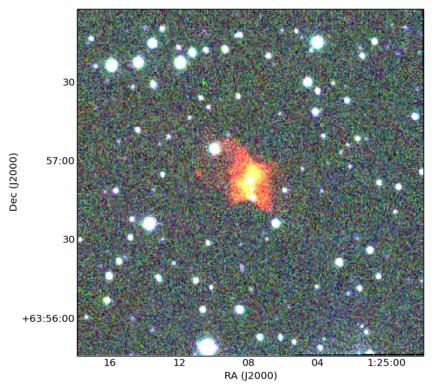}
\includegraphics[height=5.1cm]{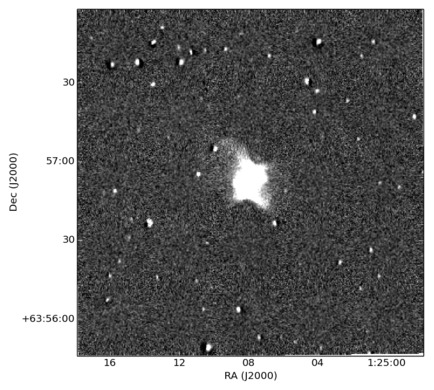}
\includegraphics[height=5.1cm]{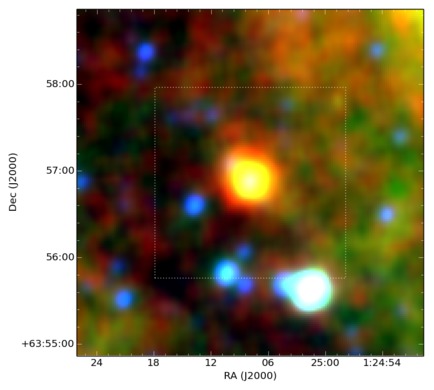}
\caption{\label{image1} Multiwavelength images of new IPHAS PNe. Columns from left to right: (1) IPHAS colour-composite image made from H$\alpha$(R), r(G) and i(B), (2) IPHAS quotient (H$\alpha$/r) and (3) WISE colour-composite made from W4(R), W3(G) and W2(B). Dashed-line boxes in WISE image represent size of a field in the IPHAS image. All images have same orientation: NE at the top left. Objects shown (from top to bottom): PN G116.0-04.8,PN G119.2+04.6,PN G122.9+00.3,PN G126.6+01.3}
\end{figure*}
\clearpage
\begin{figure*}
\includegraphics[height=5.1cm]{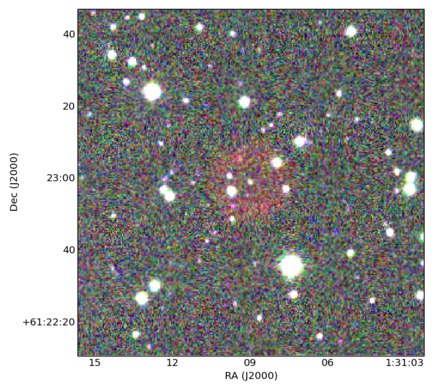}
\includegraphics[height=5.1cm]{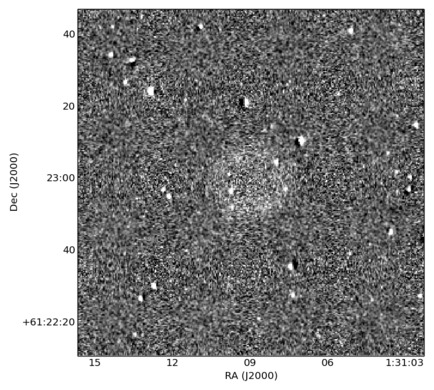}
\includegraphics[height=5.1cm]{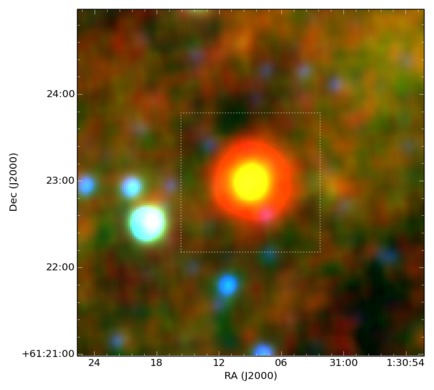}
\includegraphics[height=5.1cm]{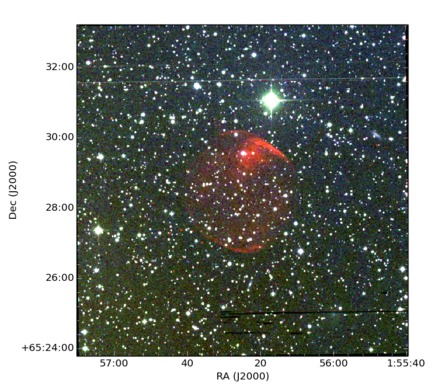}
\includegraphics[height=5.1cm]{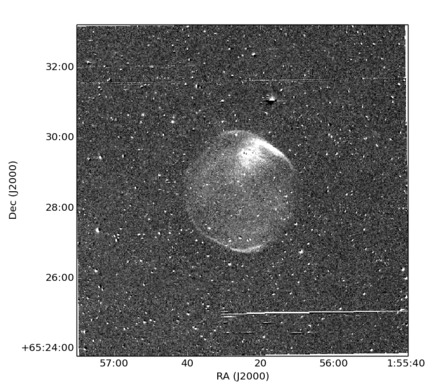}
\includegraphics[height=5.1cm]{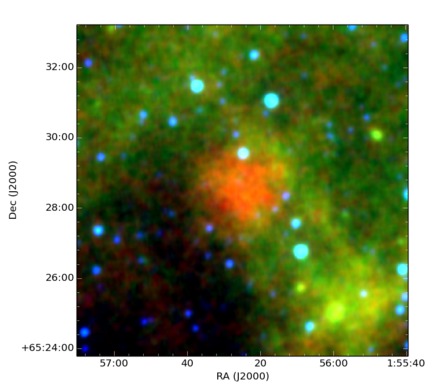}
\includegraphics[height=5.1cm]{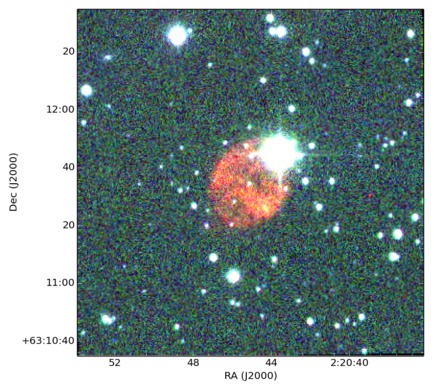}
\includegraphics[height=5.1cm]{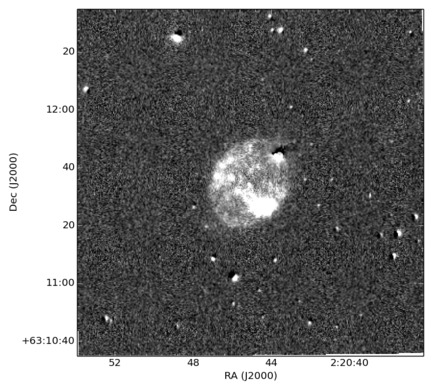}
\includegraphics[height=5.1cm]{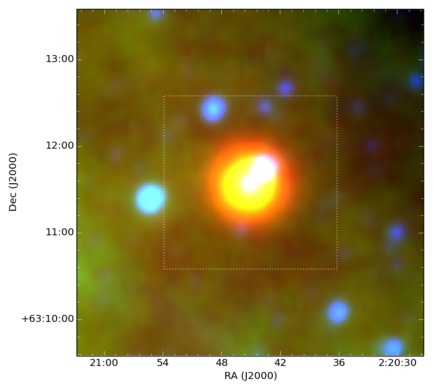}
\includegraphics[height=5.1cm]{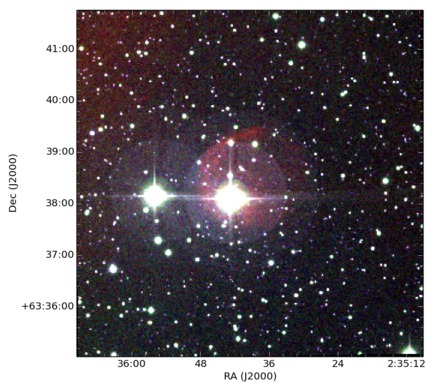}
\includegraphics[height=5.1cm]{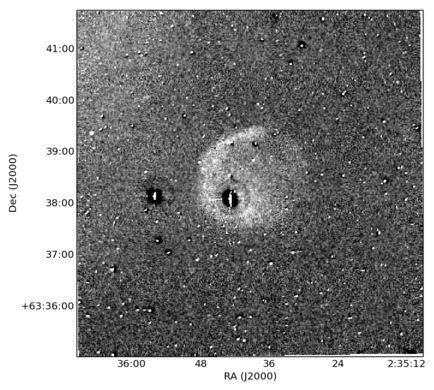}
\includegraphics[height=5.1cm]{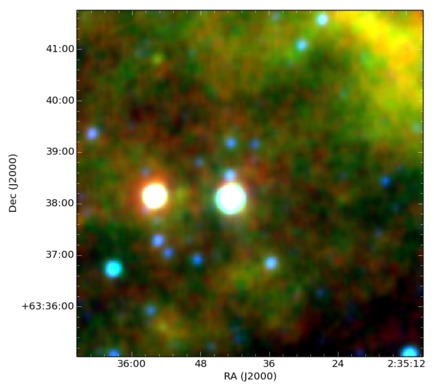}
\caption{\label{imagelabel} Same as in Fig.~\ref{image1}. Objects shown (from top to bottom):  PN G127.6-01.1,PN G129.6+03.4,PN G132.8+02.0,PN G134.1+03.0}
\end{figure*}
\clearpage
\begin{figure*}
\includegraphics[height=5.1cm]{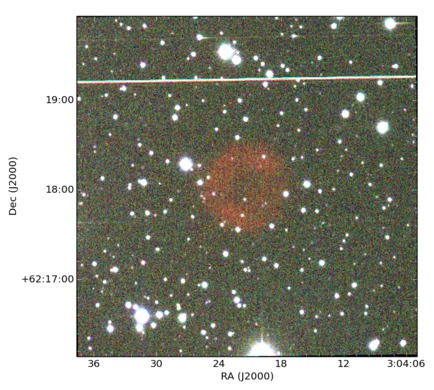}
\includegraphics[height=5.1cm]{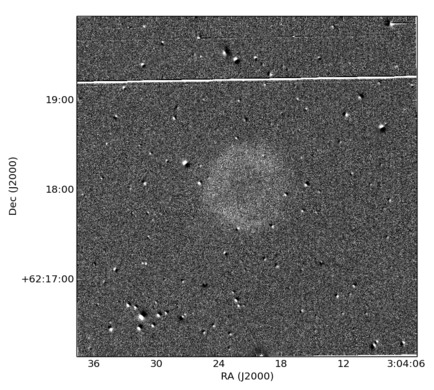}
\includegraphics[height=5.1cm]{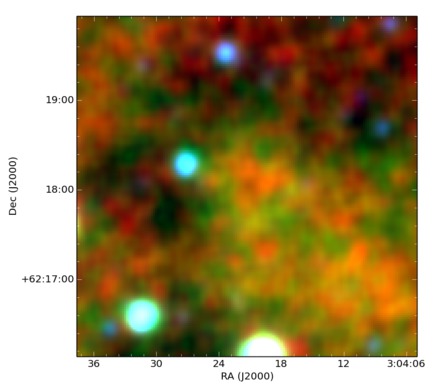}
\includegraphics[height=5.1cm]{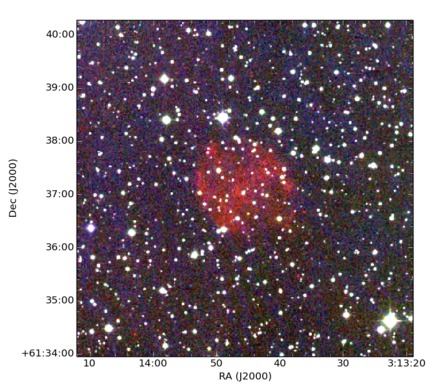}
\includegraphics[height=5.1cm]{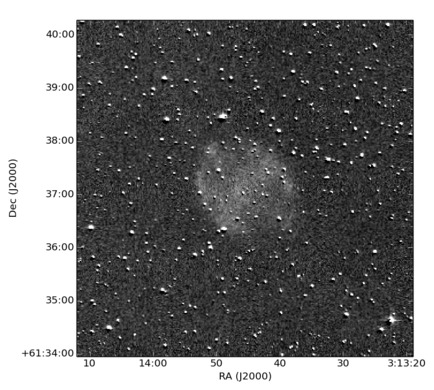}
\includegraphics[height=5.1cm]{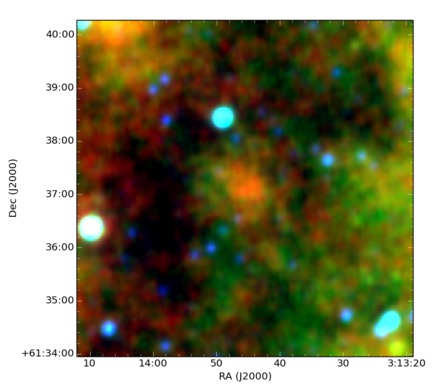}
\includegraphics[height=5.1cm]{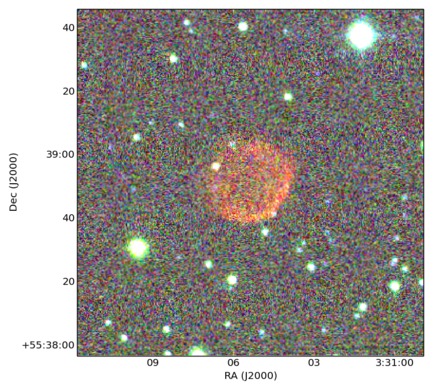}
\includegraphics[height=5.1cm]{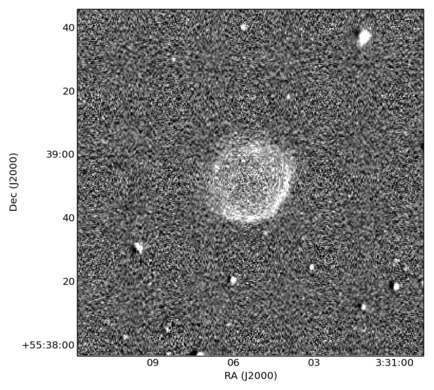}
\includegraphics[height=5.1cm]{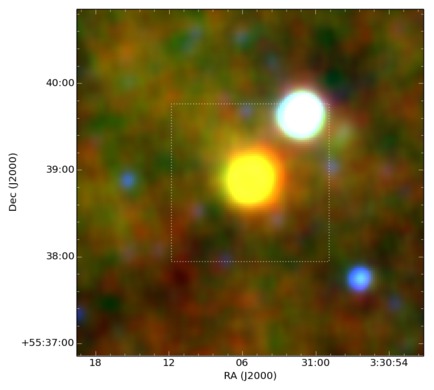}
\includegraphics[height=5.1cm]{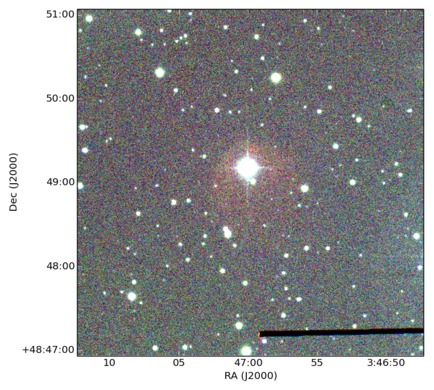}
\includegraphics[height=5.1cm]{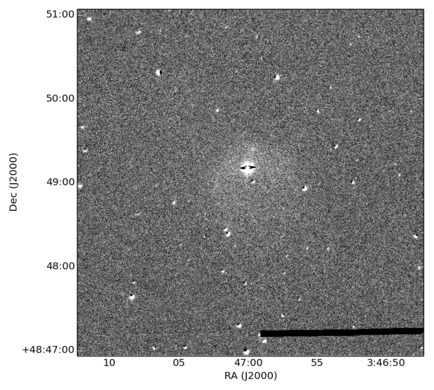}
\includegraphics[height=5.1cm]{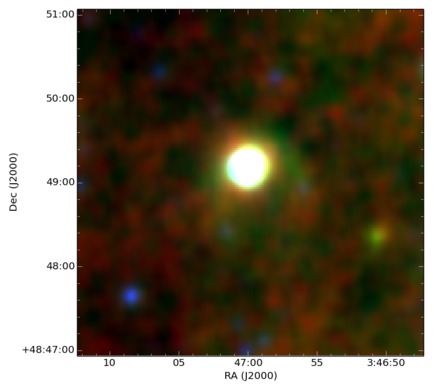}
\caption{\label{imagelabel} Same as in Fig.~\ref{image1}. Objects shown (from top to bottom):  PN G137.7+03.3,PN G139.0+03.2,PN G144.1-00.5,PN G150.1-04.5}
\end{figure*}
\clearpage
\begin{figure*}
\includegraphics[height=5.1cm]{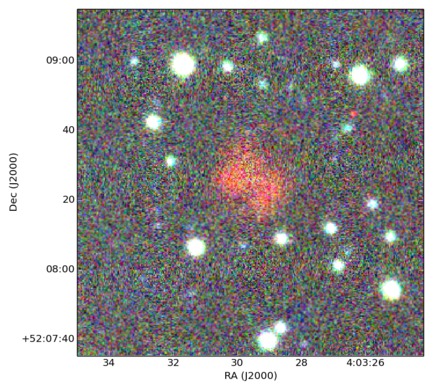}
\includegraphics[height=5.1cm]{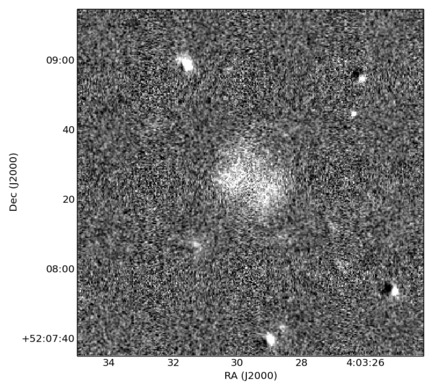}
\includegraphics[height=5.1cm]{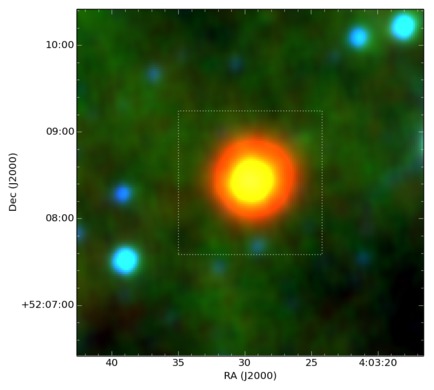}
\includegraphics[height=5.1cm]{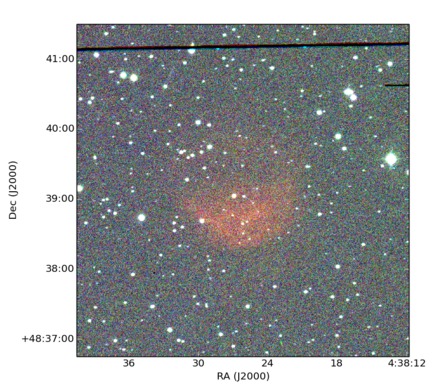}
\includegraphics[height=5.1cm]{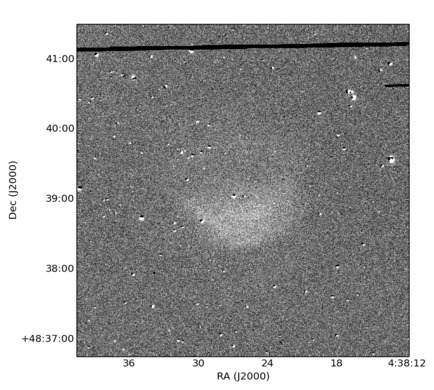}
\includegraphics[height=5.1cm]{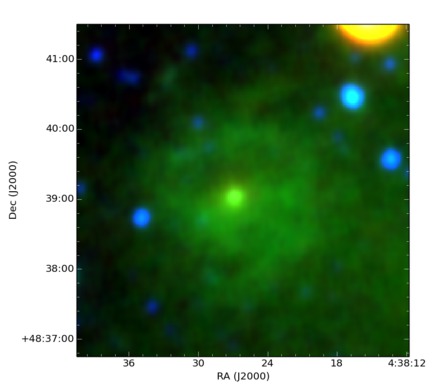}
\includegraphics[height=5.1cm]{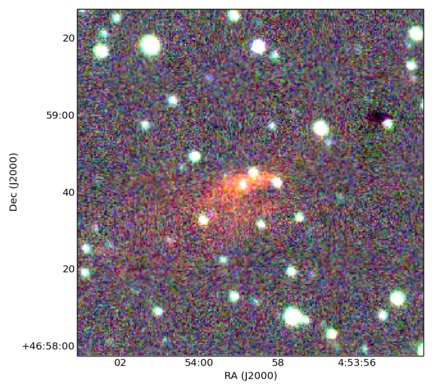}
\includegraphics[height=5.1cm]{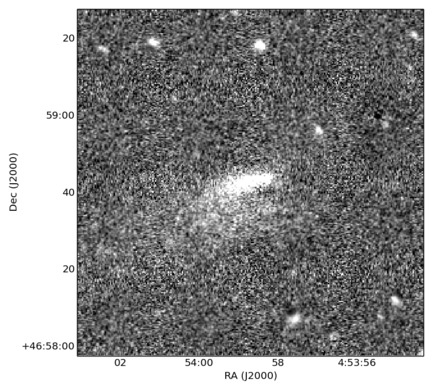}
\includegraphics[height=5.1cm]{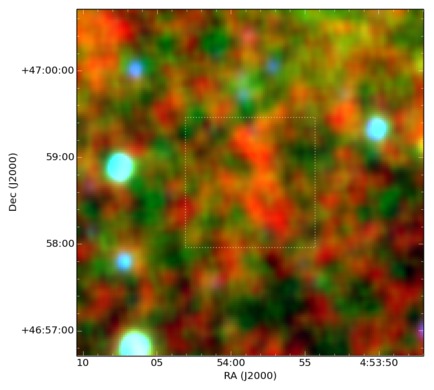}
\includegraphics[height=5.1cm]{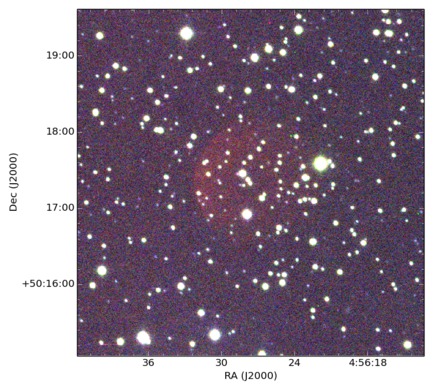}
\includegraphics[height=5.1cm]{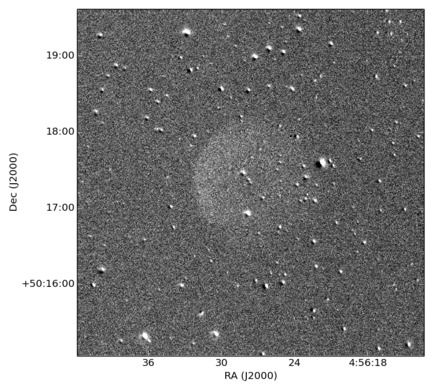}
\includegraphics[height=5.1cm]{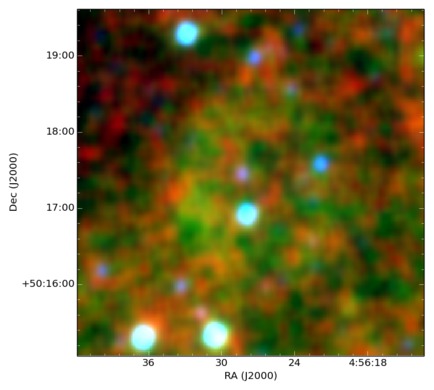}
\caption{\label{imagelabel} Same as in Fig.~\ref{image1}. Objects shown (from top to bottom):  PN G150.0-00.3,PN G156.4+01.1,PN G159.4+02.0,PN G157.1+04.4}
\end{figure*}
\clearpage
\begin{figure*}
\includegraphics[height=5.1cm]{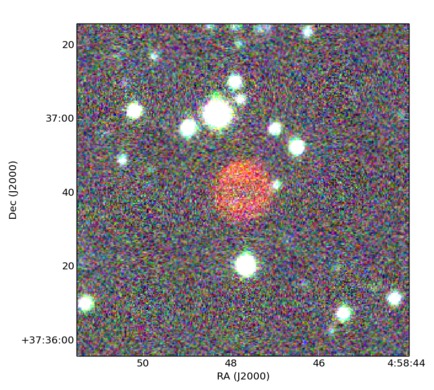}
\includegraphics[height=5.1cm]{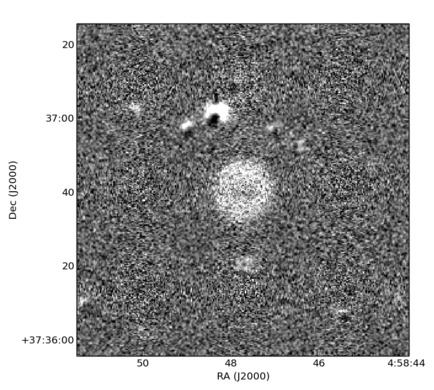}
\includegraphics[height=5.1cm]{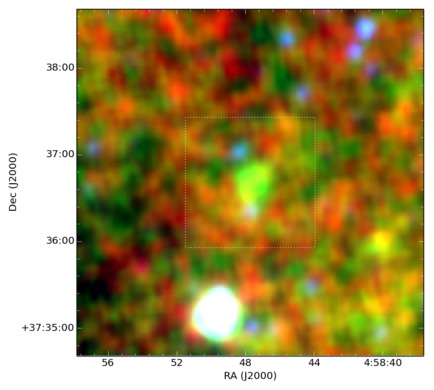}
\includegraphics[height=5.1cm]{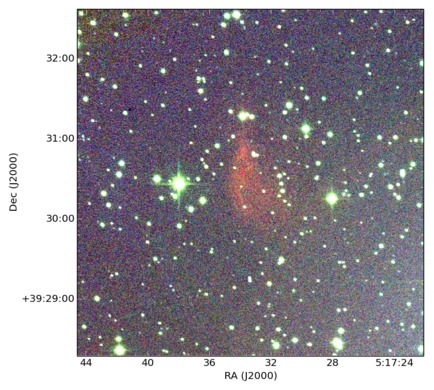}
\includegraphics[height=5.1cm]{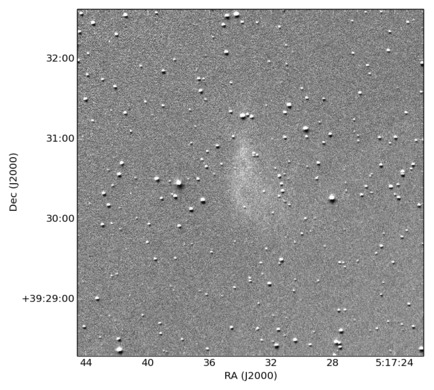}
\includegraphics[height=5.1cm]{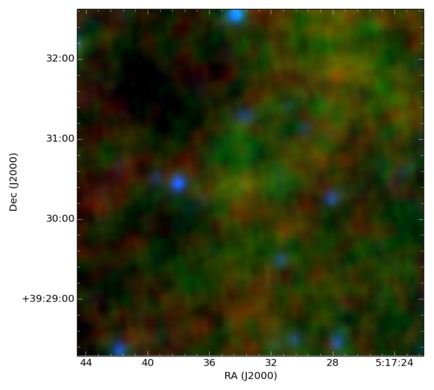}
\includegraphics[height=5.1cm]{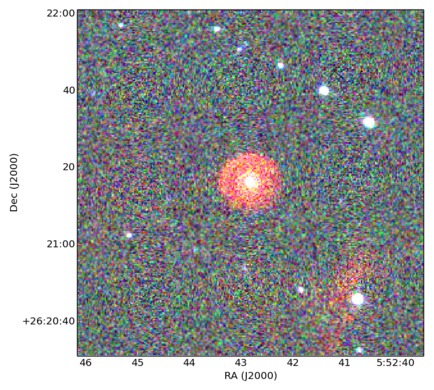}
\includegraphics[height=5.1cm]{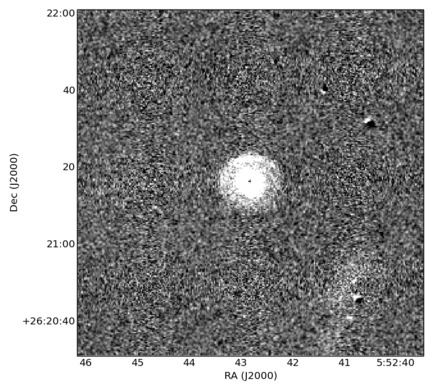}
\includegraphics[height=5.1cm]{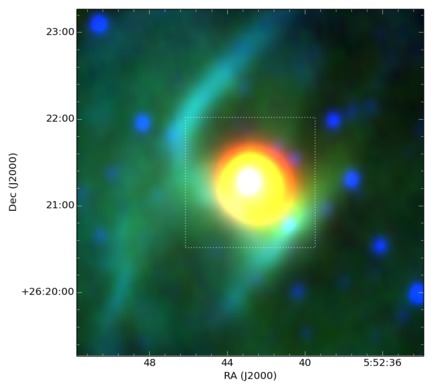}
\includegraphics[height=5.1cm]{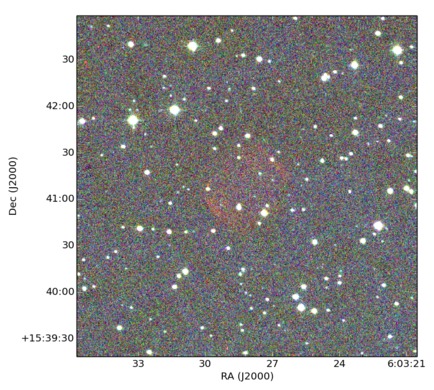}
\includegraphics[height=5.1cm]{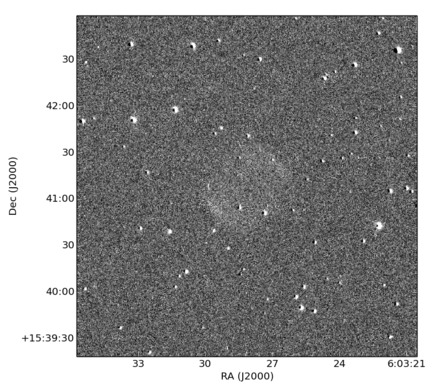}
\includegraphics[height=5.1cm]{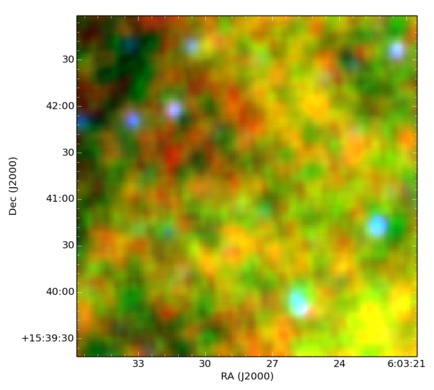}
\caption{\label{imagelabel} Same as in Fig.~\ref{image1}. Objects shown (from top to bottom):  PN G167.3-03.1,PN G167.9+00.9,PN G183.0+00.0,PN G193.5-03.1}
\end{figure*}
\clearpage
\begin{figure*}
\includegraphics[height=5.1cm]{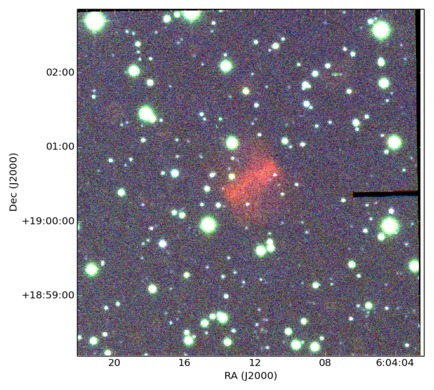}
\includegraphics[height=5.1cm]{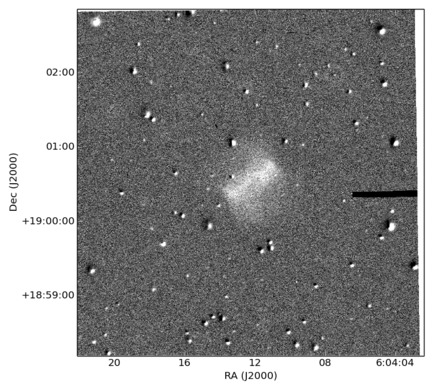}
\includegraphics[height=5.1cm]{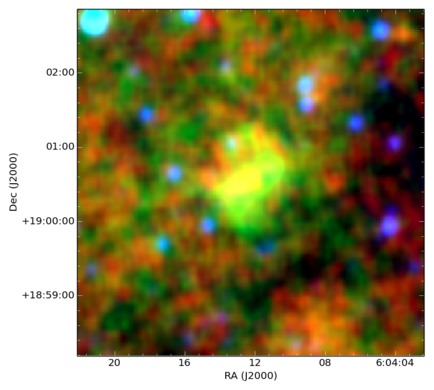}
\includegraphics[height=5.1cm]{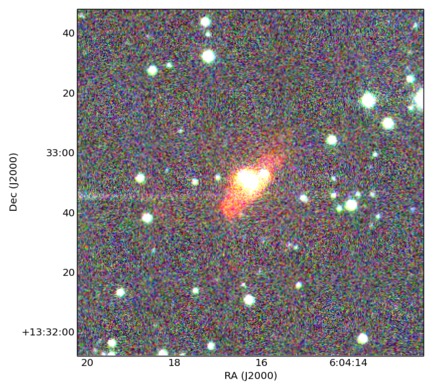}
\includegraphics[height=5.1cm]{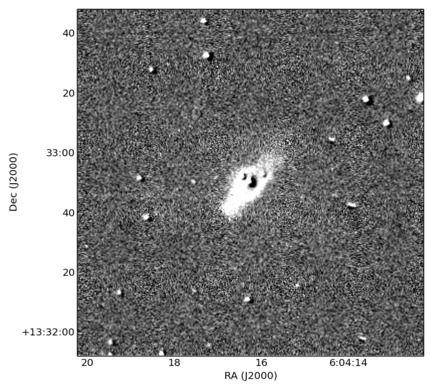}
\includegraphics[height=5.1cm]{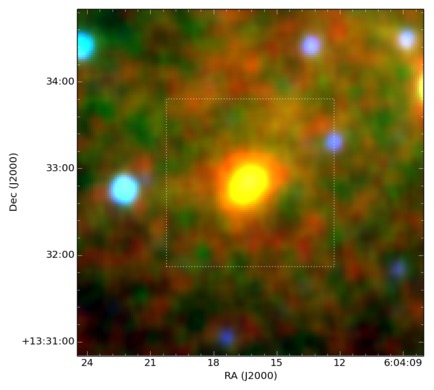}
\includegraphics[height=5.1cm]{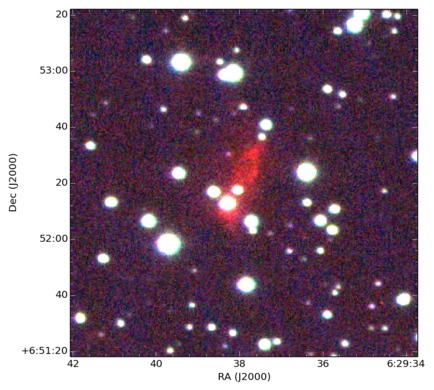}
\includegraphics[height=5.1cm]{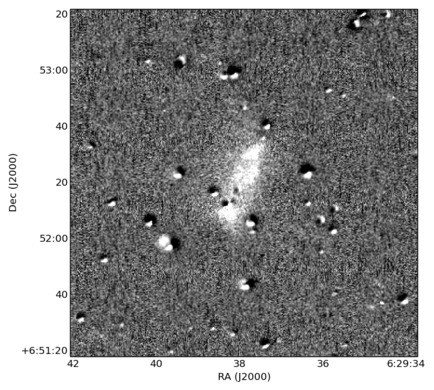}
\includegraphics[height=5.1cm]{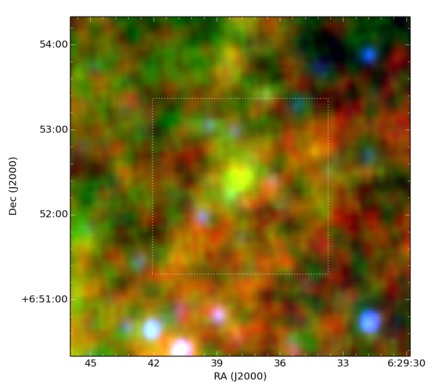}
\includegraphics[height=5.1cm]{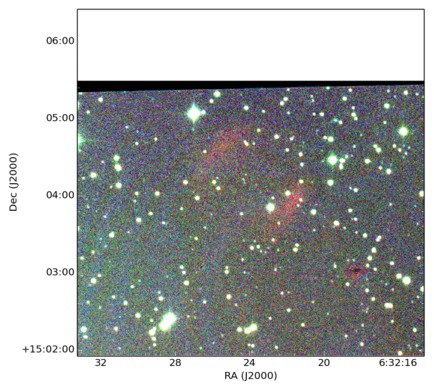}
\includegraphics[height=5.1cm]{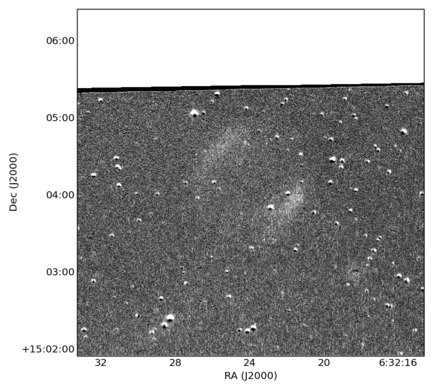}
\includegraphics[height=5.1cm]{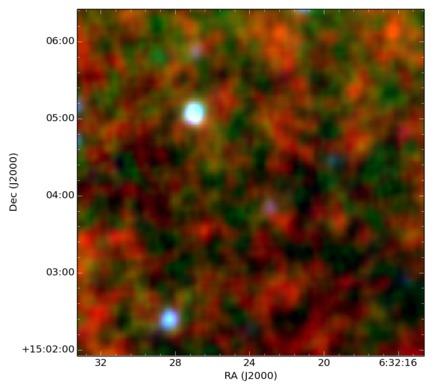}
\caption{\label{imagelabel} Same as in Fig.~\ref{image1}. Objects shown (from top to bottom):  PN G190.7-01.3,PN G195.4-04.0,PN G204.3-01.6,PN G197.3+02.7}
\end{figure*}
\clearpage
\begin{figure*}
\includegraphics[height=5.1cm]{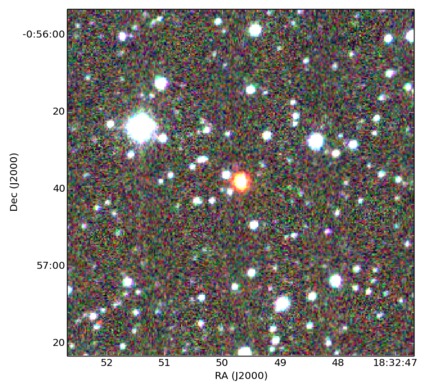}
\includegraphics[height=5.1cm]{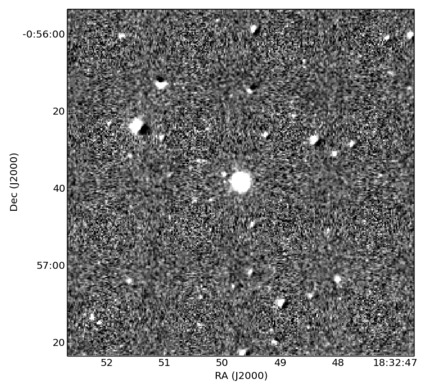}
\includegraphics[height=5.1cm]{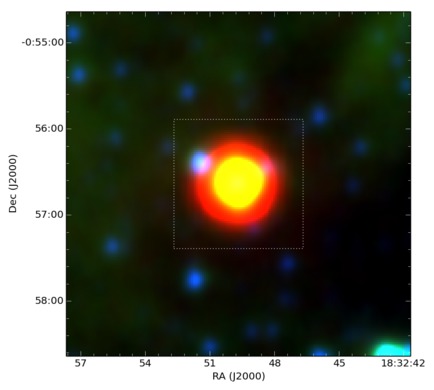}
\includegraphics[height=5.1cm]{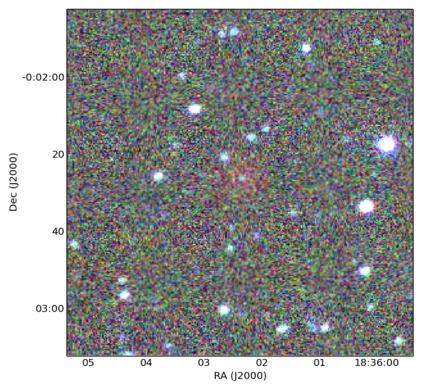}
\includegraphics[height=5.1cm]{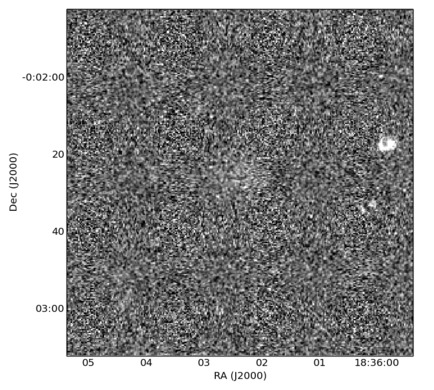}
\includegraphics[height=5.1cm]{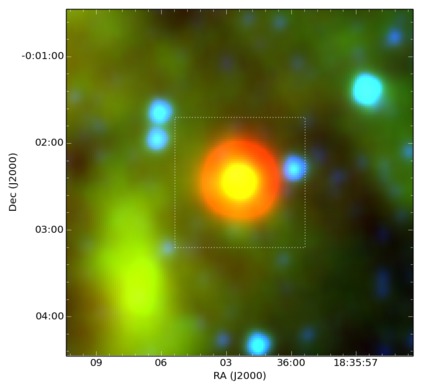}
\includegraphics[height=5.1cm]{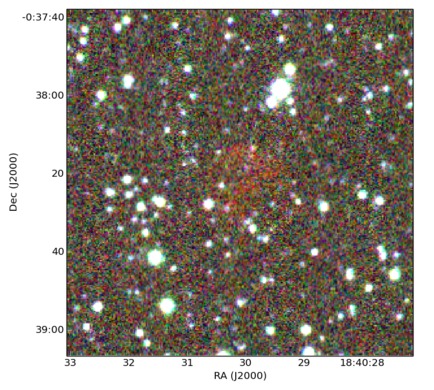}
\includegraphics[height=5.1cm]{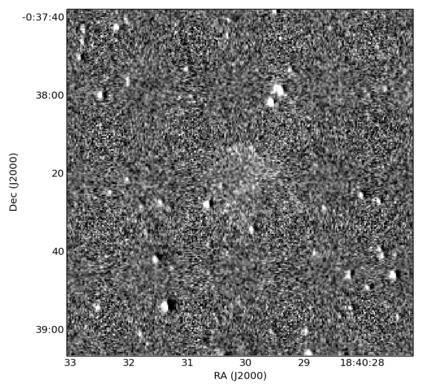}
\includegraphics[height=5.1cm]{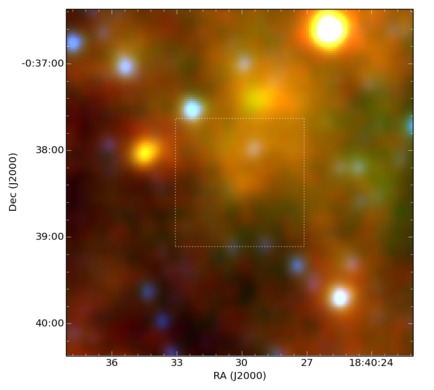}
\includegraphics[height=5.1cm]{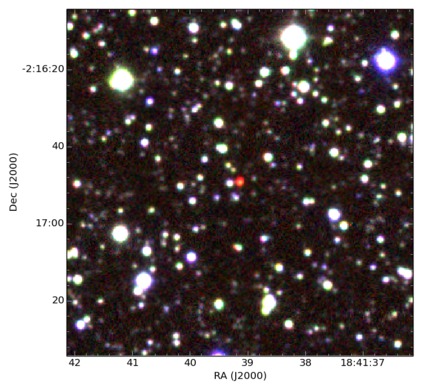}
\includegraphics[height=5.1cm]{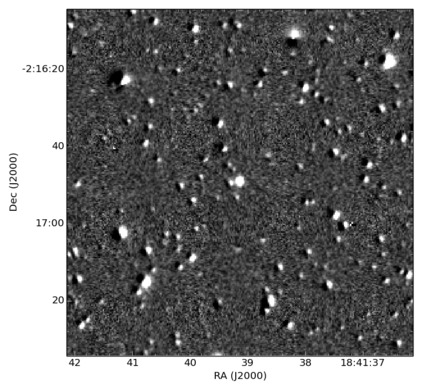}
\includegraphics[height=5.1cm]{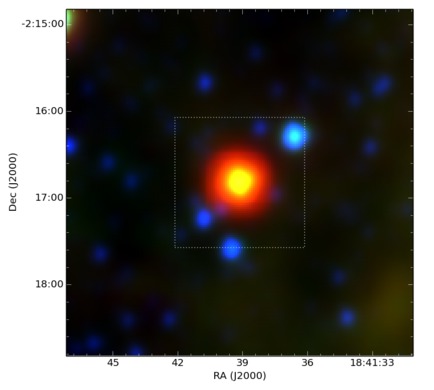}
\caption{\label{imagelabel} Same as in Fig.~\ref{image1}. Objects shown (from top to bottom):  PN G029.9+03.7,PN G031.1+03.4,PN G031.1+02.1,PN G029.7+01.1}
\end{figure*}
\clearpage
\begin{figure*}
\includegraphics[height=5.1cm]{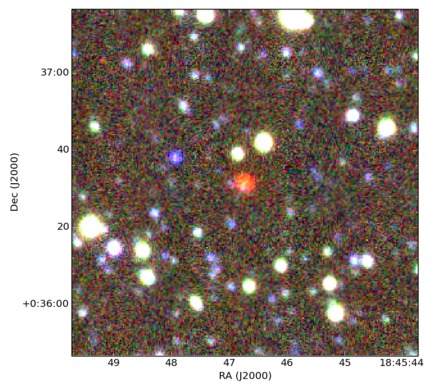}
\includegraphics[height=5.1cm]{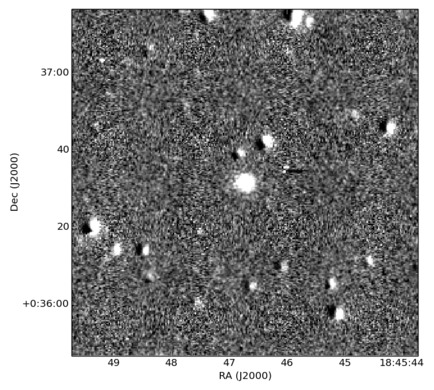}
\includegraphics[height=5.1cm]{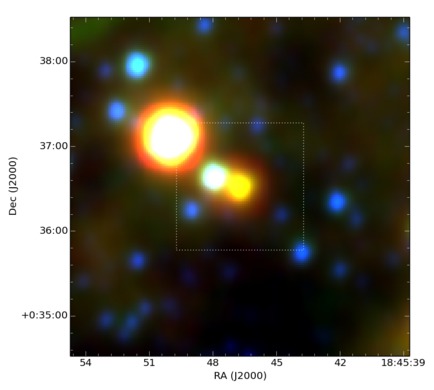}
\includegraphics[height=5.1cm]{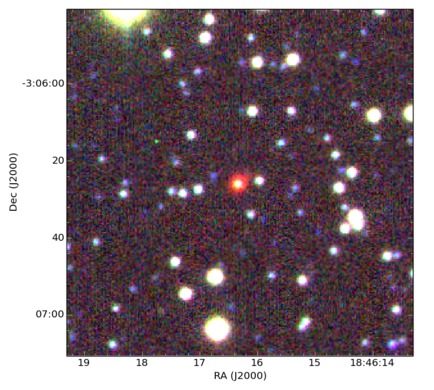}
\includegraphics[height=5.1cm]{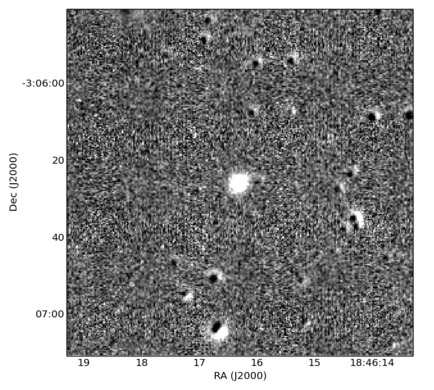}
\includegraphics[height=5.1cm]{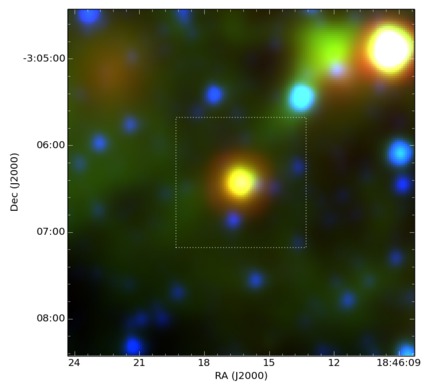}
\includegraphics[height=5.1cm]{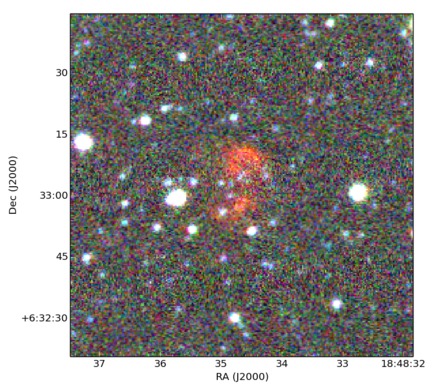}
\includegraphics[height=5.1cm]{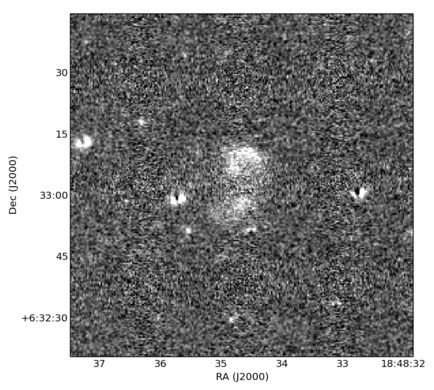}
\includegraphics[height=5.1cm]{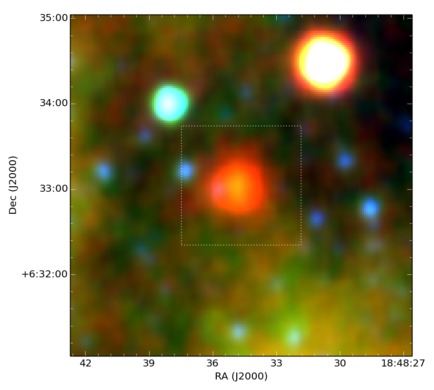}
\includegraphics[height=5.1cm]{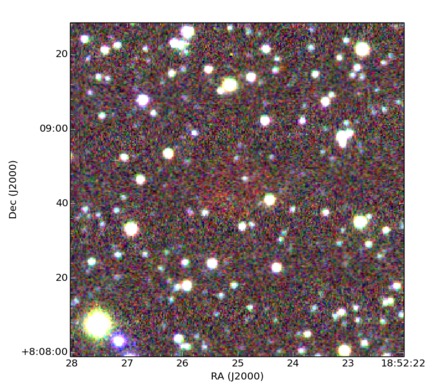}
\includegraphics[height=5.1cm]{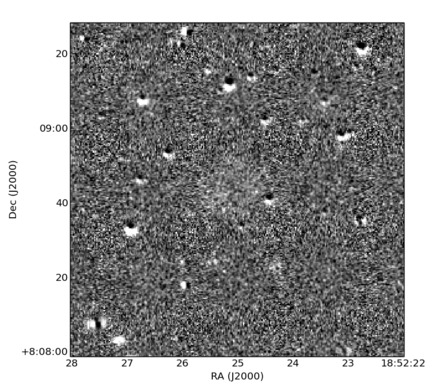}
\includegraphics[height=5.1cm]{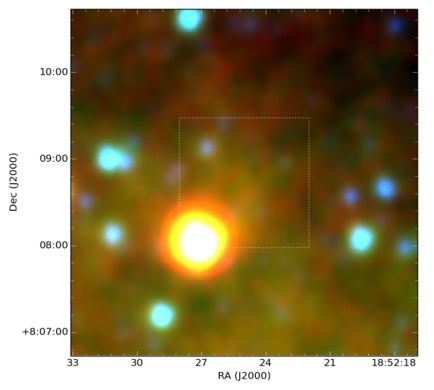}
\caption{\label{imagelabel} Same as in Fig.~\ref{image1}. Objects shown (from top to bottom):  PN G032.8+01.5,PN G029.5-00.2,PN G038.4+03.6,PN G040.3+03.4}
\end{figure*}
\clearpage
\begin{figure*}
\includegraphics[height=5.1cm]{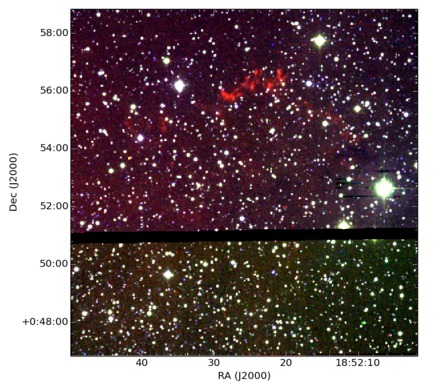}
\includegraphics[height=5.1cm]{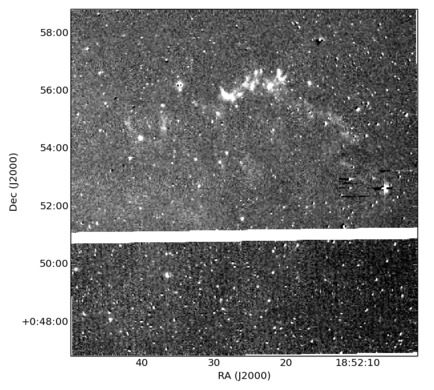}
\includegraphics[height=5.1cm]{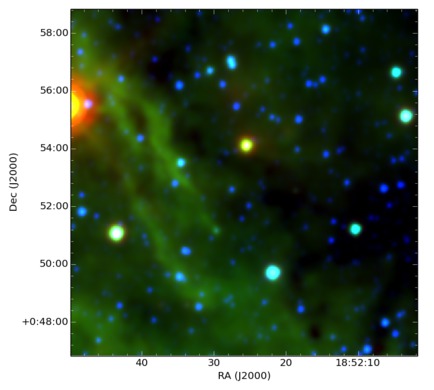}
\includegraphics[height=5.1cm]{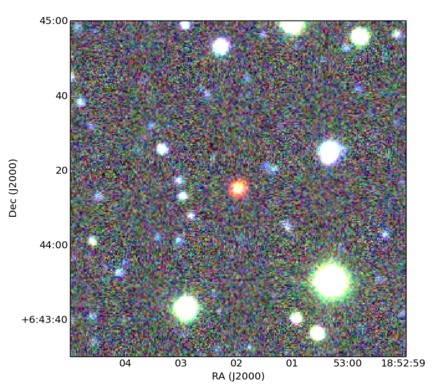}
\includegraphics[height=5.1cm]{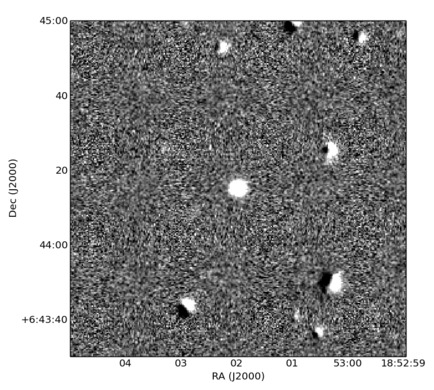}
\includegraphics[height=5.1cm]{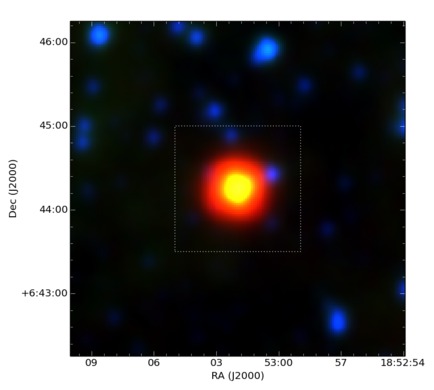}
\includegraphics[height=5.1cm]{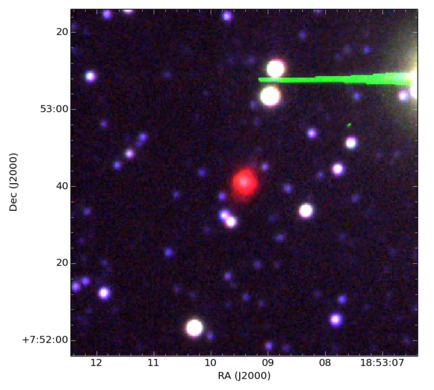}
\includegraphics[height=5.1cm]{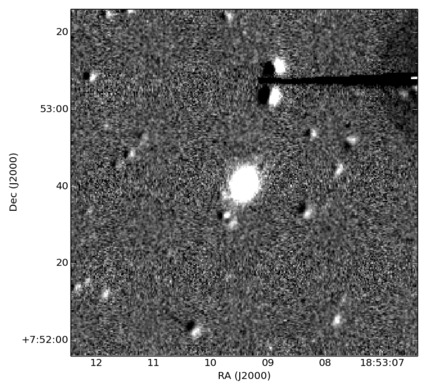}
\includegraphics[height=5.1cm]{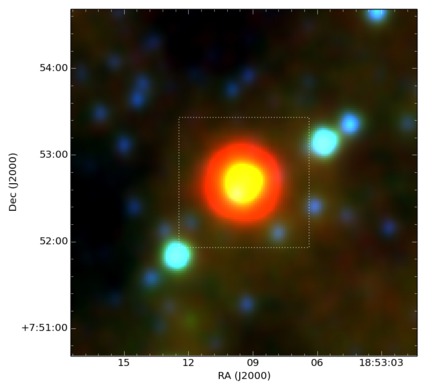}
\includegraphics[height=5.1cm]{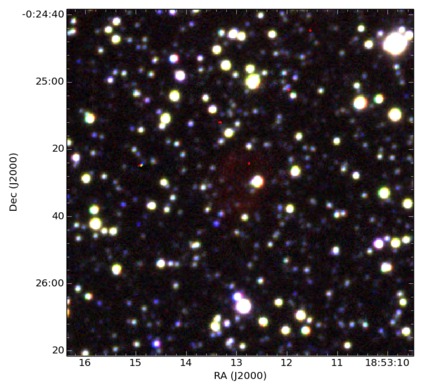}
\includegraphics[height=5.1cm]{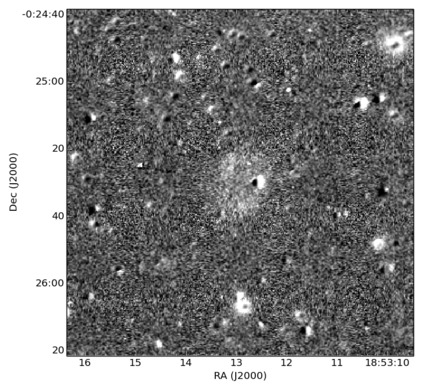}
\includegraphics[height=5.1cm]{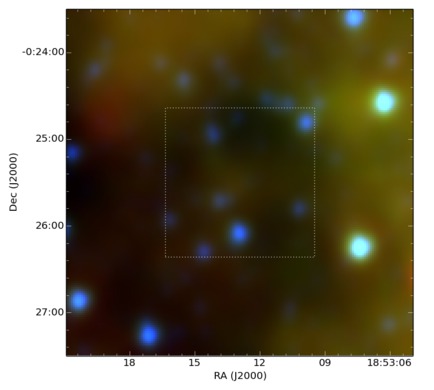}
\caption{\label{imagelabel} Same as in Fig.~\ref{image1}. Objects shown (from top to bottom):  PN G033.8+00.1,PN G039.1+02.7,PN G040.1+03.2,PN G032.7-00.5}
\end{figure*}
\clearpage
\begin{figure*}
\includegraphics[height=5.1cm]{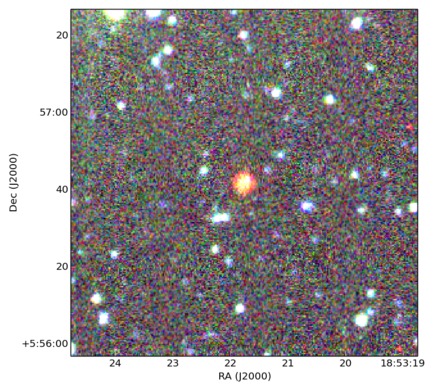}
\includegraphics[height=5.1cm]{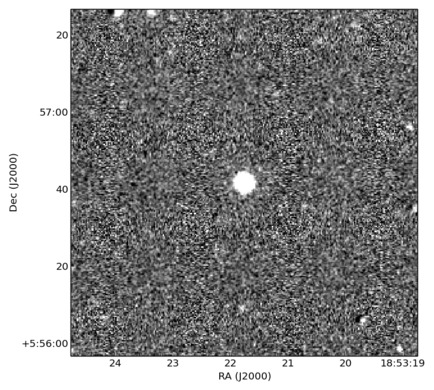}
\includegraphics[height=5.1cm]{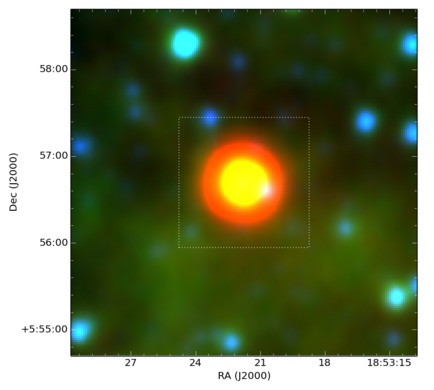}
\includegraphics[height=5.1cm]{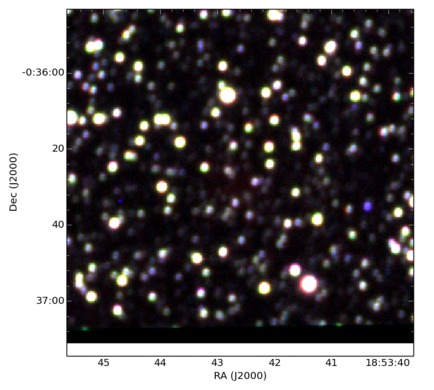}
\includegraphics[height=5.1cm]{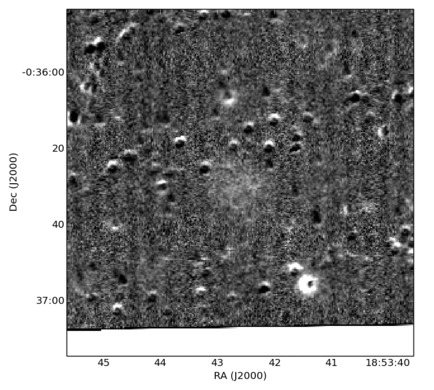}
\includegraphics[height=5.1cm]{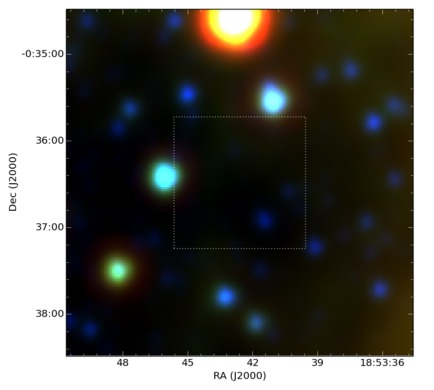}
\includegraphics[height=5.1cm]{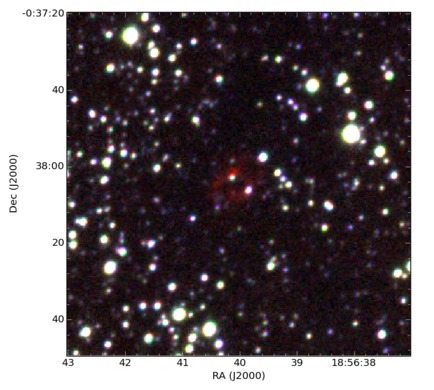}
\includegraphics[height=5.1cm]{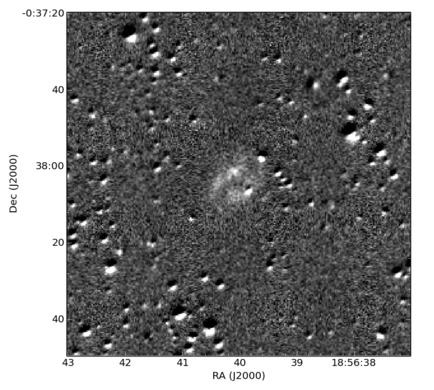}
\includegraphics[height=5.1cm]{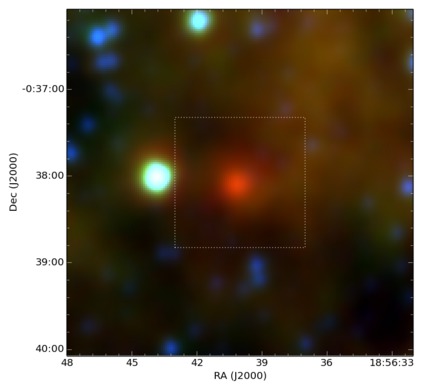}
\includegraphics[height=5.1cm]{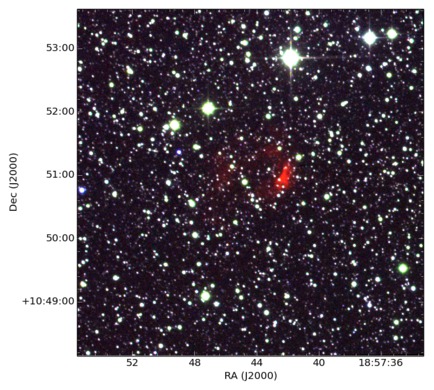}
\includegraphics[height=5.1cm]{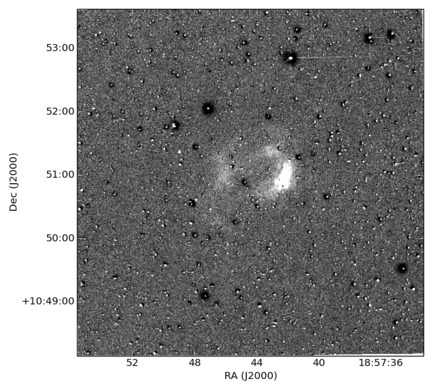}
\includegraphics[height=5.1cm]{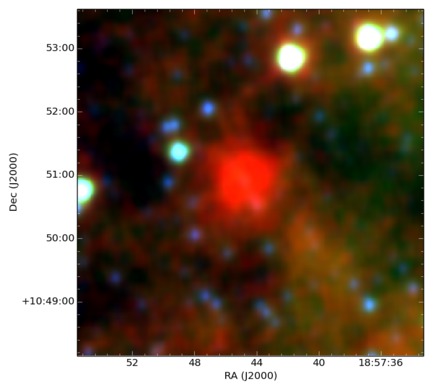}
\caption{\label{imagelabel} Same as in Fig.~\ref{image1}. Objects shown (from top to bottom):  PN G038.4+02.2,PN G032.6-00.7,PN G032.9-01.4,PN G043.3+03.5}
\end{figure*}
\clearpage
\begin{figure*}
\includegraphics[height=5.1cm]{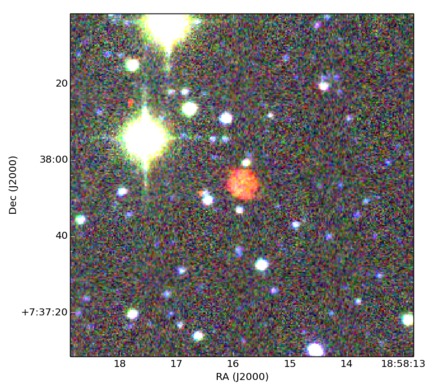}
\includegraphics[height=5.1cm]{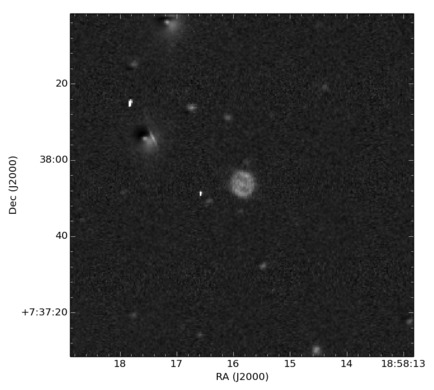}
\includegraphics[height=5.1cm]{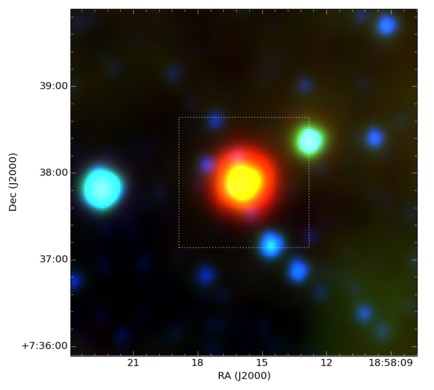}
\includegraphics[height=5.1cm]{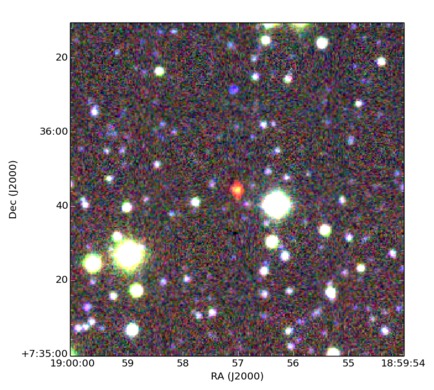}
\includegraphics[height=5.1cm]{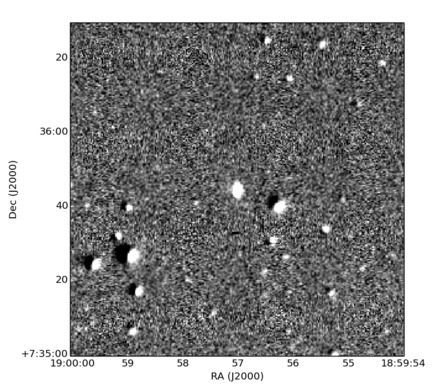}
\includegraphics[height=5.1cm]{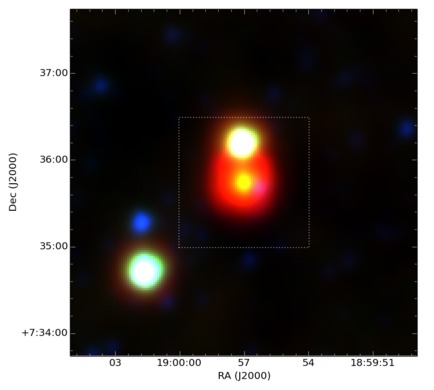}
\includegraphics[height=5.1cm]{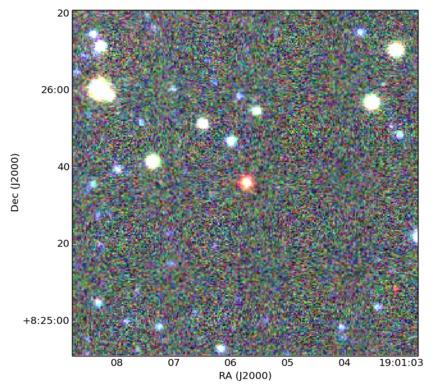}
\includegraphics[height=5.1cm]{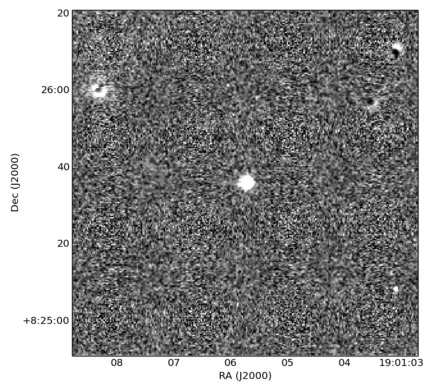}
\includegraphics[height=5.1cm]{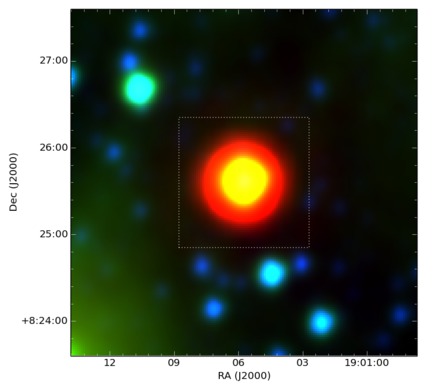}
\includegraphics[height=5.1cm]{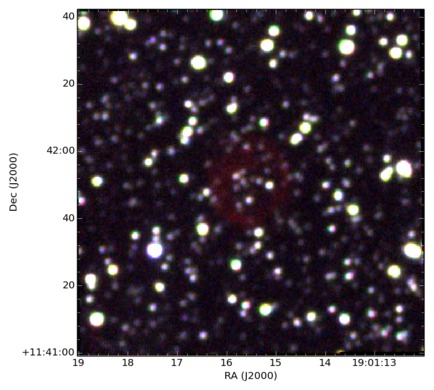}
\includegraphics[height=5.1cm]{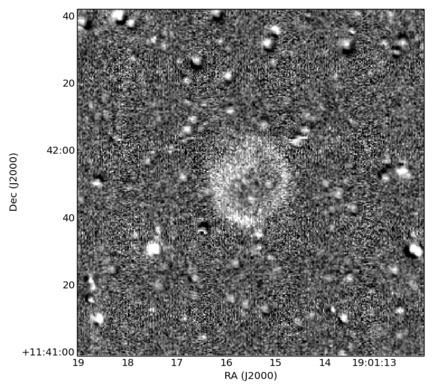}
\includegraphics[height=5.1cm]{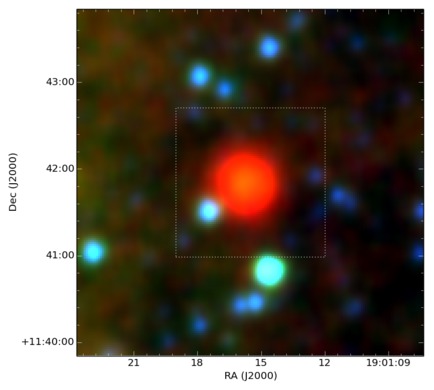}
\caption{\label{imagelabel} Same as in Fig.~\ref{image1}. Objects shown (from top to bottom):  PN G040.5+01.9,PN G040.6+01.5,PN G041.5+01.7,PN G044.4+03.1}
\end{figure*}
\clearpage
\begin{figure*}
\includegraphics[height=5.1cm]{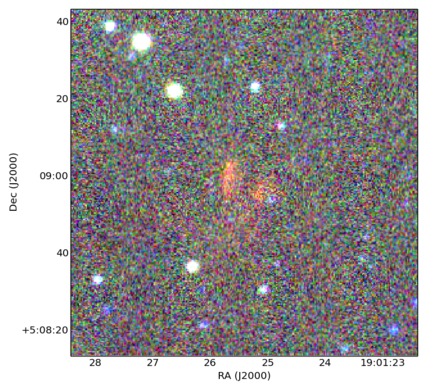}
\includegraphics[height=5.1cm]{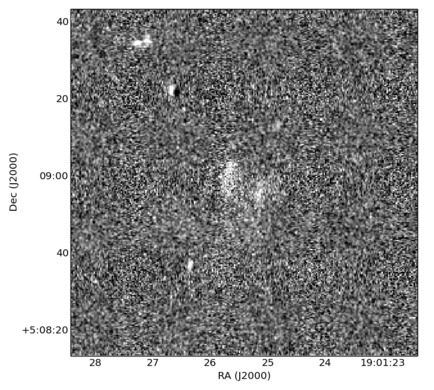}
\includegraphics[height=5.1cm]{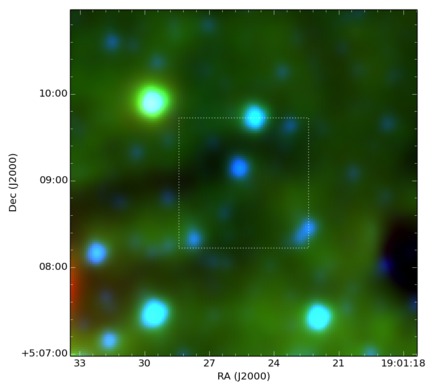}
\includegraphics[height=5.1cm]{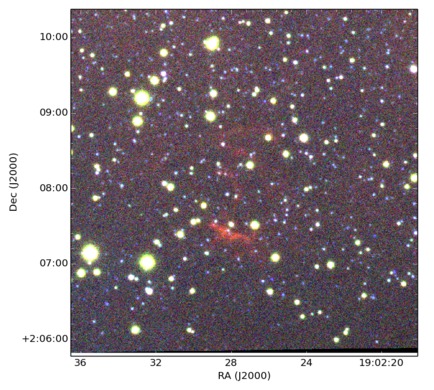}
\includegraphics[height=5.1cm]{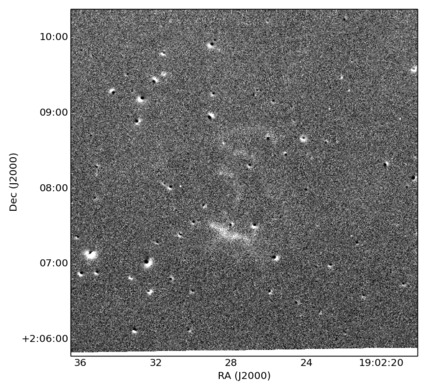}
\includegraphics[height=5.1cm]{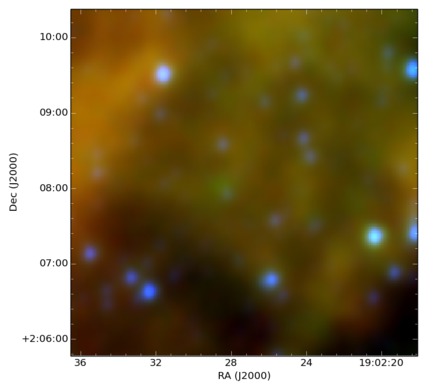}
\includegraphics[height=5.1cm]{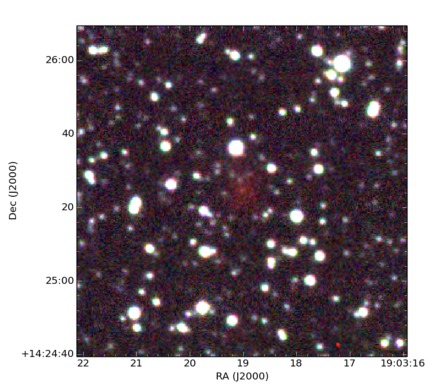}
\includegraphics[height=5.1cm]{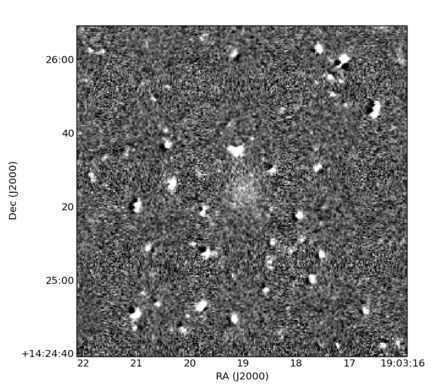}
\includegraphics[height=5.1cm]{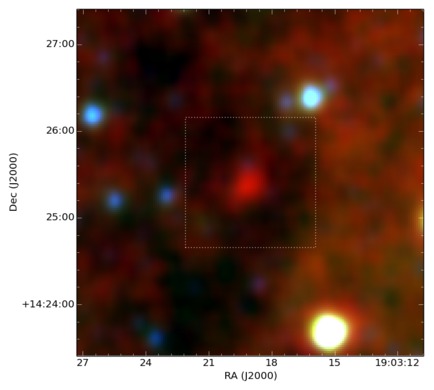}
\includegraphics[height=5.1cm]{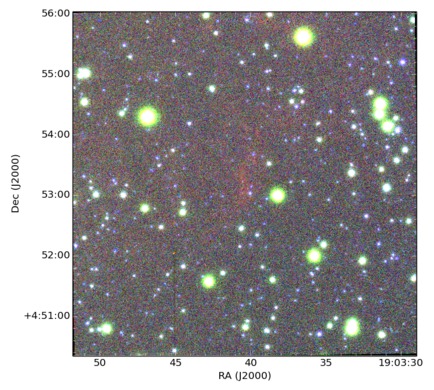}
\includegraphics[height=5.1cm]{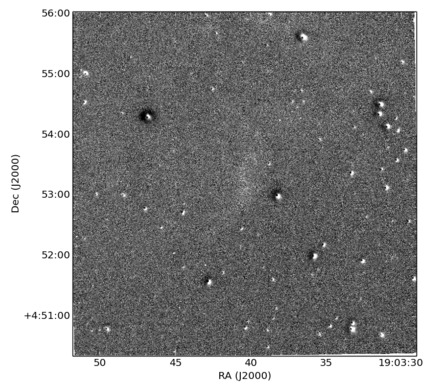}
\includegraphics[height=5.1cm]{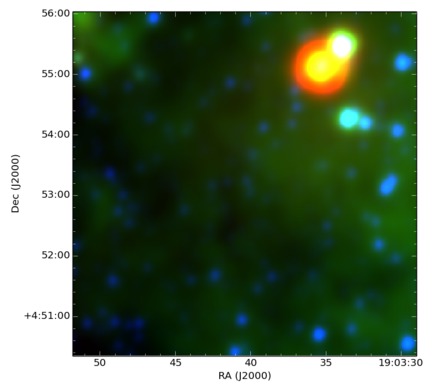}
\caption{\label{imagelabel} Same as in Fig.~\ref{image1}. Objects shown (from top to bottom):  PN G038.6+00.1,PN G036.0-01.4,PN G047.1+03.9,PN G038.6-00.4}
\end{figure*}
\clearpage
\begin{figure*}
\includegraphics[height=5.1cm]{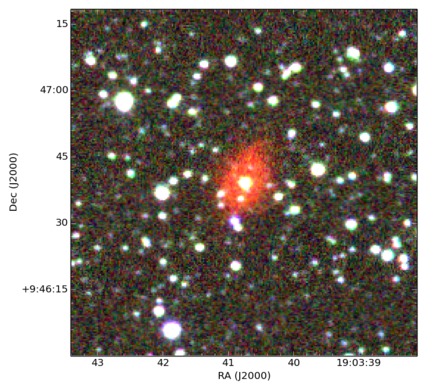}
\includegraphics[height=5.1cm]{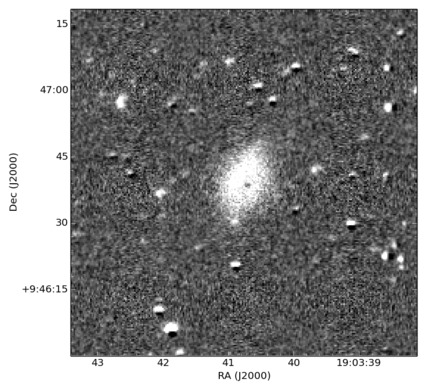}
\includegraphics[height=5.1cm]{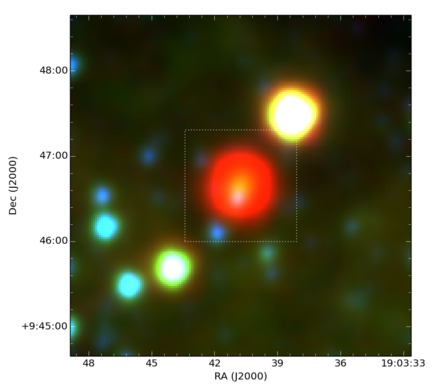}
\includegraphics[height=5.1cm]{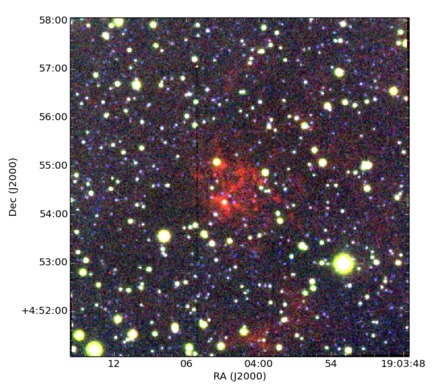}
\includegraphics[height=5.1cm]{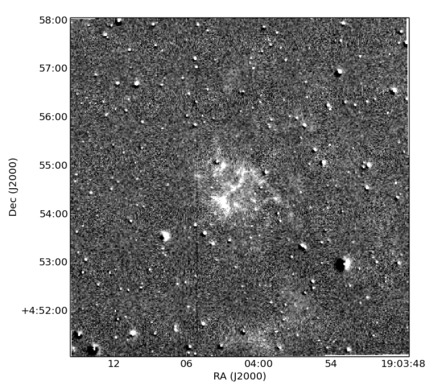}
\includegraphics[height=5.1cm]{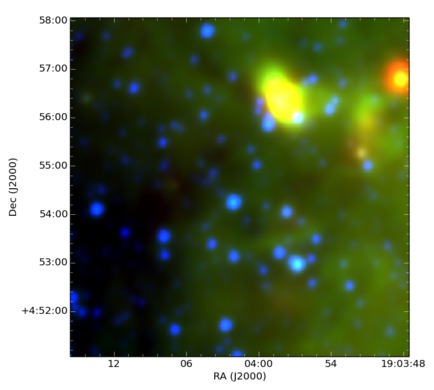}
\includegraphics[height=5.1cm]{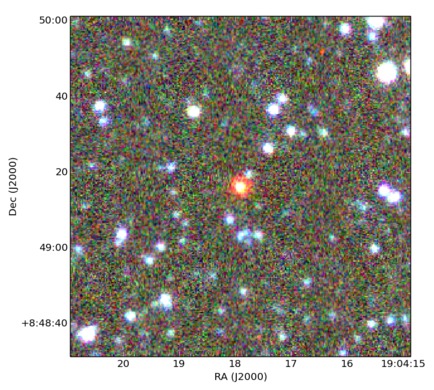}
\includegraphics[height=5.1cm]{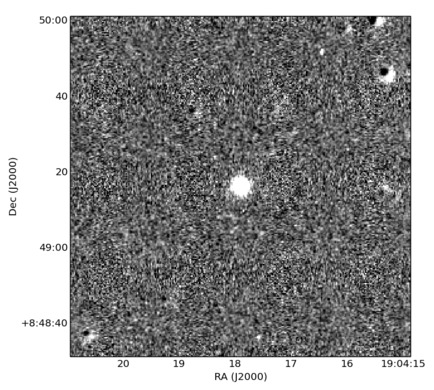}
\includegraphics[height=5.1cm]{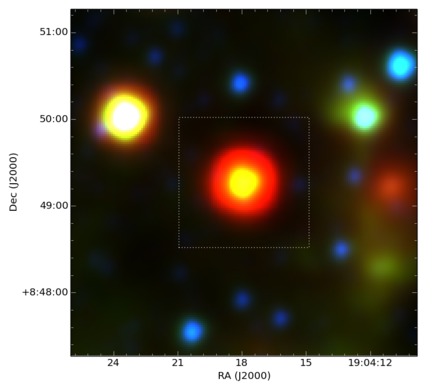}
\includegraphics[height=5.1cm]{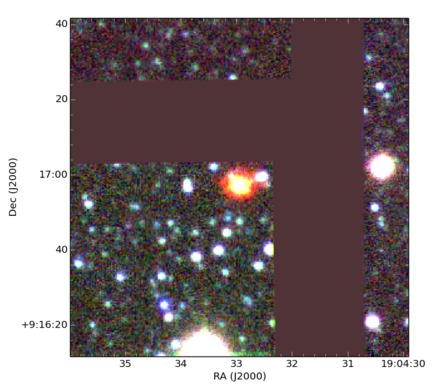}
\includegraphics[height=5.1cm]{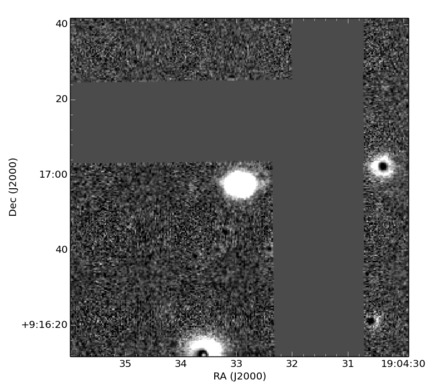}
\includegraphics[height=5.1cm]{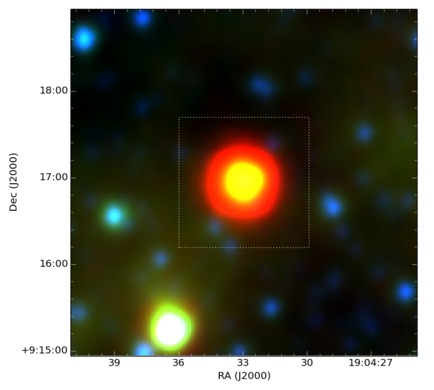}
\caption{\label{imagelabel} Same as in Fig.~\ref{image1}. Objects shown (from top to bottom):  PN G043.0+01.7,PN G038.7-00.5,PN G042.2+01.1,PN G042.6+01.3}
\end{figure*}
\clearpage
\begin{figure*}
\includegraphics[height=5.1cm]{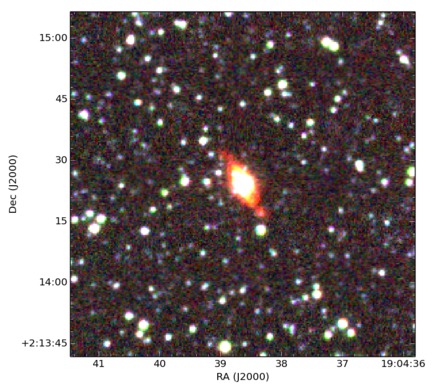}
\includegraphics[height=5.1cm]{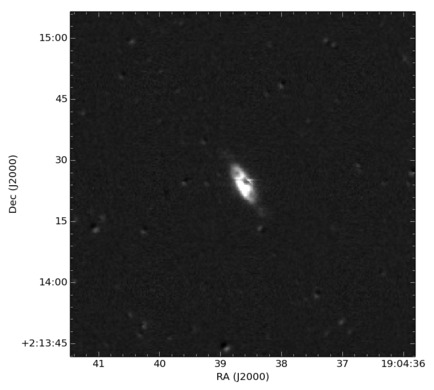}
\includegraphics[height=5.1cm]{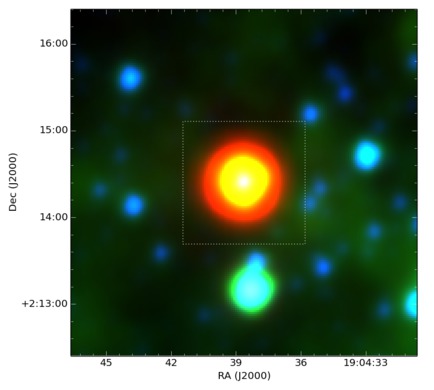}
\includegraphics[height=5.1cm]{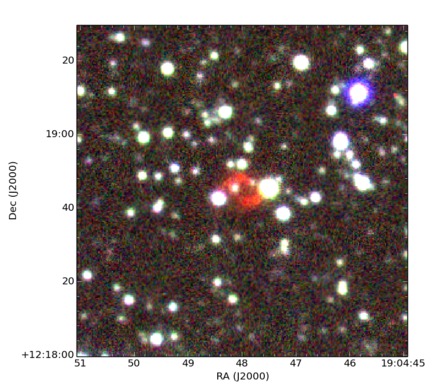}
\includegraphics[height=5.1cm]{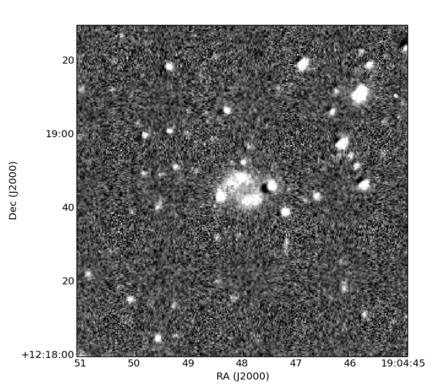}
\includegraphics[height=5.1cm]{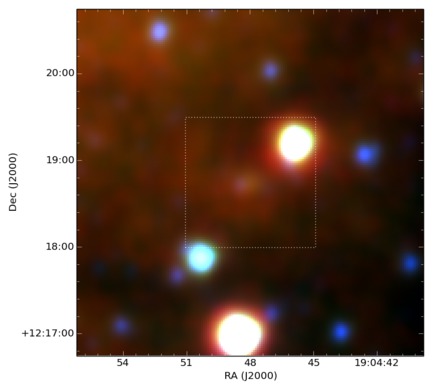}
\includegraphics[height=5.1cm]{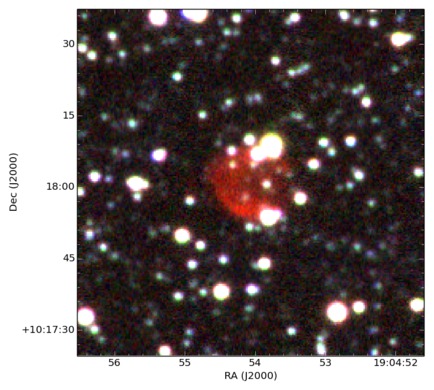}
\includegraphics[height=5.1cm]{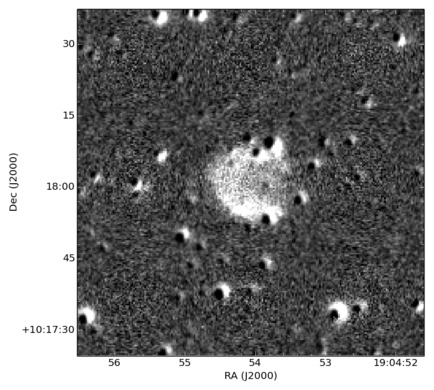}
\includegraphics[height=5.1cm]{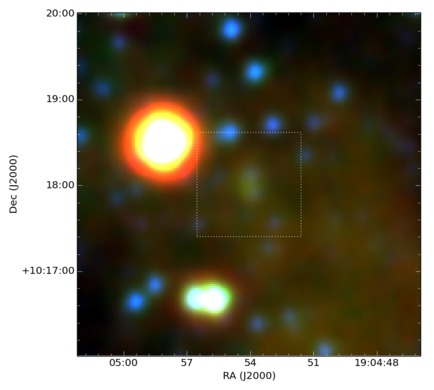}
\includegraphics[height=5.1cm]{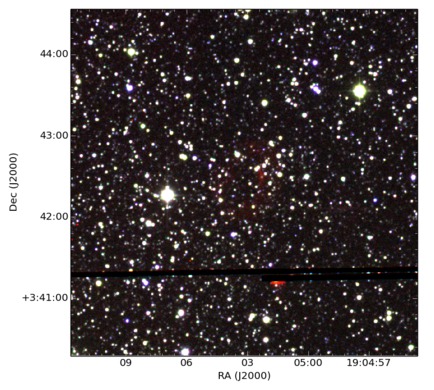}
\includegraphics[height=5.1cm]{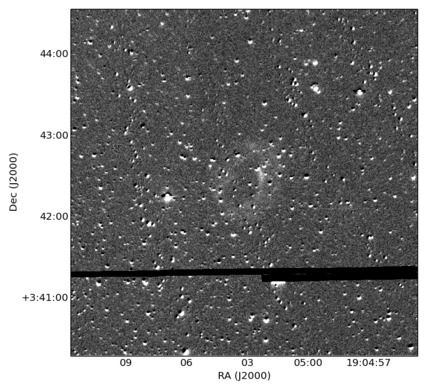}
\includegraphics[height=5.1cm]{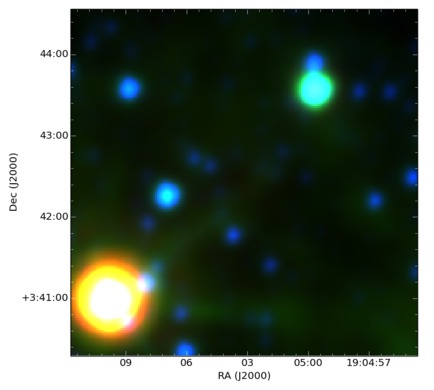}
\caption{\label{imagelabel} Same as in Fig.~\ref{image1}. Objects shown (from top to bottom):  PN G036.4-01.9,PN G045.4+02.6,PN G043.6+01.7,PN G037.7-01.3}
\end{figure*}
\clearpage
\begin{figure*}
\includegraphics[height=5.1cm]{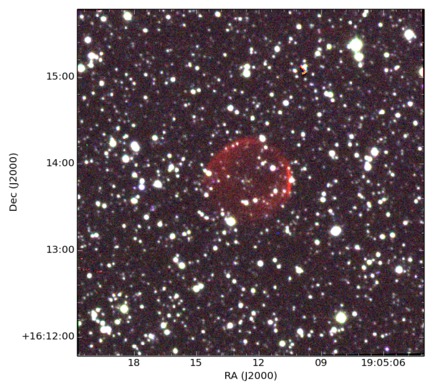}
\includegraphics[height=5.1cm]{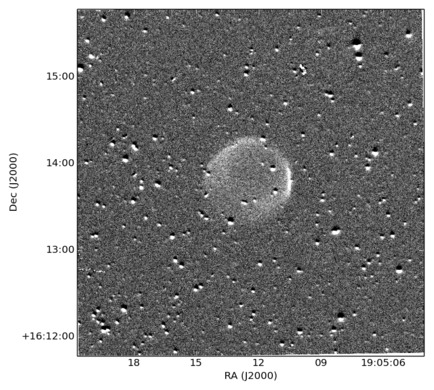}
\includegraphics[height=5.1cm]{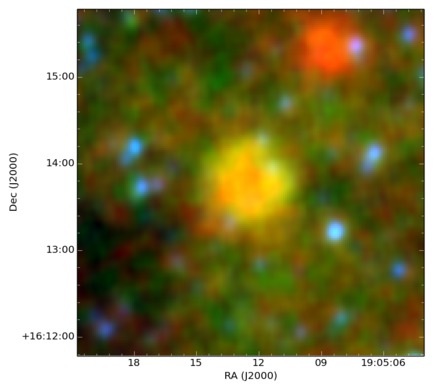}
\includegraphics[height=5.1cm]{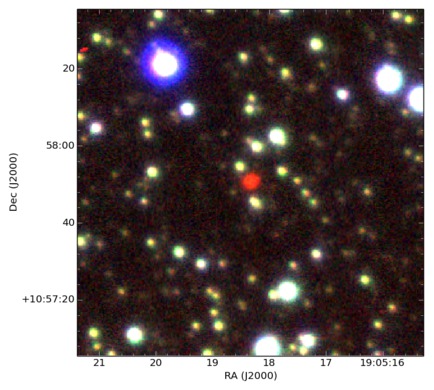}
\includegraphics[height=5.1cm]{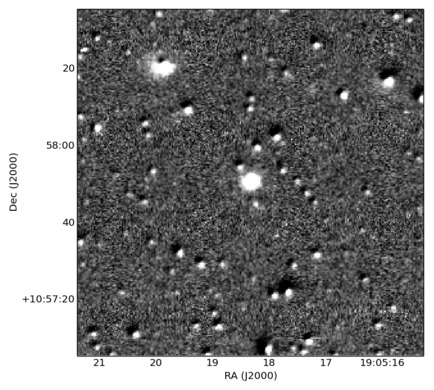}
\includegraphics[height=5.1cm]{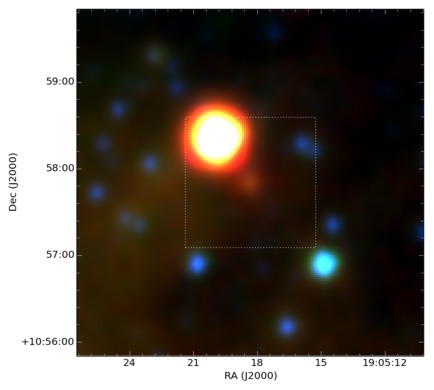}
\includegraphics[height=5.1cm]{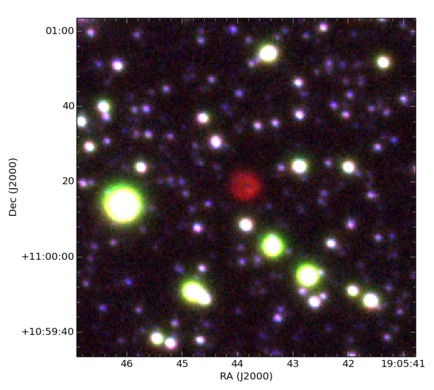}
\includegraphics[height=5.1cm]{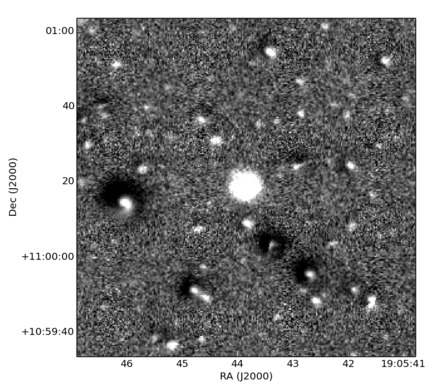}
\includegraphics[height=5.1cm]{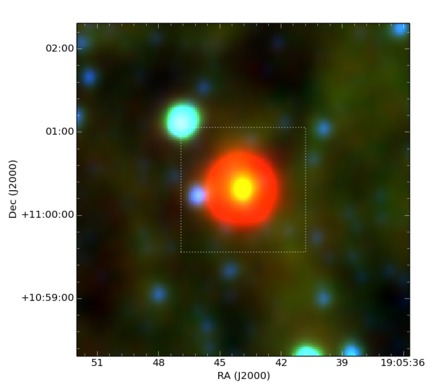}
\includegraphics[height=5.1cm]{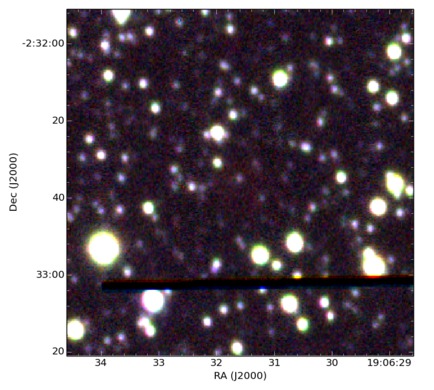}
\includegraphics[height=5.1cm]{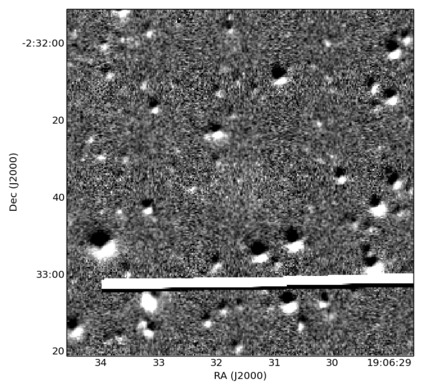}
\includegraphics[height=5.1cm]{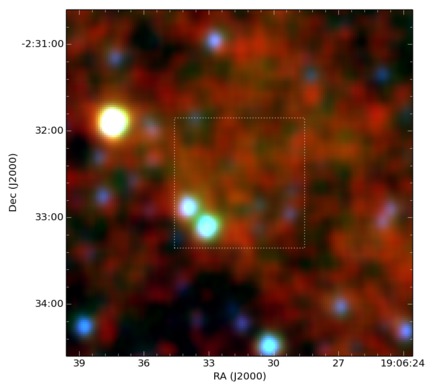}
\caption{\label{imagelabel} Same as in Fig.~\ref{image1}. Objects shown (from top to bottom):  PN G048.9+04.3,PN G044.2+01.9,PN G044.3+01.8,PN G032.3-04.5}
\end{figure*}
\clearpage
\begin{figure*}
\includegraphics[height=5.1cm]{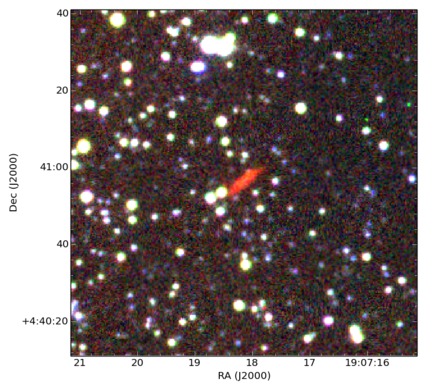}
\includegraphics[height=5.1cm]{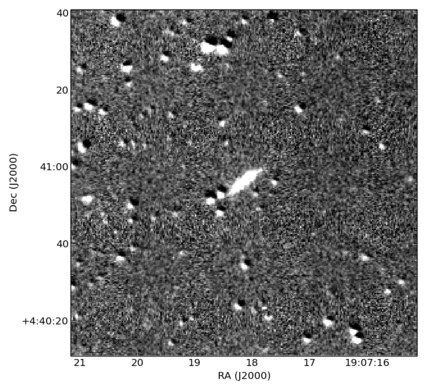}
\includegraphics[height=5.1cm]{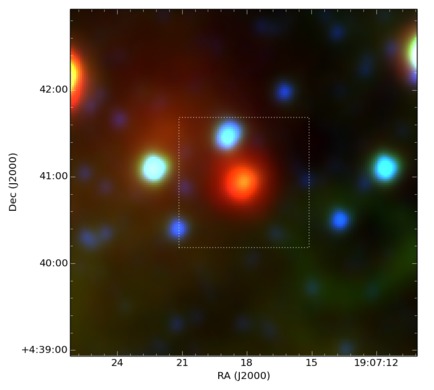}
\includegraphics[height=5.1cm]{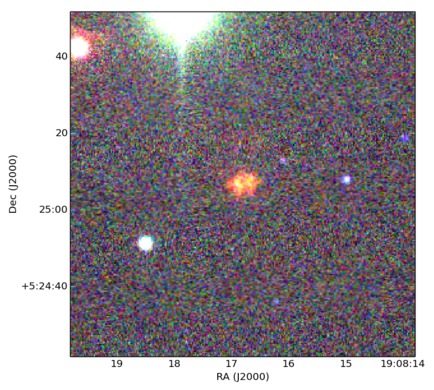}
\includegraphics[height=5.1cm]{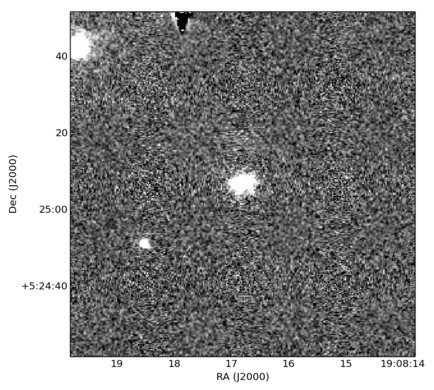}
\includegraphics[height=5.1cm]{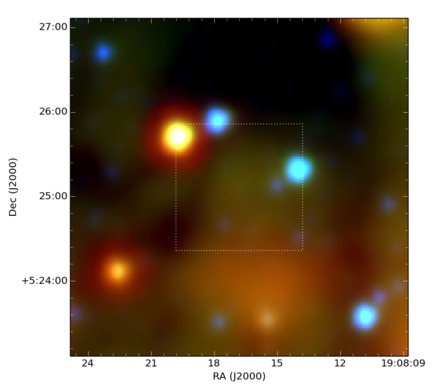}
\includegraphics[height=5.1cm]{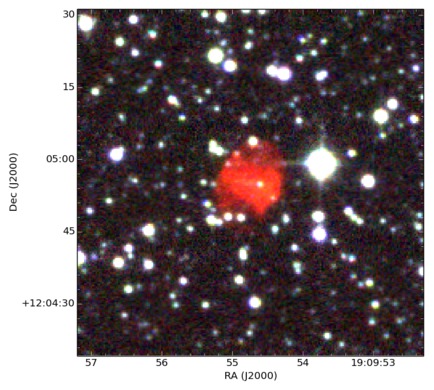}
\includegraphics[height=5.1cm]{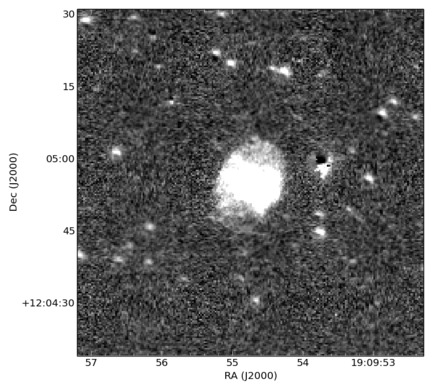}
\includegraphics[height=5.1cm]{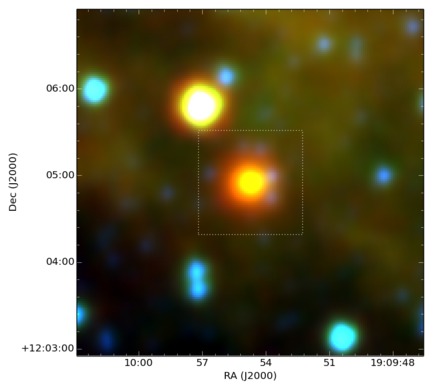}
\includegraphics[height=5.1cm]{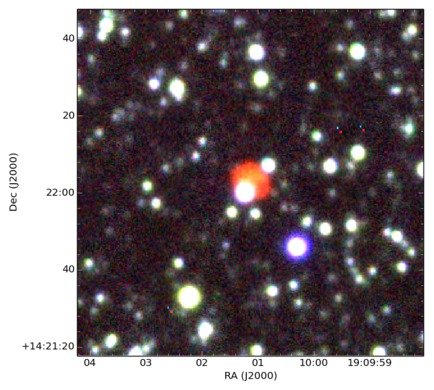}
\includegraphics[height=5.1cm]{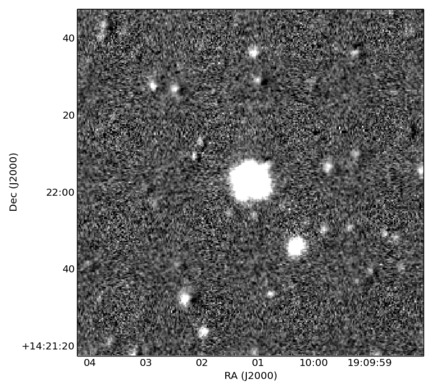}
\includegraphics[height=5.1cm]{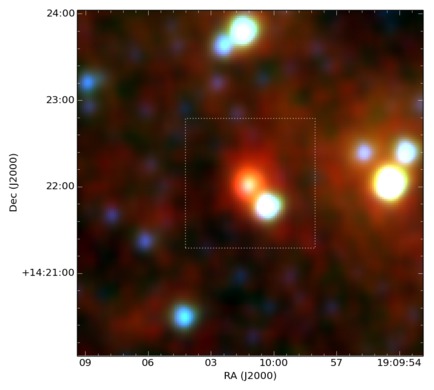}
\caption{\label{imagelabel} Same as in Fig.~\ref{image1}. Objects shown (from top to bottom):  PN G038.9-01.3,PN G039.6-01.2,PN G045.7+01.4,PN G047.8+02.4}
\end{figure*}
\clearpage
\begin{figure*}
\includegraphics[height=5.1cm]{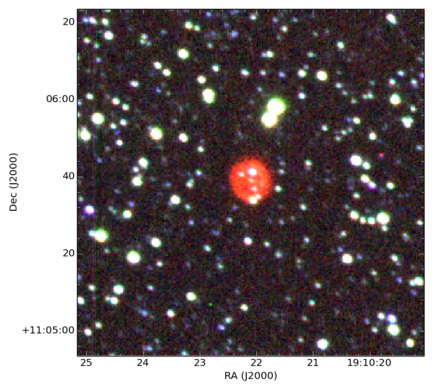}
\includegraphics[height=5.1cm]{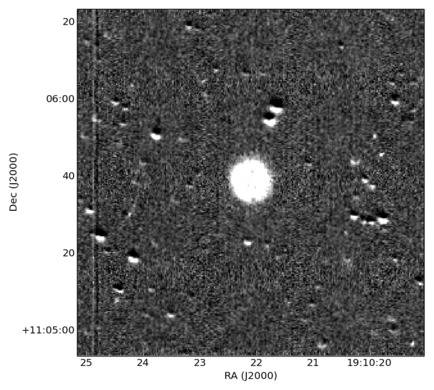}
\includegraphics[height=5.1cm]{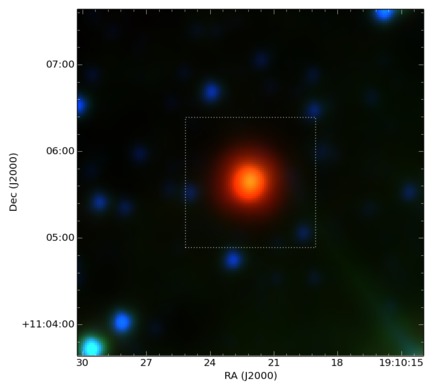}
\includegraphics[height=5.1cm]{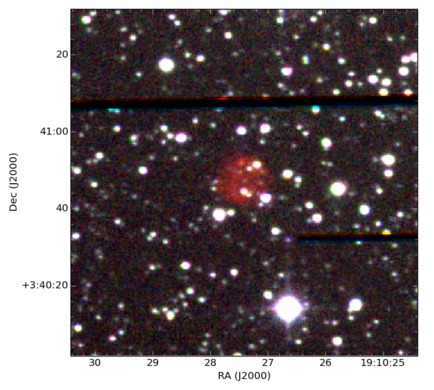}
\includegraphics[height=5.1cm]{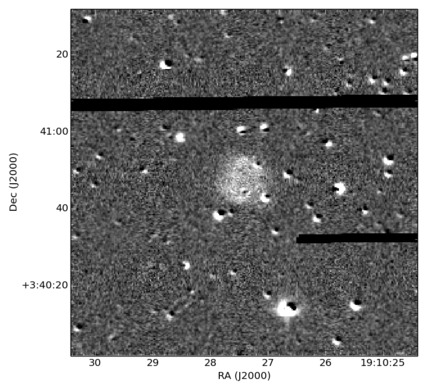}
\includegraphics[height=5.1cm]{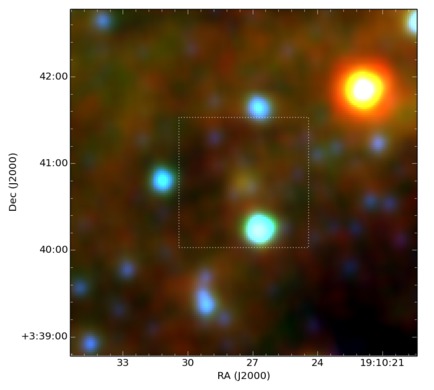}
\includegraphics[height=5.1cm]{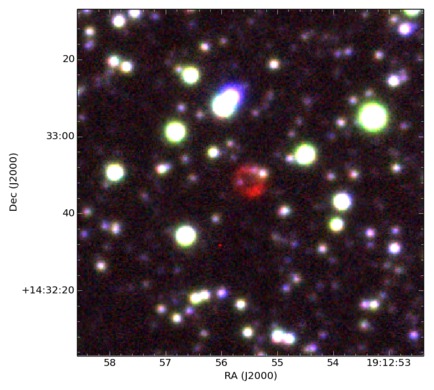}
\includegraphics[height=5.1cm]{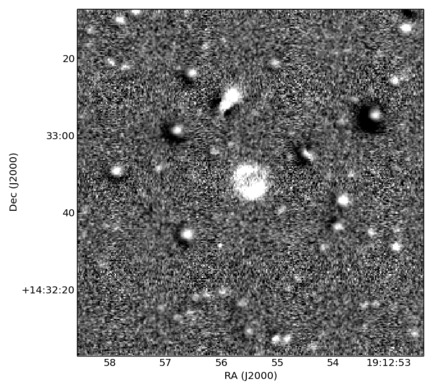}
\includegraphics[height=5.1cm]{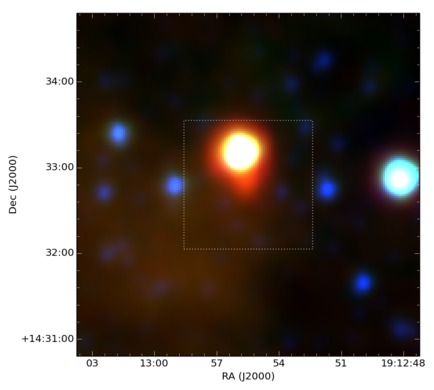}
\includegraphics[height=5.1cm]{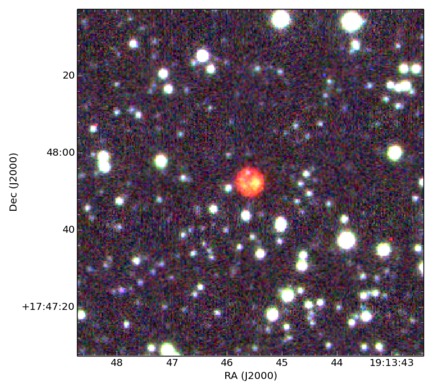}
\includegraphics[height=5.1cm]{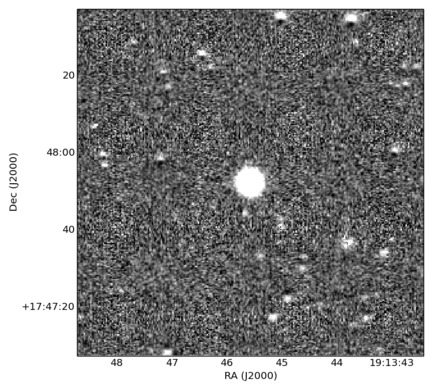}
\includegraphics[height=5.1cm]{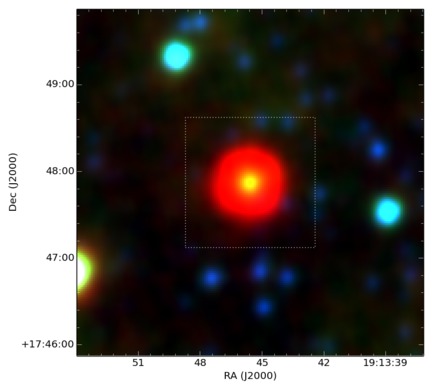}
\caption{\label{imagelabel} Same as in Fig.~\ref{image1}. Objects shown (from top to bottom):  PN G044.9+00.8,PN G038.3-02.5,PN G048.2+01.9,PN G051.2+03.2}
\end{figure*}
\clearpage
\begin{figure*}
\includegraphics[height=5.1cm]{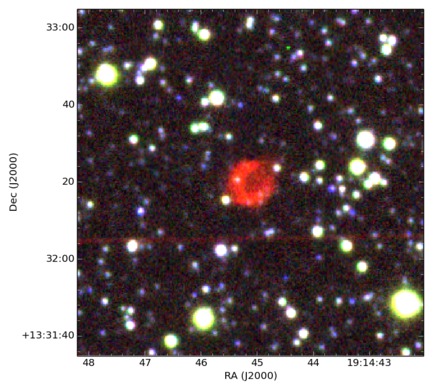}
\includegraphics[height=5.1cm]{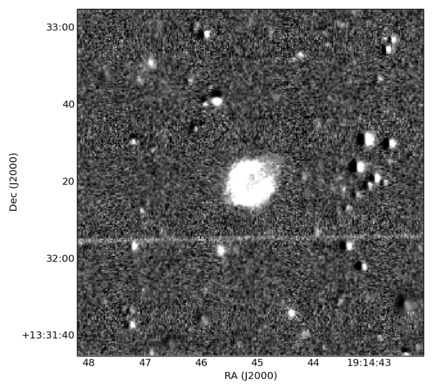}
\includegraphics[height=5.1cm]{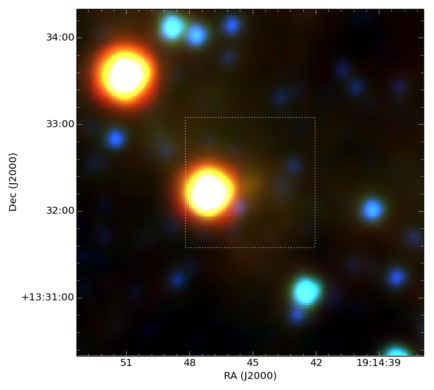}
\includegraphics[height=5.1cm]{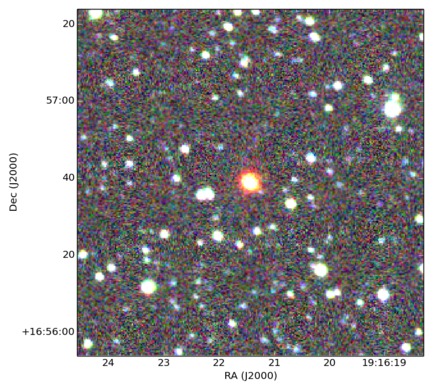}
\includegraphics[height=5.1cm]{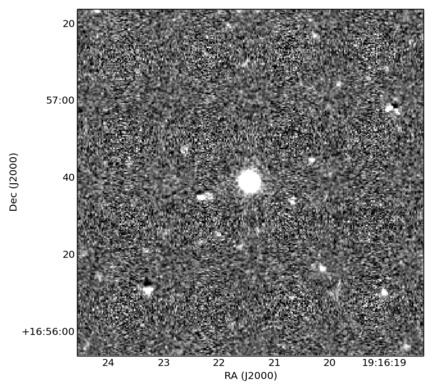}
\includegraphics[height=5.1cm]{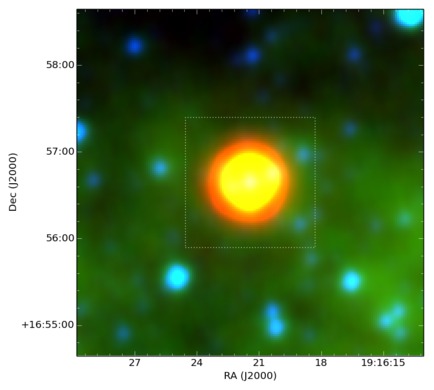}
\includegraphics[height=5.1cm]{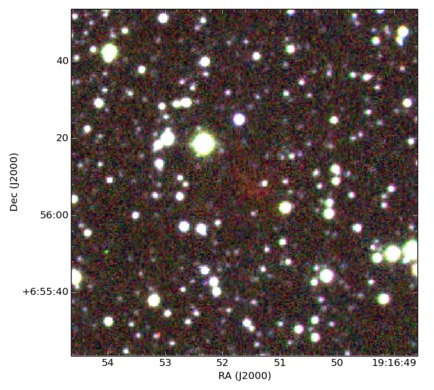}
\includegraphics[height=5.1cm]{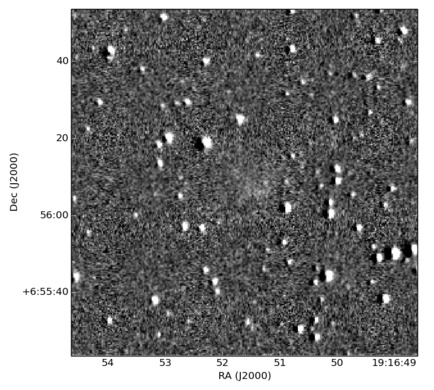}
\includegraphics[height=5.1cm]{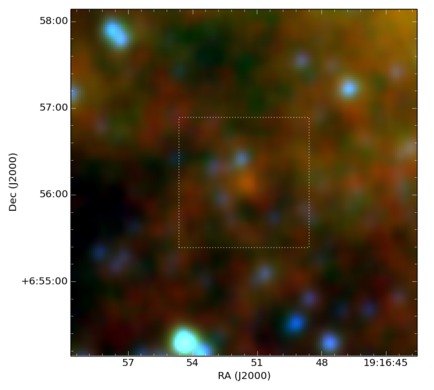}
\includegraphics[height=5.1cm]{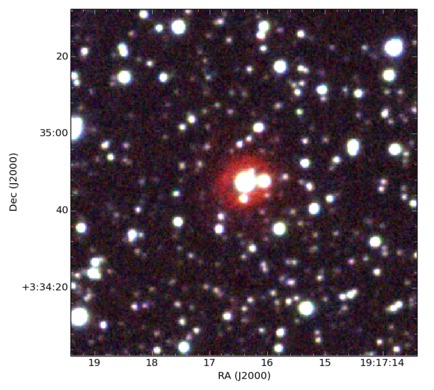}
\includegraphics[height=5.1cm]{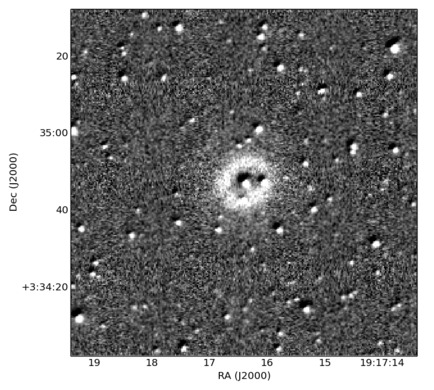}
\includegraphics[height=5.1cm]{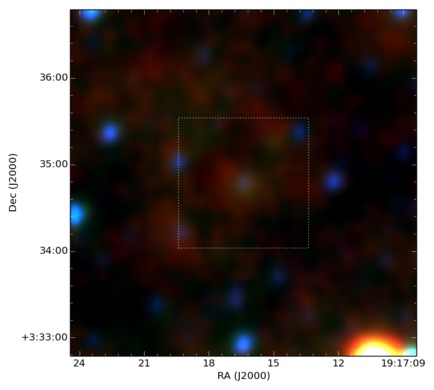}
\caption{\label{imagelabel} Same as in Fig.~\ref{image1}. Objects shown (from top to bottom):  PN G047.6+01.0,PN G050.8+02.3,PN G042.0-02.4,PN G039.0-04.0}
\end{figure*}
\clearpage
\begin{figure*}
\includegraphics[height=5.1cm]{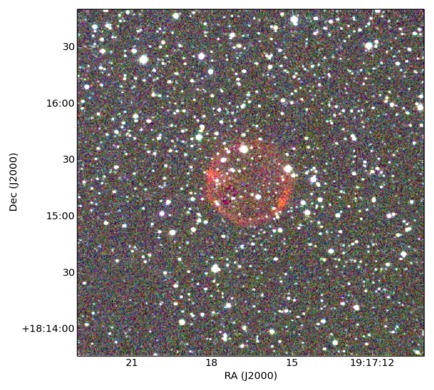}
\includegraphics[height=5.1cm]{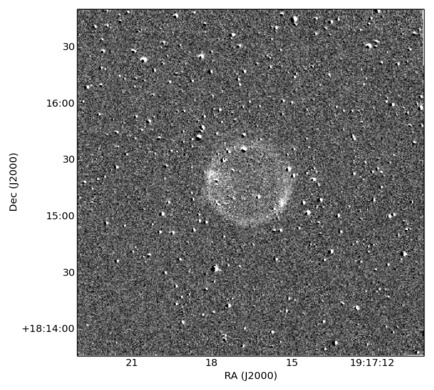}
\includegraphics[height=5.1cm]{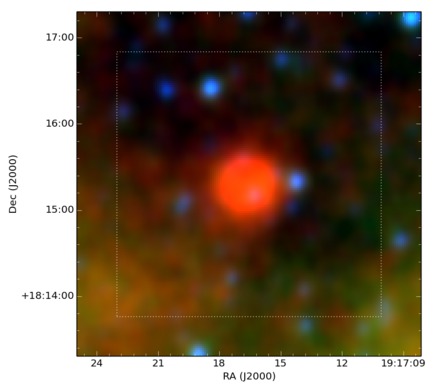}
\includegraphics[height=5.1cm]{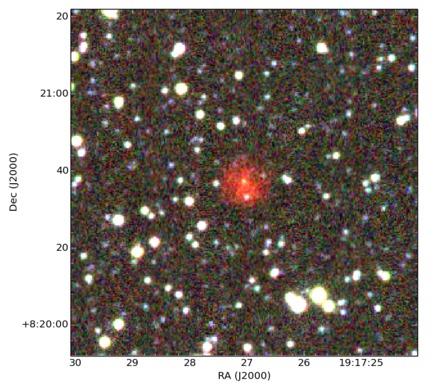}
\includegraphics[height=5.1cm]{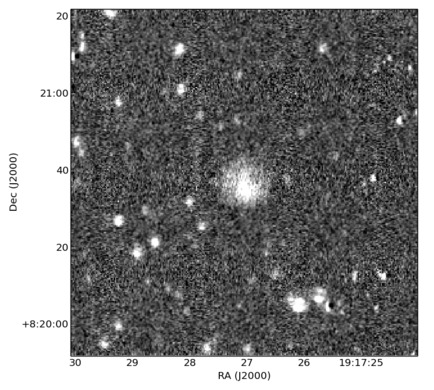}
\includegraphics[height=5.1cm]{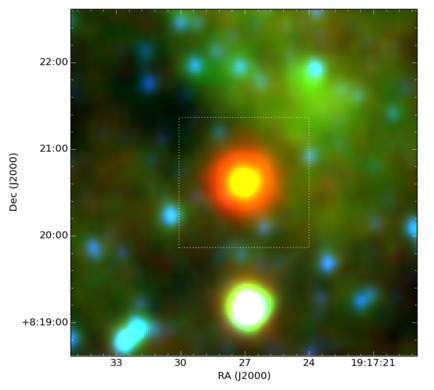}
\includegraphics[height=5.1cm]{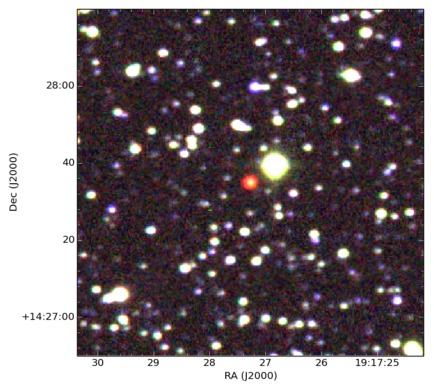}
\includegraphics[height=5.1cm]{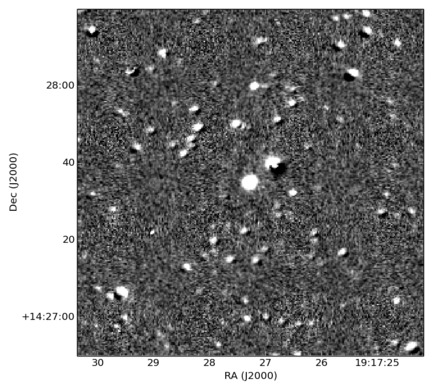}
\includegraphics[height=5.1cm]{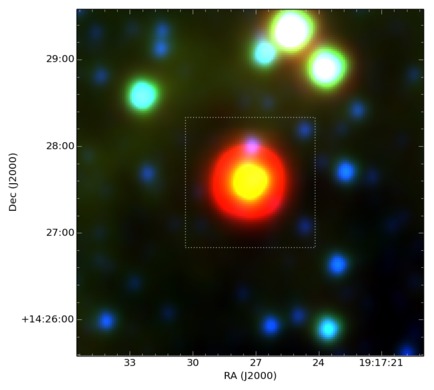}
\includegraphics[height=5.1cm]{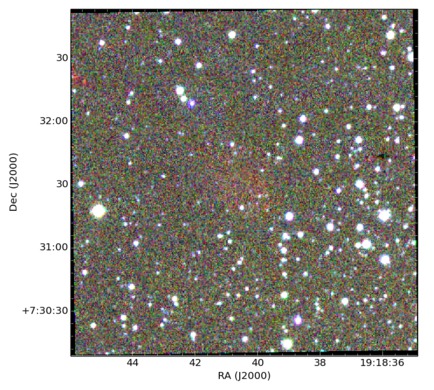}
\includegraphics[height=5.1cm]{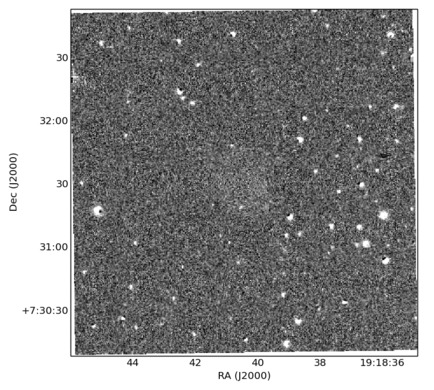}
\includegraphics[height=5.1cm]{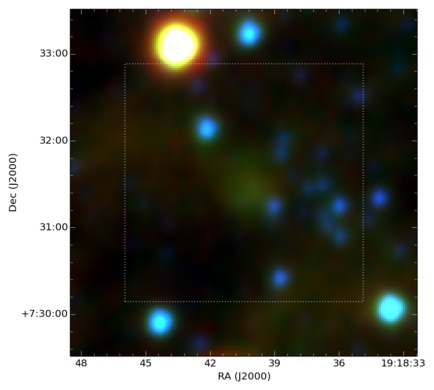}
\caption{\label{imagelabel} Same as in Fig.~\ref{image1}. Objects shown (from top to bottom):  PN G052.0+02.7,PN G043.3-01.9,PN G048.7+00.9,PN G042.7-02.5}
\end{figure*}
\clearpage
\begin{figure*}
\includegraphics[height=5.1cm]{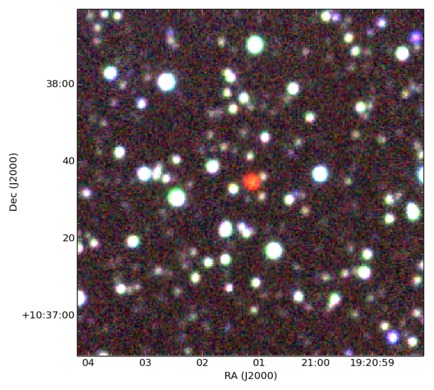}
\includegraphics[height=5.1cm]{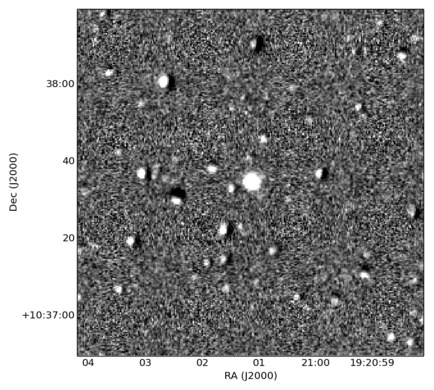}
\includegraphics[height=5.1cm]{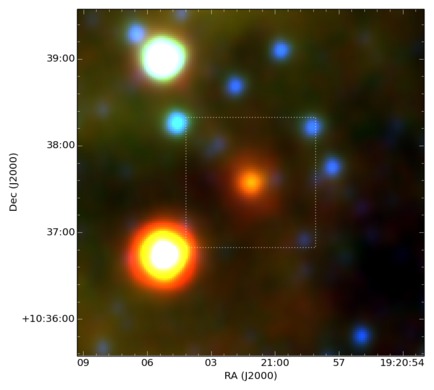}
\includegraphics[height=5.1cm]{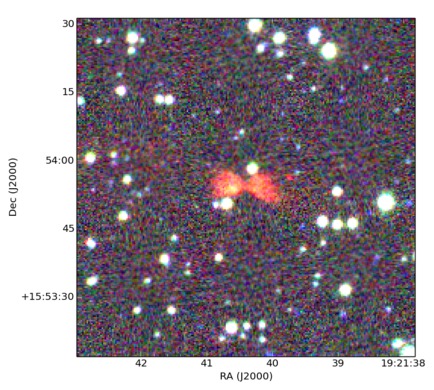}
\includegraphics[height=5.1cm]{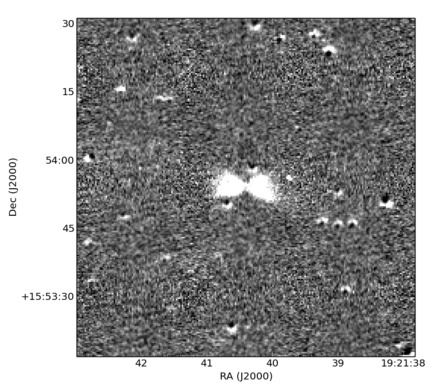}
\includegraphics[height=5.1cm]{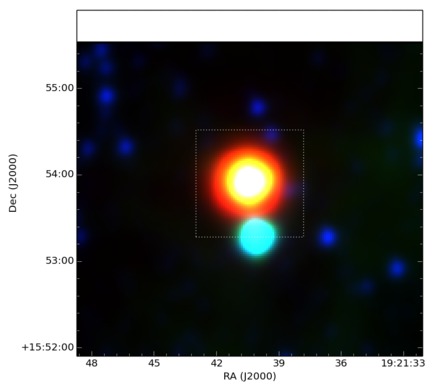}
\includegraphics[height=5.1cm]{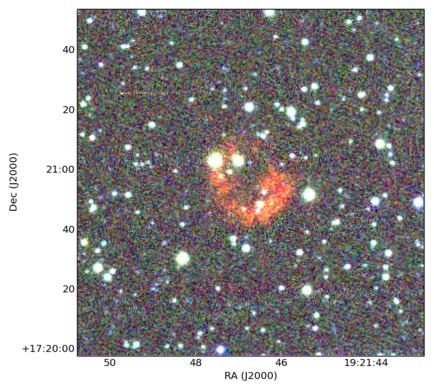}
\includegraphics[height=5.1cm]{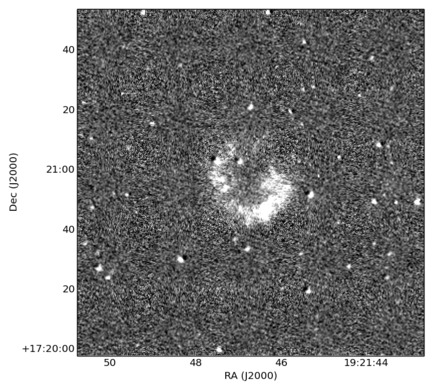}
\includegraphics[height=5.1cm]{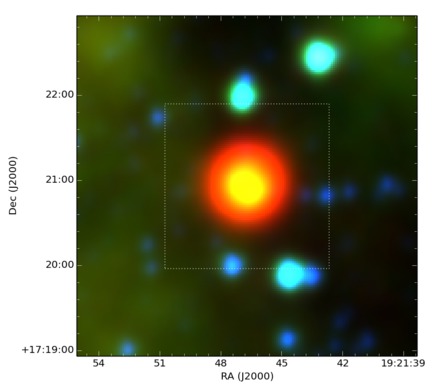}
\includegraphics[height=5.1cm]{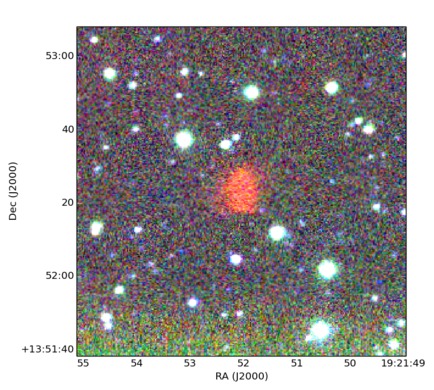}
\includegraphics[height=5.1cm]{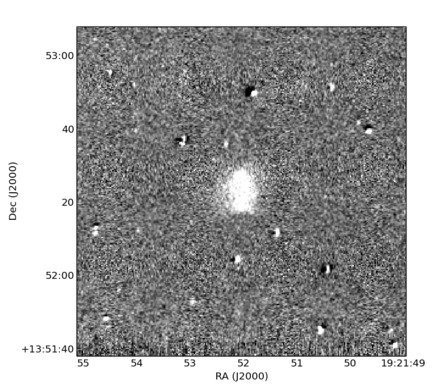}
\includegraphics[height=5.1cm]{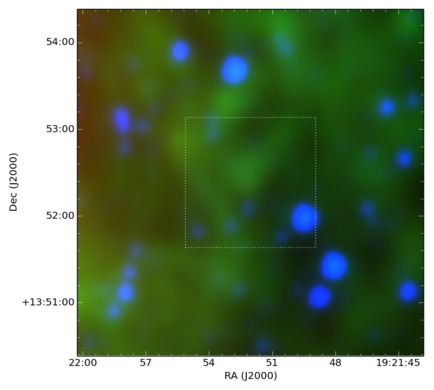}
\caption{\label{imagelabel} Same as in Fig.~\ref{image1}. Objects shown (from top to bottom):  PN G045.7-01.6,PN G050.4+00.7,PN G051.7+01.3,PN G048.7-00.2}
\end{figure*}
\clearpage
\begin{figure*}
\includegraphics[height=5.1cm]{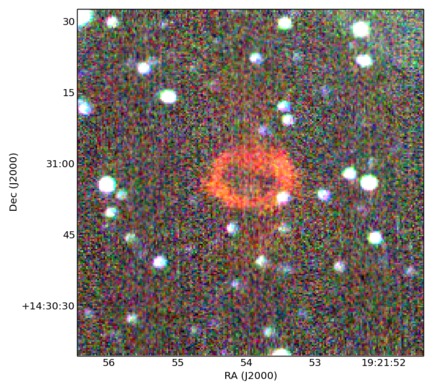}
\includegraphics[height=5.1cm]{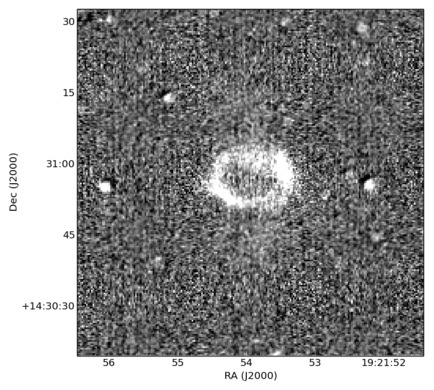}
\includegraphics[height=5.1cm]{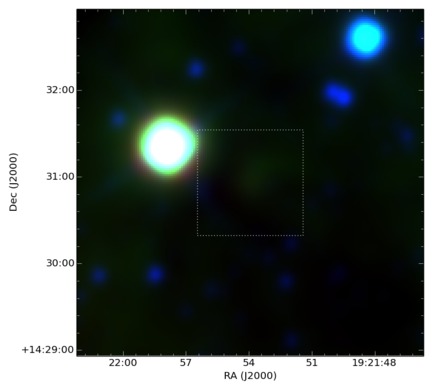}
\includegraphics[height=5.1cm]{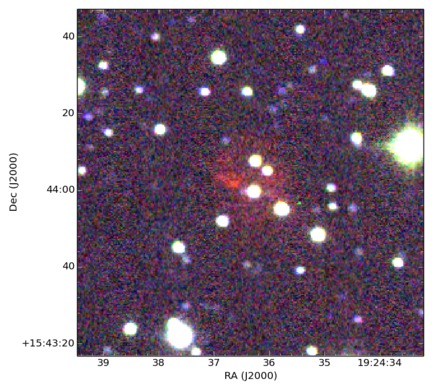}
\includegraphics[height=5.1cm]{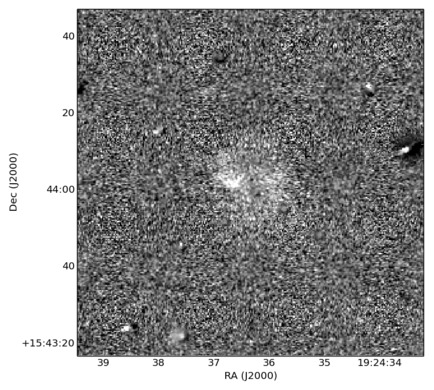}
\includegraphics[height=5.1cm]{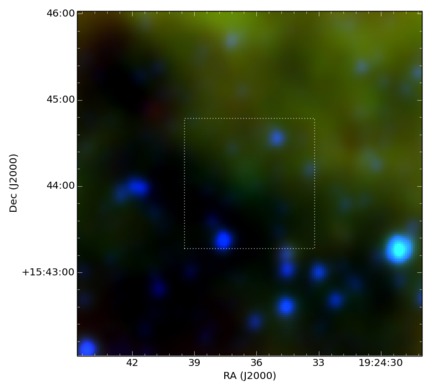}
\includegraphics[height=5.1cm]{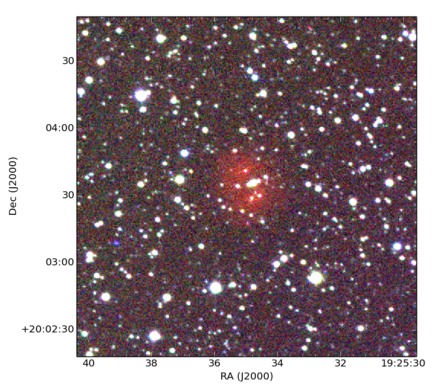}
\includegraphics[height=5.1cm]{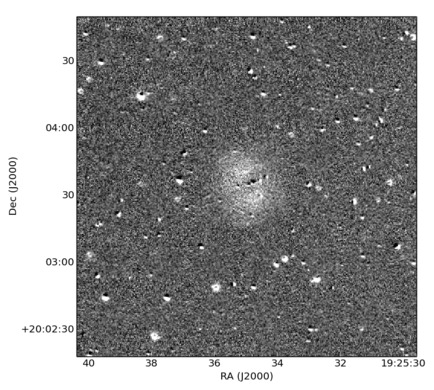}
\includegraphics[height=5.1cm]{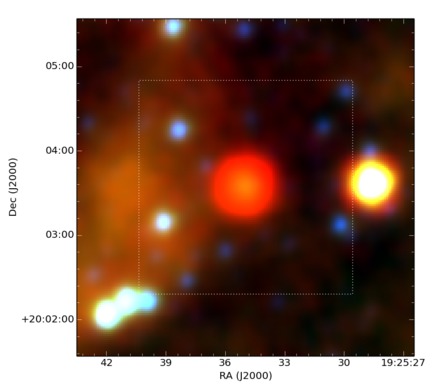}
\includegraphics[height=5.1cm]{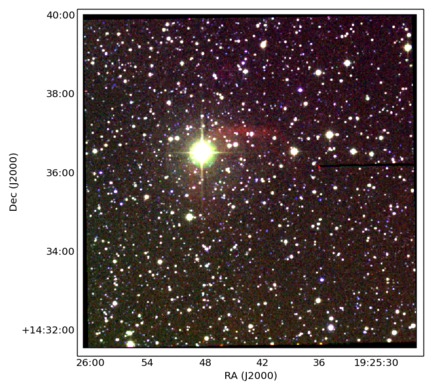}
\includegraphics[height=5.1cm]{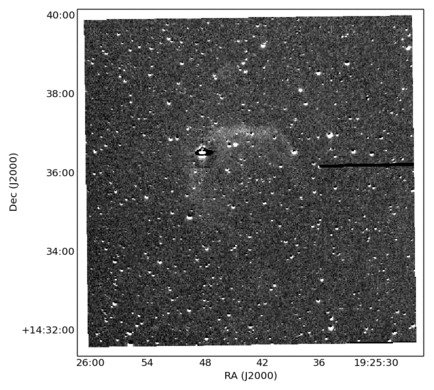}
\includegraphics[height=5.1cm]{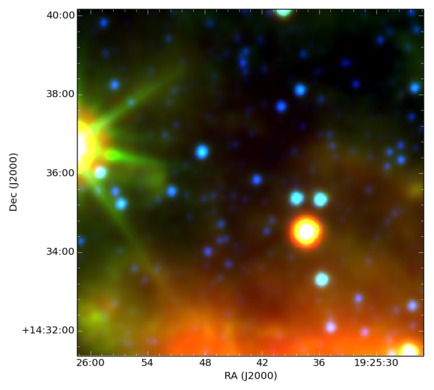}
\caption{\label{imagelabel} Same as in Fig.~\ref{image1}. Objects shown (from top to bottom):  PN G049.2+00.0,PN G050.6+00.0,PN G054.5+01.8,PN G049.7-00.7}
\end{figure*}
\clearpage
\begin{figure*}
\includegraphics[height=5.1cm]{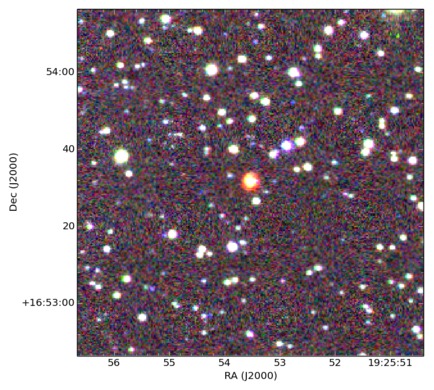}
\includegraphics[height=5.1cm]{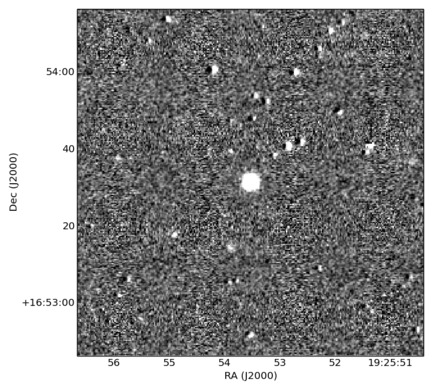}
\includegraphics[height=5.1cm]{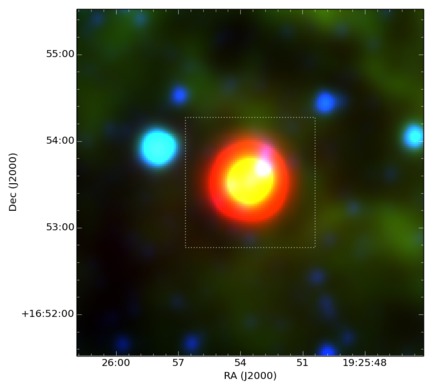}
\includegraphics[height=5.1cm]{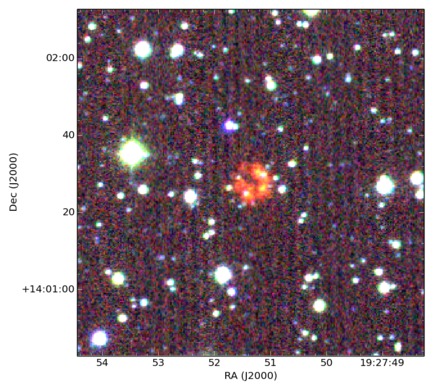}
\includegraphics[height=5.1cm]{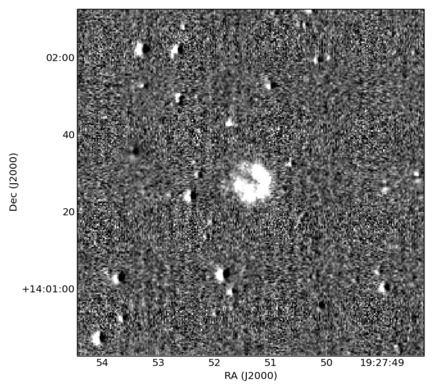}
\includegraphics[height=5.1cm]{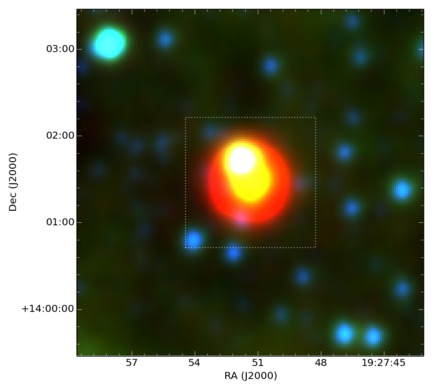}
\includegraphics[height=5.1cm]{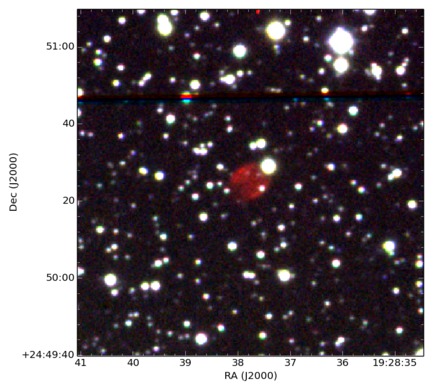}
\includegraphics[height=5.1cm]{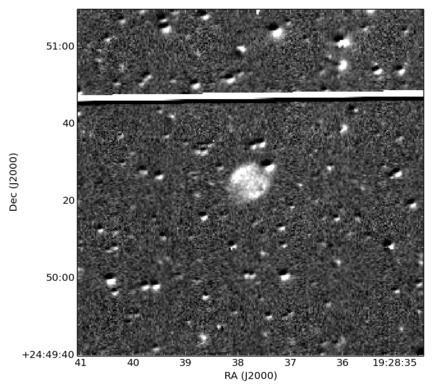}
\includegraphics[height=5.1cm]{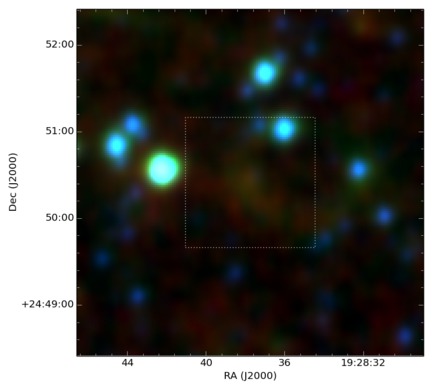}
\includegraphics[height=5.1cm]{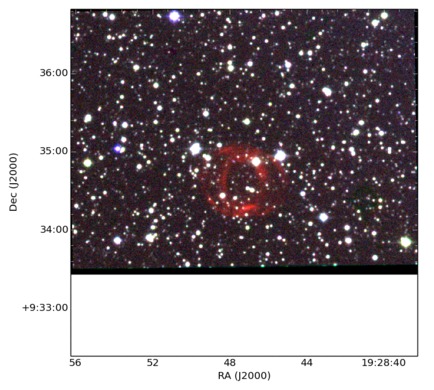}
\includegraphics[height=5.1cm]{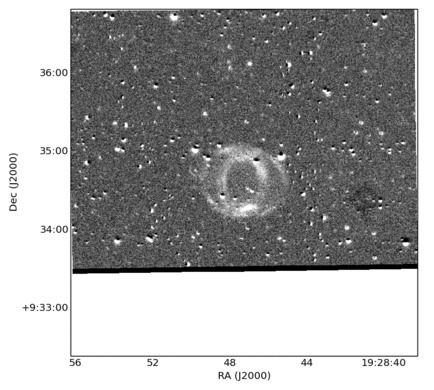}
\includegraphics[height=5.1cm]{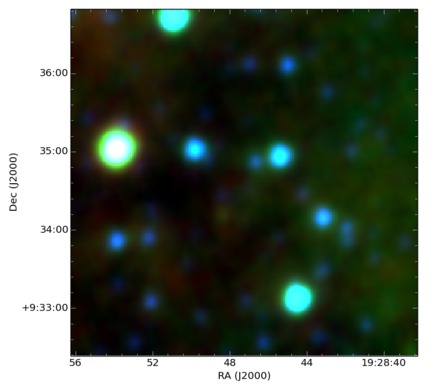}
\caption{\label{imagelabel} Same as in Fig.~\ref{image1}. Objects shown (from top to bottom):  PN G051.8+00.2,PN G049.5-01.4,PN G059.1+03.5,PN G045.7-03.8}
\end{figure*}
\clearpage
\begin{figure*}
\includegraphics[height=5.1cm]{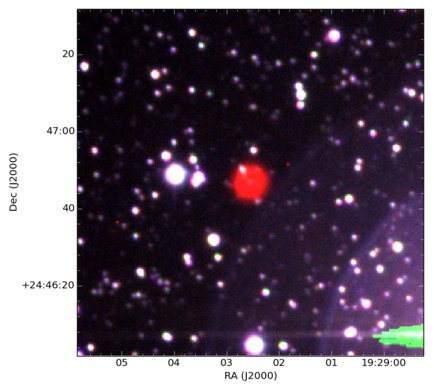}
\includegraphics[height=5.1cm]{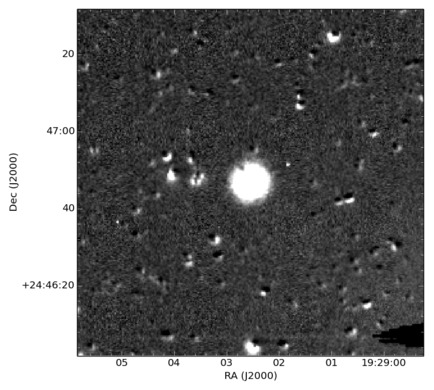}
\includegraphics[height=5.1cm]{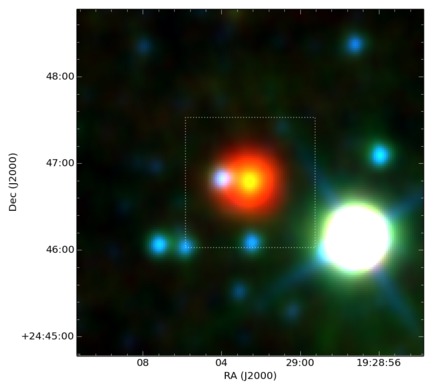}
\includegraphics[height=5.1cm]{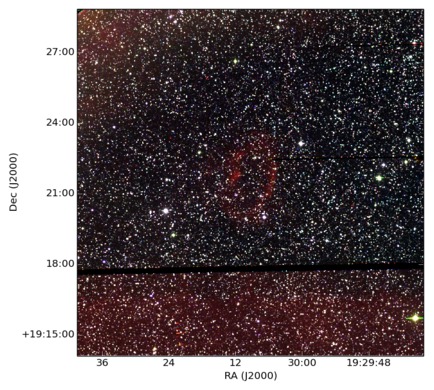}
\includegraphics[height=5.1cm]{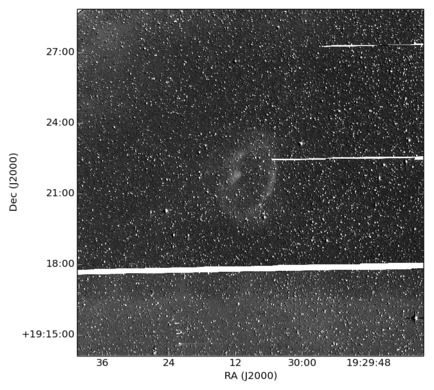}
\includegraphics[height=5.1cm]{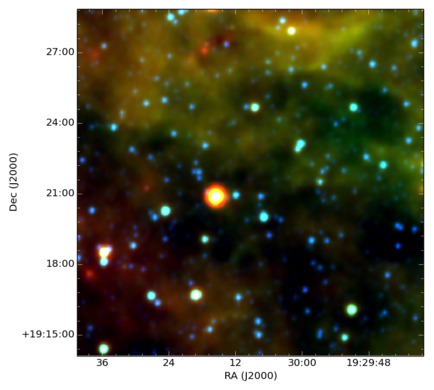}
\includegraphics[height=5.1cm]{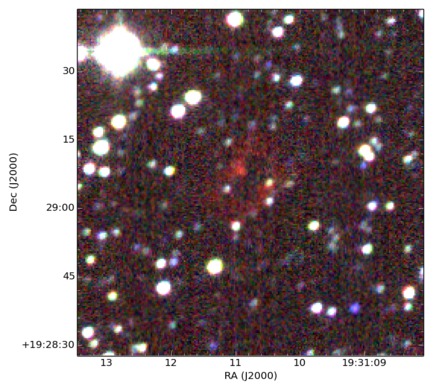}
\includegraphics[height=5.1cm]{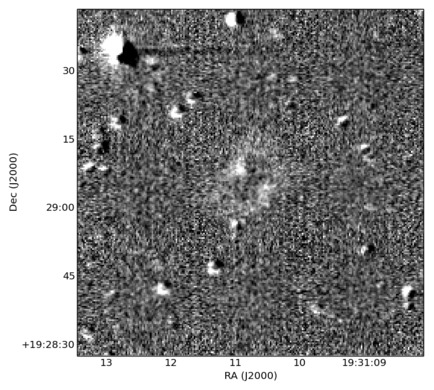}
\includegraphics[height=5.1cm]{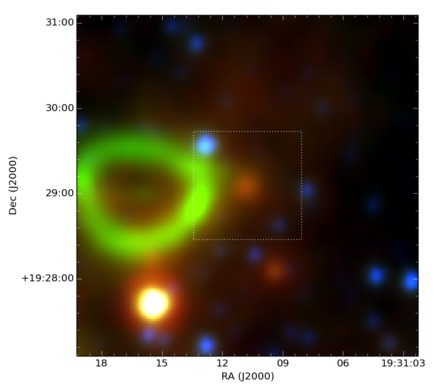}
\includegraphics[height=5.1cm]{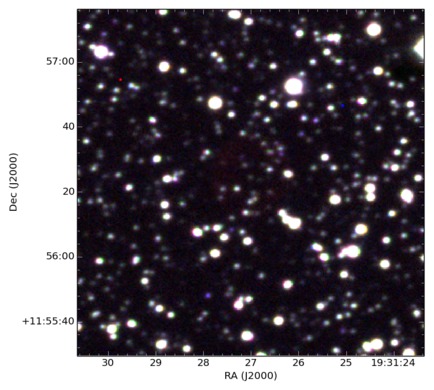}
\includegraphics[height=5.1cm]{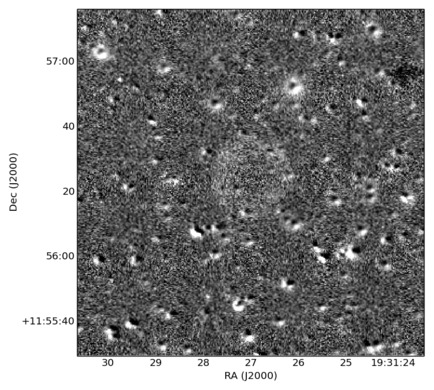}
\includegraphics[height=5.1cm]{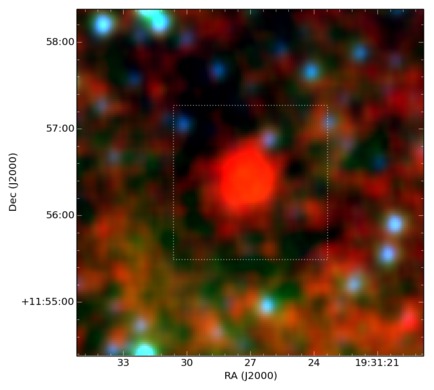}
\caption{\label{imagelabel} Same as in Fig.~\ref{image1}. Objects shown (from top to bottom):  PN G059.1+03.3,PN G054.4+00.5,PN G054.7+00.4,PN G048.1-03.2}
\end{figure*}
\clearpage
\begin{figure*}
\includegraphics[height=5.1cm]{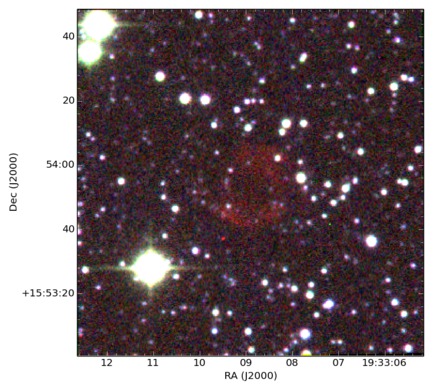}
\includegraphics[height=5.1cm]{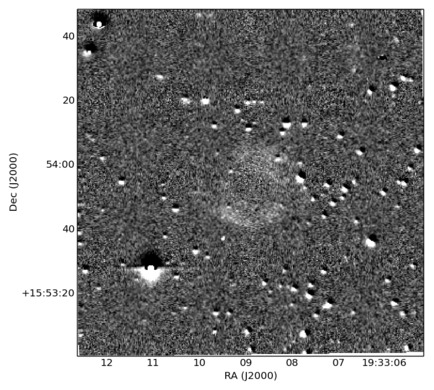}
\includegraphics[height=5.1cm]{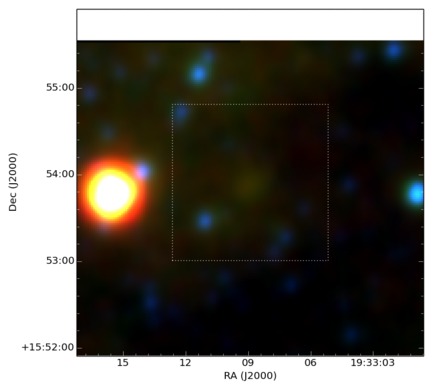}
\includegraphics[height=5.1cm]{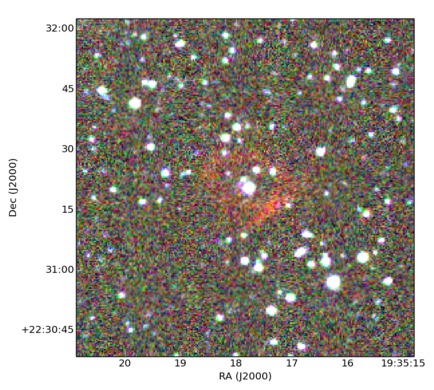}
\includegraphics[height=5.1cm]{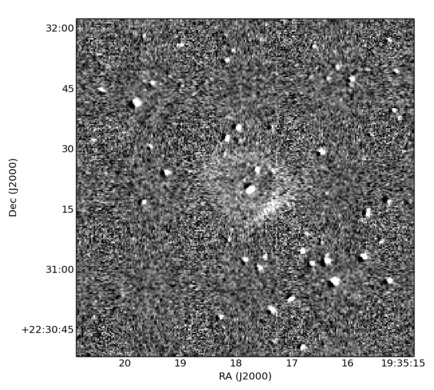}
\includegraphics[height=5.1cm]{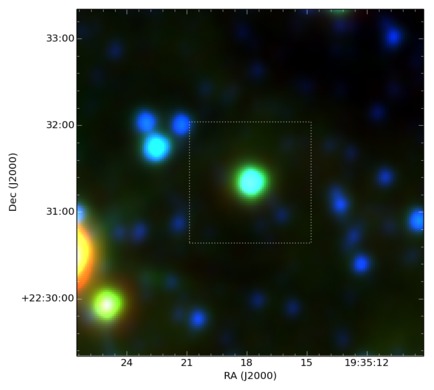}
\includegraphics[height=5.1cm]{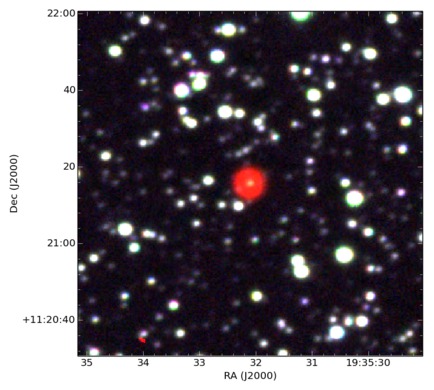}
\includegraphics[height=5.1cm]{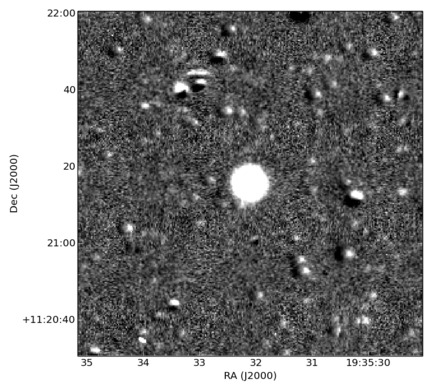}
\includegraphics[height=5.1cm]{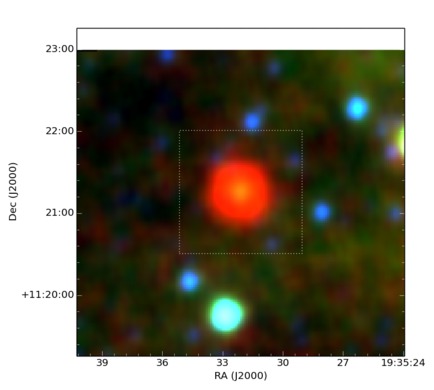}
\includegraphics[height=5.1cm]{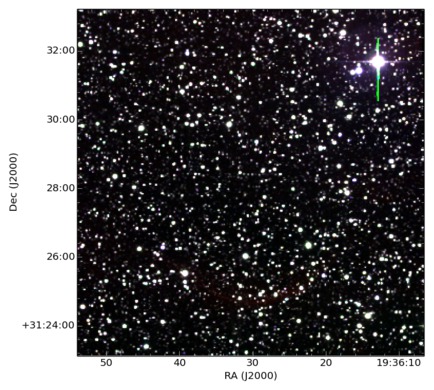}
\includegraphics[height=5.1cm]{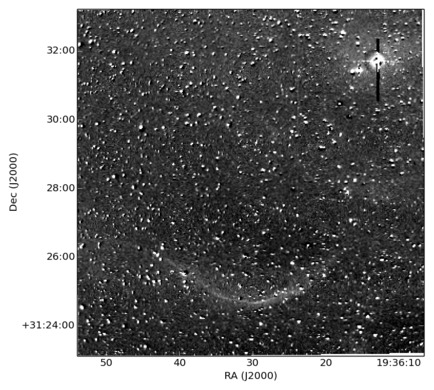}
\includegraphics[height=5.1cm]{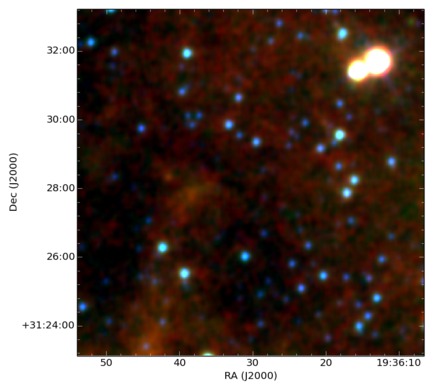}
\caption{\label{imagelabel} Same as in Fig.~\ref{image1}. Objects shown (from top to bottom):  PN G051.7-01.7,PN G057.8+01.0,PN G048.0-04.4,PN G065.8+05.1}
\end{figure*}
\clearpage
\begin{figure*}
\includegraphics[height=5.1cm]{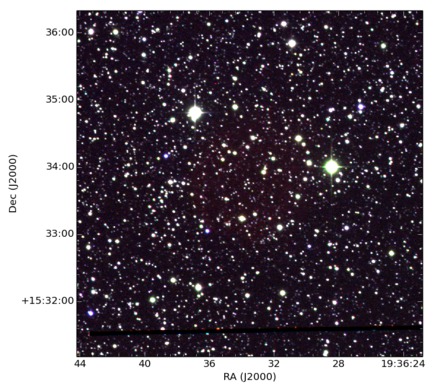}
\includegraphics[height=5.1cm]{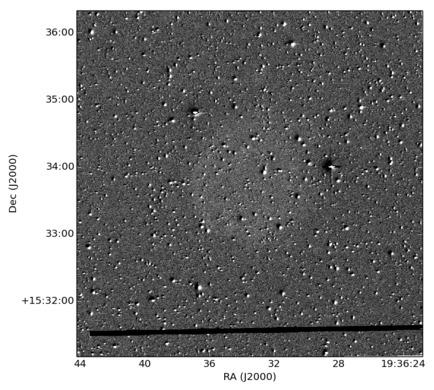}
\includegraphics[height=5.1cm]{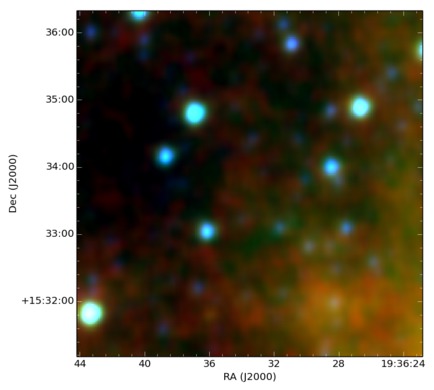}
\includegraphics[height=5.1cm]{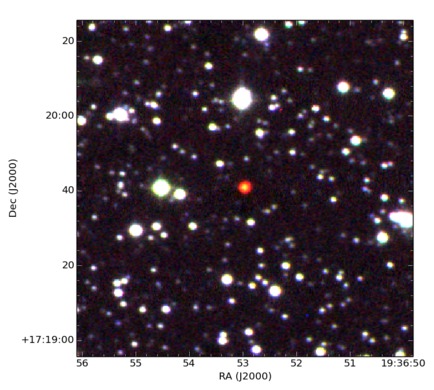}
\includegraphics[height=5.1cm]{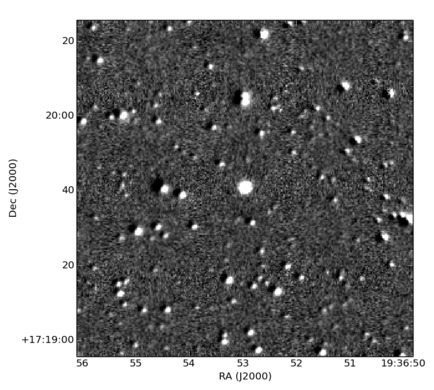}
\includegraphics[height=5.1cm]{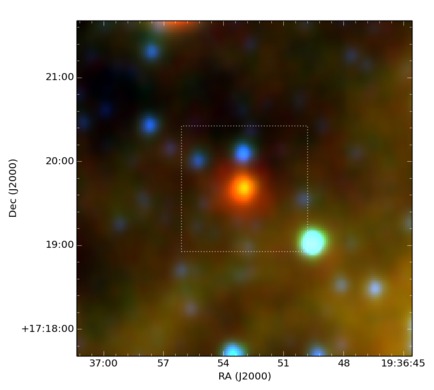}
\includegraphics[height=5.1cm]{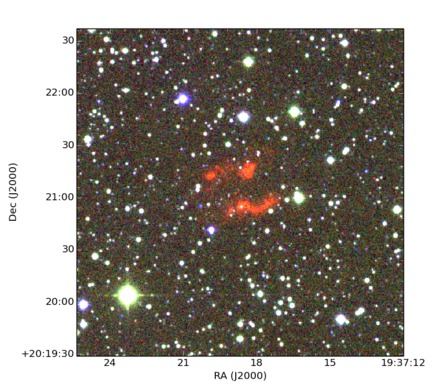}
\includegraphics[height=5.1cm]{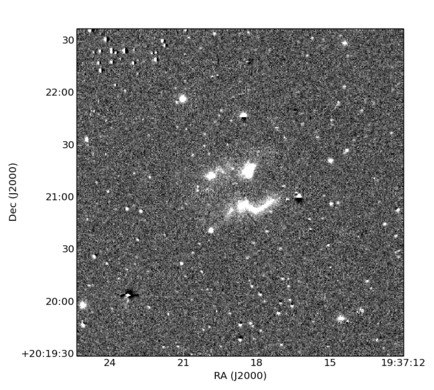}
\includegraphics[height=5.1cm]{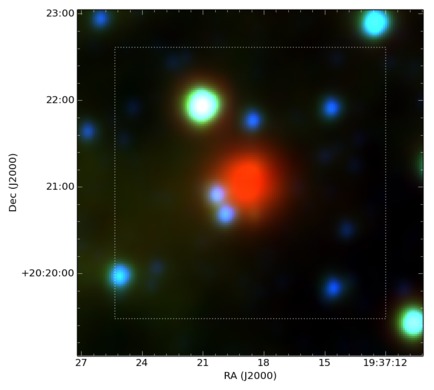}
\includegraphics[height=5.1cm]{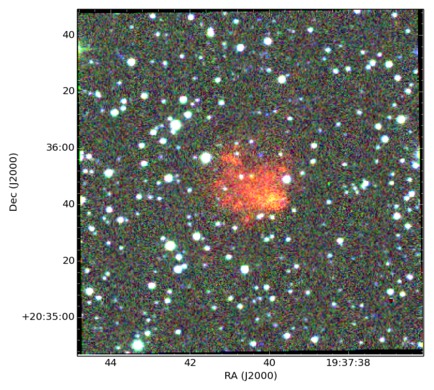}
\includegraphics[height=5.1cm]{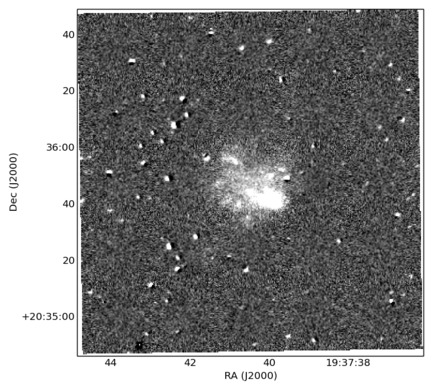}
\includegraphics[height=5.1cm]{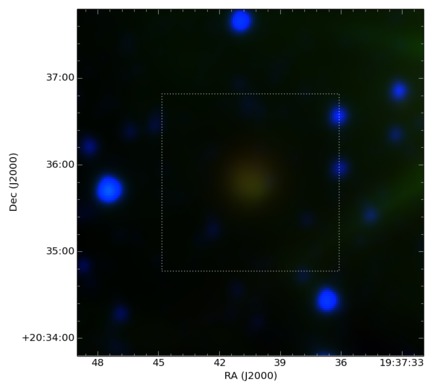}
\caption{\label{imagelabel} Same as in Fig.~\ref{image1}. Objects shown (from top to bottom):  PN G051.9-02.5,PN G053.4-01.8,PN G056.1-00.4,PN G056.4-00.3}
\end{figure*}
\clearpage
\begin{figure*}
\includegraphics[height=5.1cm]{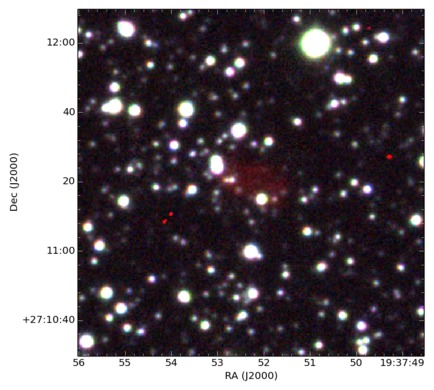}
\includegraphics[height=5.1cm]{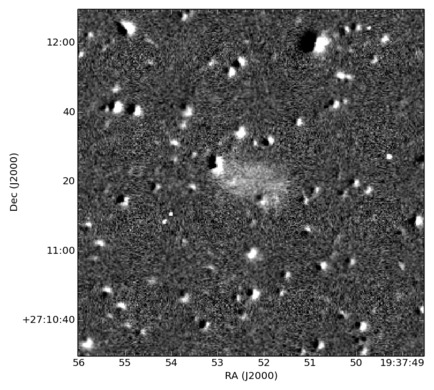}
\includegraphics[height=5.1cm]{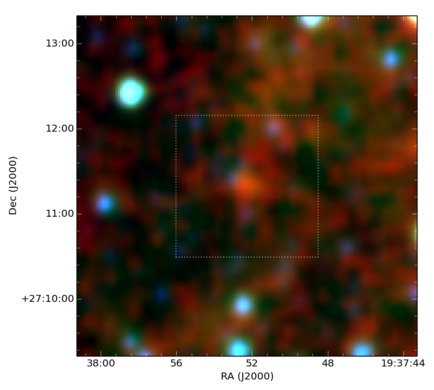}
\includegraphics[height=5.1cm]{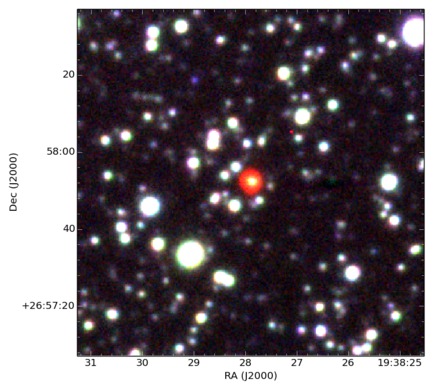}
\includegraphics[height=5.1cm]{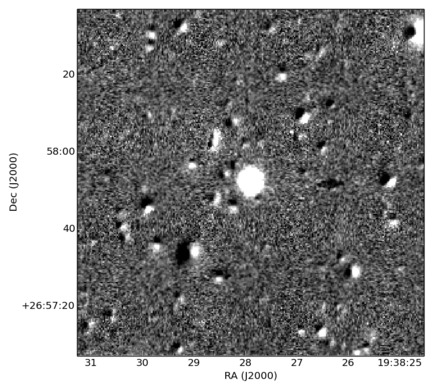}
\includegraphics[height=5.1cm]{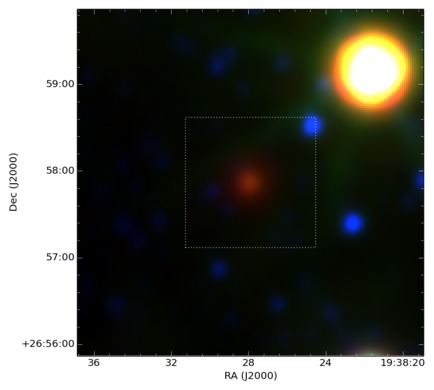}
\includegraphics[height=5.1cm]{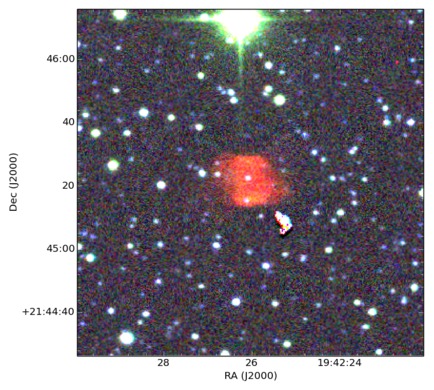}
\includegraphics[height=5.1cm]{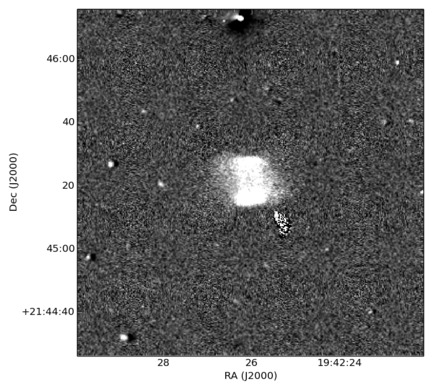}
\includegraphics[height=5.1cm]{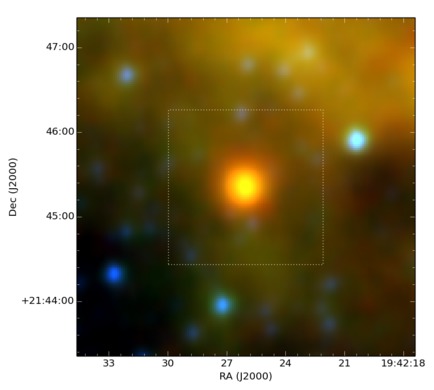}
\includegraphics[height=5.1cm]{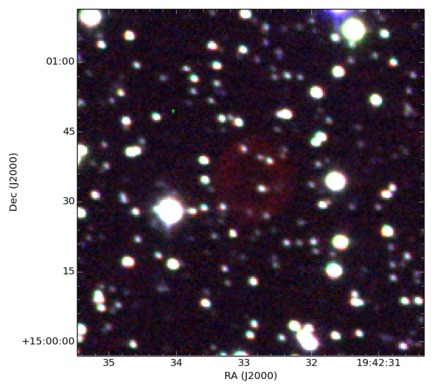}
\includegraphics[height=5.1cm]{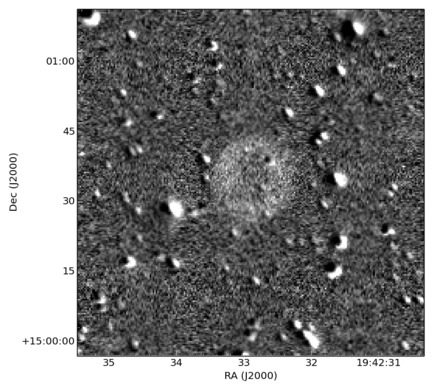}
\includegraphics[height=5.1cm]{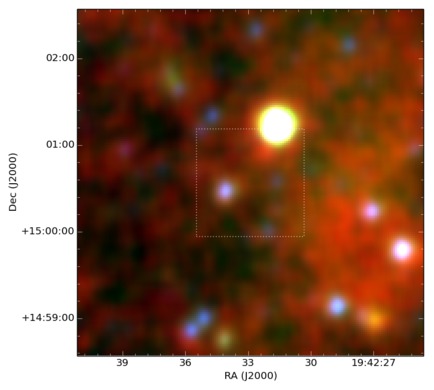}
\caption{\label{imagelabel} Same as in Fig.~\ref{image1}. Objects shown (from top to bottom):  PN G062.1+02.8,PN G062.0+02.5,PN G057.9-00.7,PN G052.1-04.1}
\end{figure*}
\clearpage
\begin{figure*}
\includegraphics[height=5.1cm]{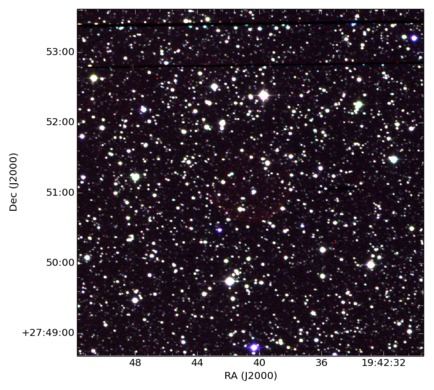}
\includegraphics[height=5.1cm]{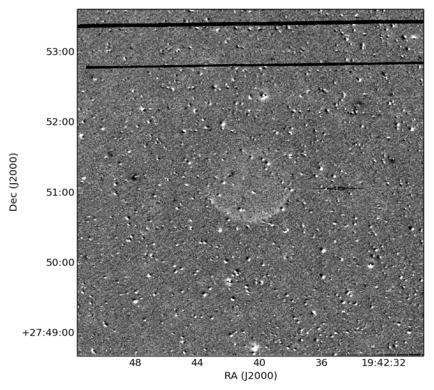}
\includegraphics[height=5.1cm]{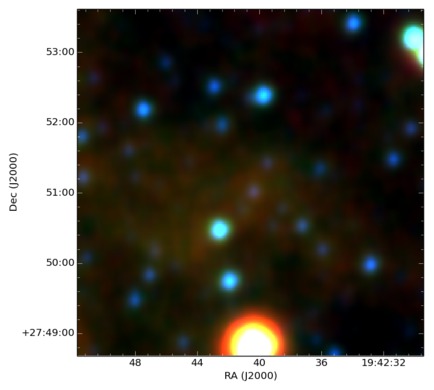}
\includegraphics[height=5.1cm]{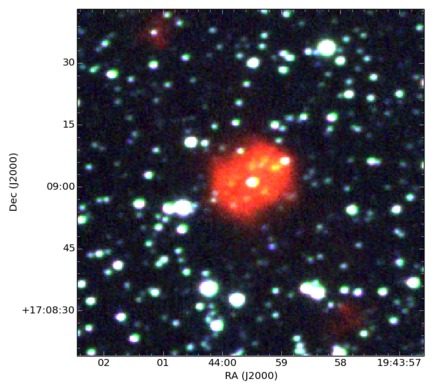}
\includegraphics[height=5.1cm]{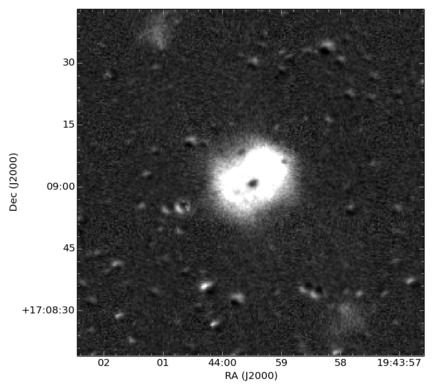}
\includegraphics[height=5.1cm]{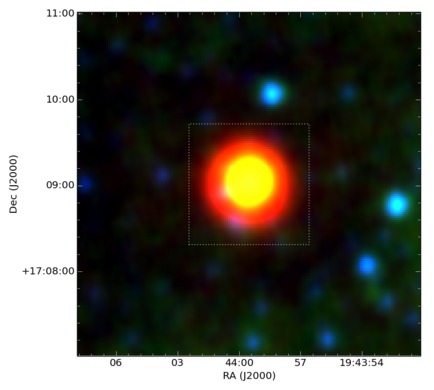}
\includegraphics[height=5.1cm]{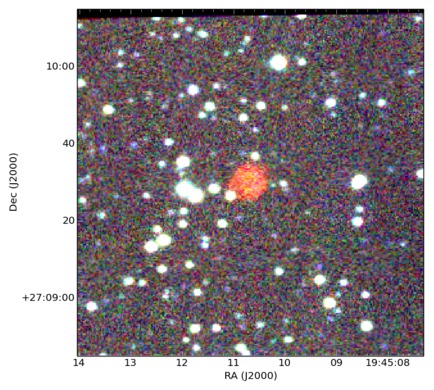}
\includegraphics[height=5.1cm]{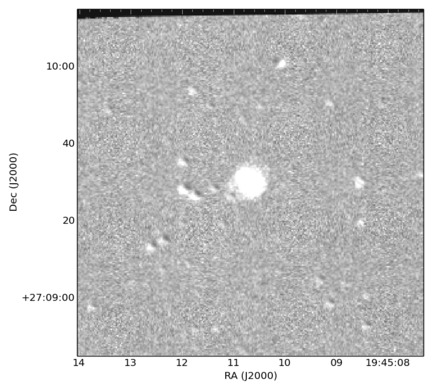}
\includegraphics[height=5.1cm]{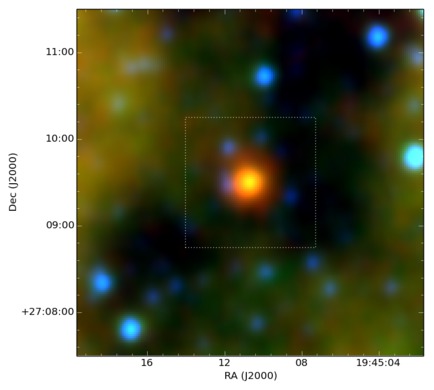}
\includegraphics[height=5.1cm]{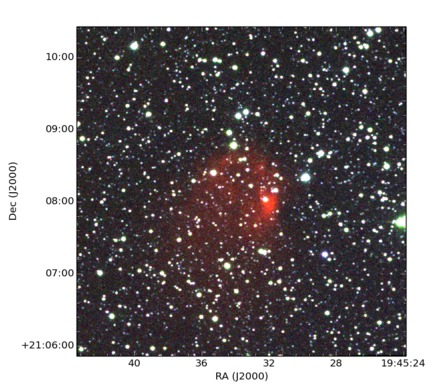}
\includegraphics[height=5.1cm]{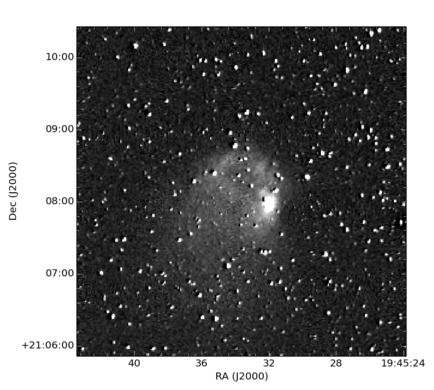}
\includegraphics[height=5.1cm]{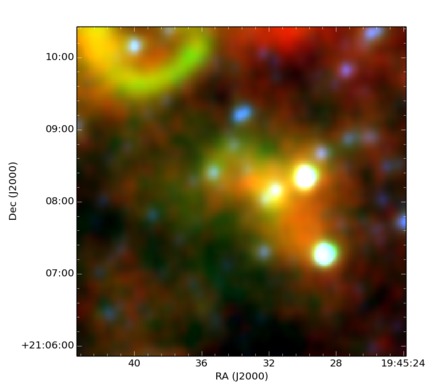}
\caption{\label{imagelabel} Same as in Fig.~\ref{image1}. Objects shown (from top to bottom):  PN G063.3+02.2,PN G054.2-03.4,PN G062.9+01.3,PN G057.8-01.7}
\end{figure*}
\clearpage
\begin{figure*}
\includegraphics[height=5.1cm]{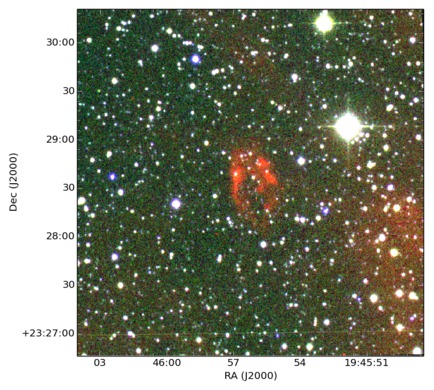}
\includegraphics[height=5.1cm]{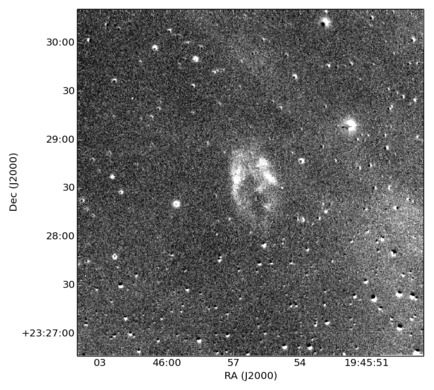}
\includegraphics[height=5.1cm]{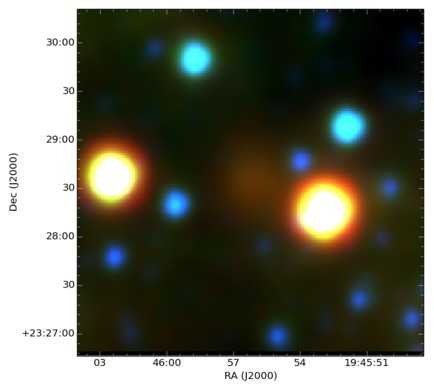}
\includegraphics[height=5.1cm]{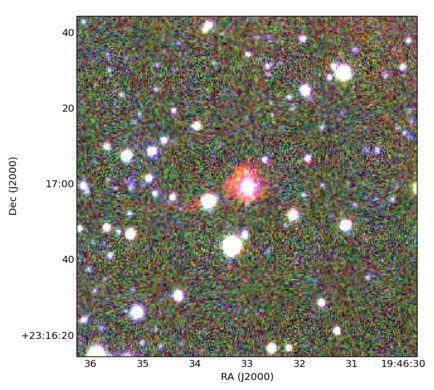}
\includegraphics[height=5.1cm]{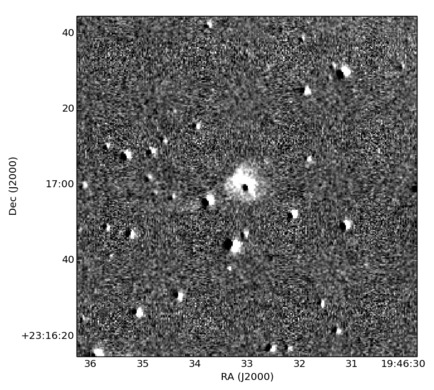}
\includegraphics[height=5.1cm]{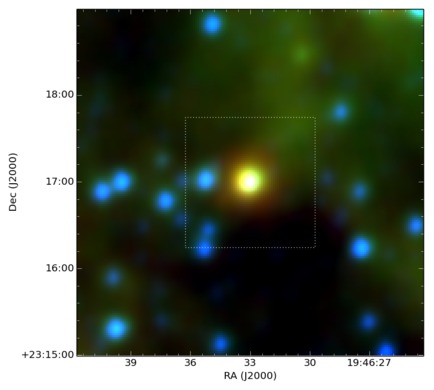}
\includegraphics[height=5.1cm]{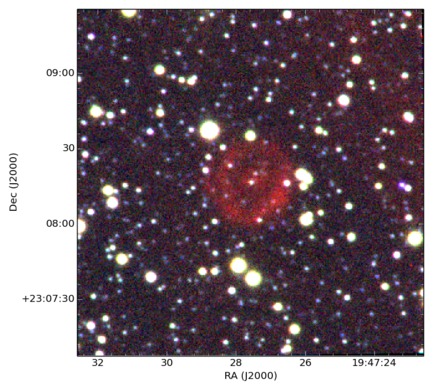}
\includegraphics[height=5.1cm]{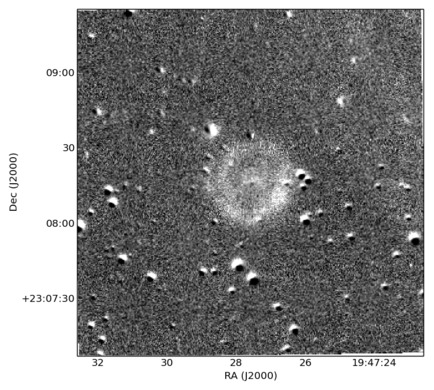}
\includegraphics[height=5.1cm]{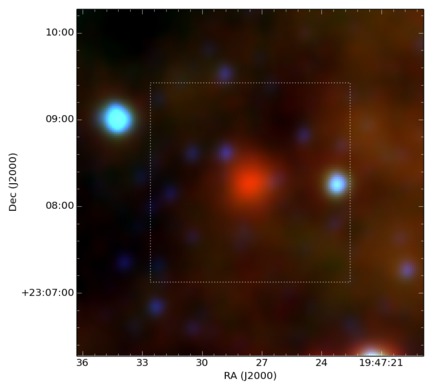}
\includegraphics[height=5.1cm]{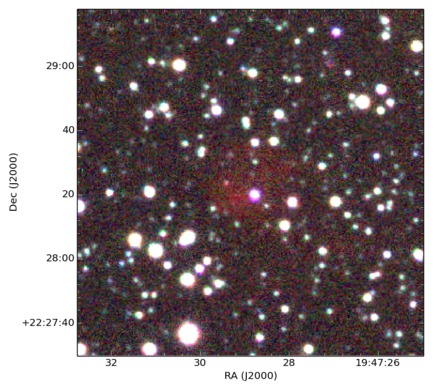}
\includegraphics[height=5.1cm]{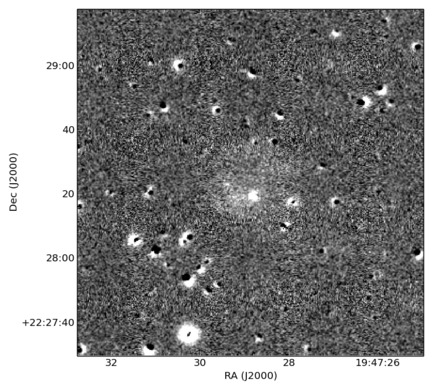}
\includegraphics[height=5.1cm]{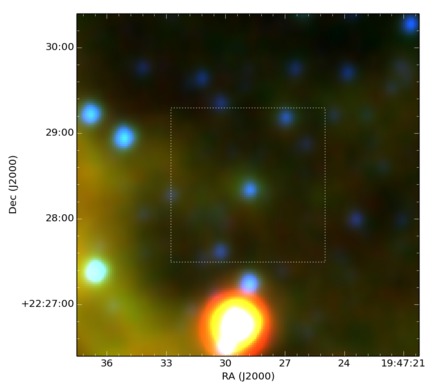}
\caption{\label{imagelabel} Same as in Fig.~\ref{image1}. Objects shown (from top to bottom):  PN G059.8-00.6,PN G059.7-00.8,PN G059.7-01.0,PN G059.1-01.4}
\end{figure*}
\clearpage
\begin{figure*}
\includegraphics[height=5.1cm]{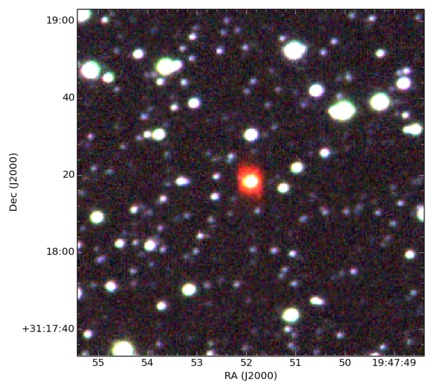}
\includegraphics[height=5.1cm]{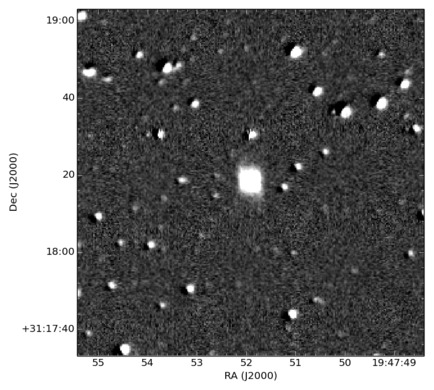}
\includegraphics[height=5.1cm]{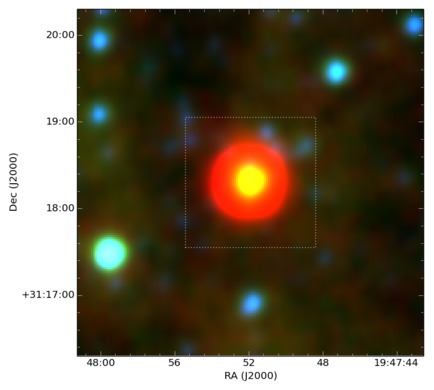}
\includegraphics[height=5.1cm]{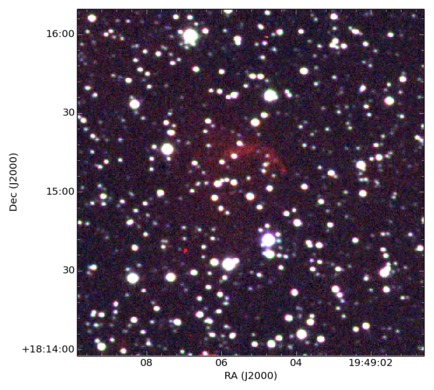}
\includegraphics[height=5.1cm]{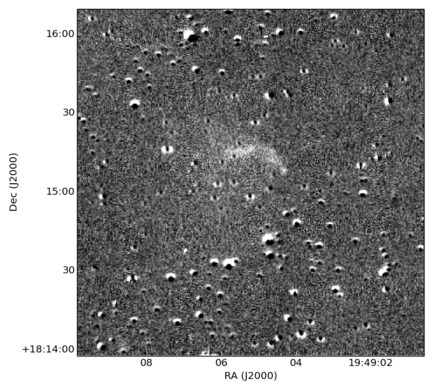}
\includegraphics[height=5.1cm]{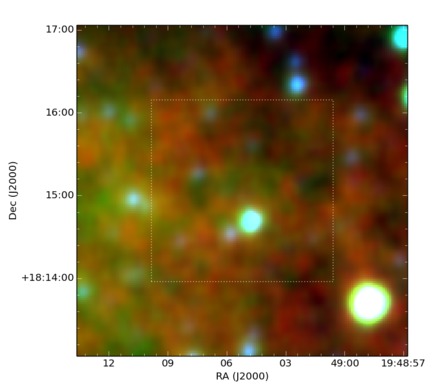}
\includegraphics[height=5.1cm]{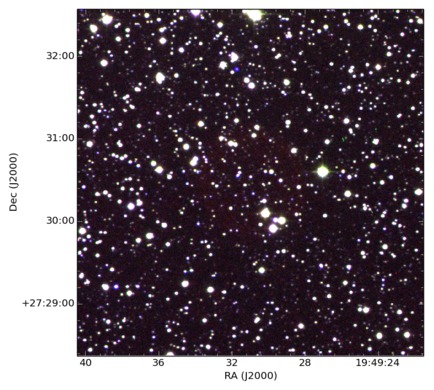}
\includegraphics[height=5.1cm]{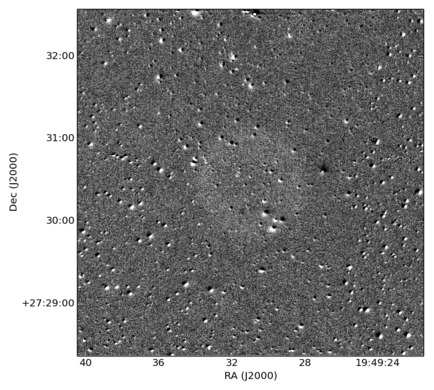}
\includegraphics[height=5.1cm]{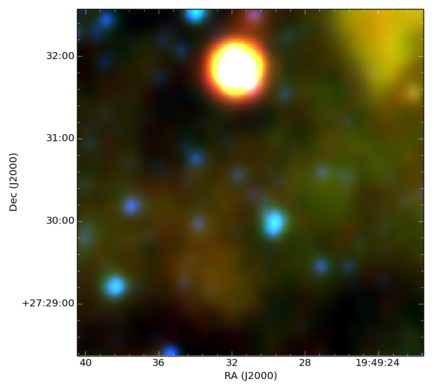}
\includegraphics[height=5.1cm]{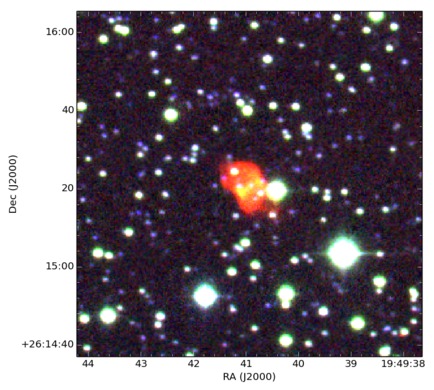}
\includegraphics[height=5.1cm]{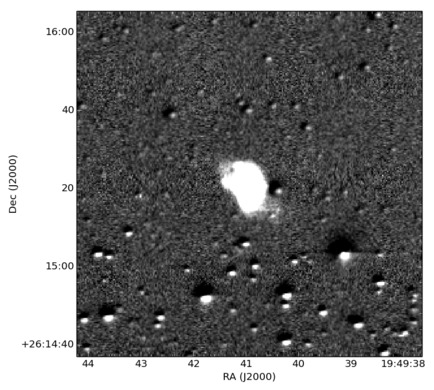}
\includegraphics[height=5.1cm]{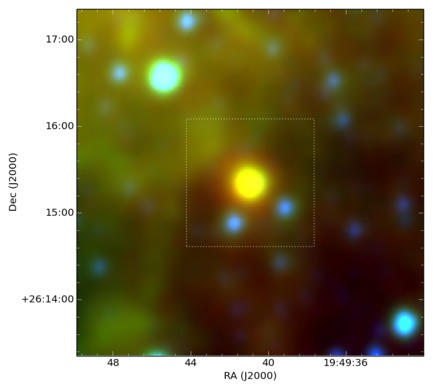}
\caption{\label{imagelabel} Same as in Fig.~\ref{image1}. Objects shown (from top to bottom):  PN G066.8+02.9,PN G055.7-03.8,PN G063.7+00.7,PN G062.7+00.0}
\end{figure*}
\clearpage
\begin{figure*}
\includegraphics[height=5.1cm]{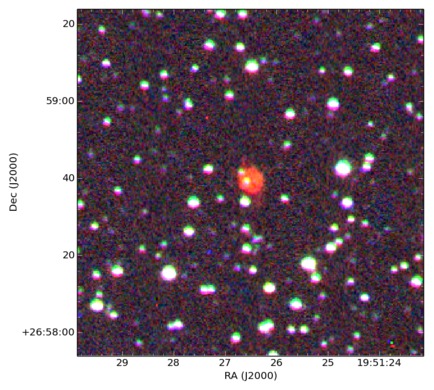}
\includegraphics[height=5.1cm]{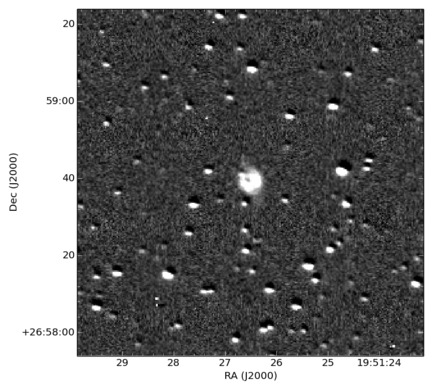}
\includegraphics[height=5.1cm]{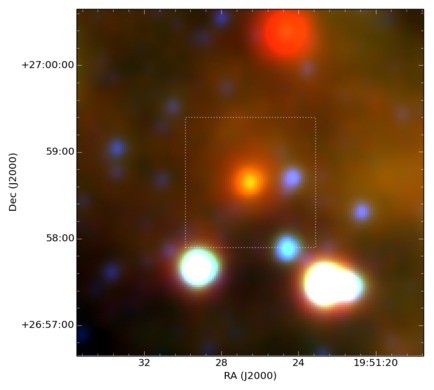}
\includegraphics[height=5.1cm]{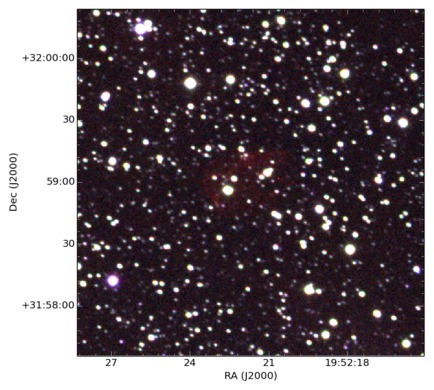}
\includegraphics[height=5.1cm]{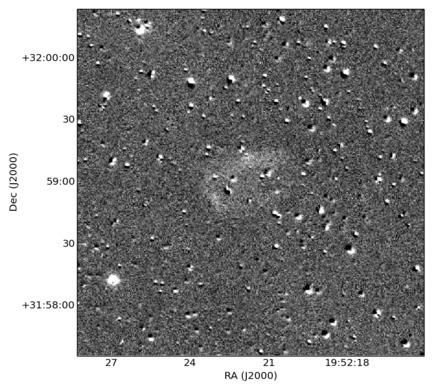}
\includegraphics[height=5.1cm]{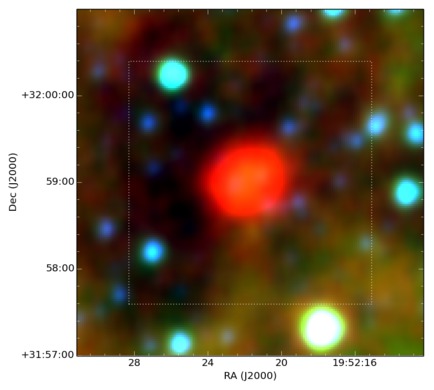}
\includegraphics[height=5.1cm]{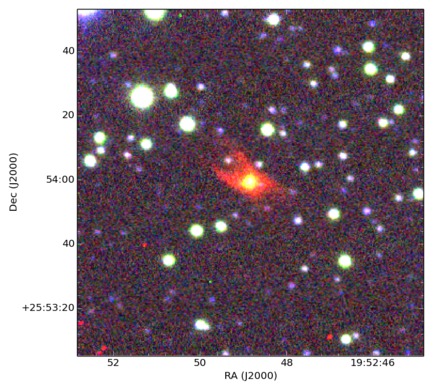}
\includegraphics[height=5.1cm]{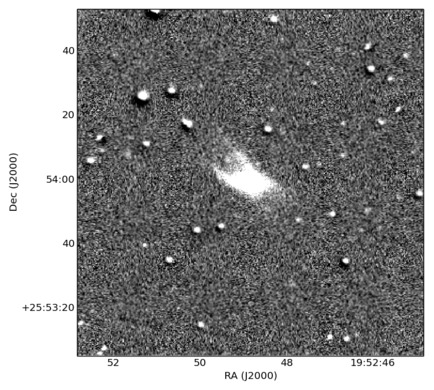}
\includegraphics[height=5.1cm]{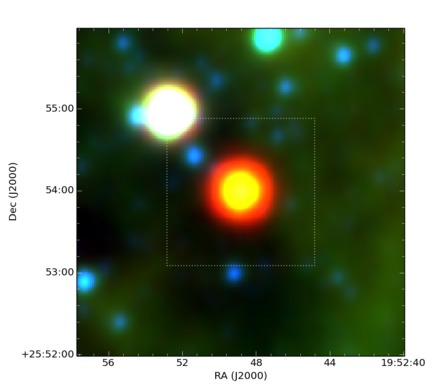}
\includegraphics[height=5.1cm]{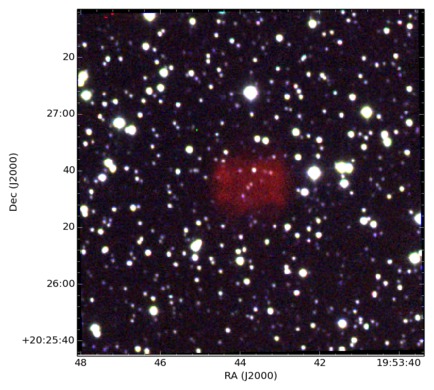}
\includegraphics[height=5.1cm]{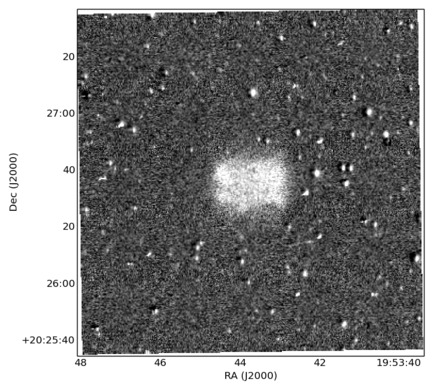}
\includegraphics[height=5.1cm]{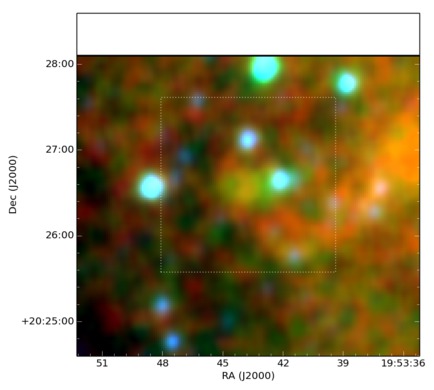}
\caption{\label{imagelabel} Same as in Fig.~\ref{image1}. Objects shown (from top to bottom):  PN G063.5+00.0,PN G067.9+02.4,PN G062.7-00.7,PN G058.1-03.7}
\end{figure*}
\clearpage
\begin{figure*}
\includegraphics[height=5.1cm]{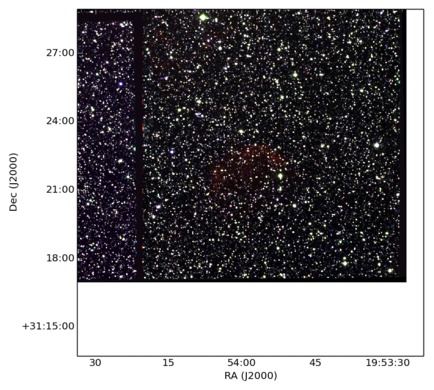}
\includegraphics[height=5.1cm]{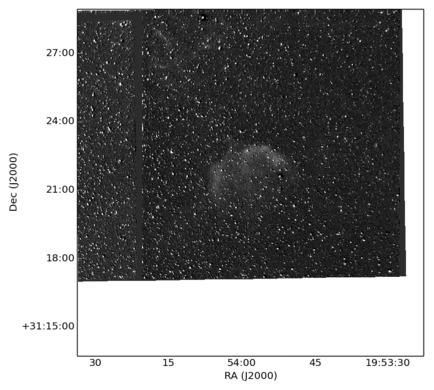}
\includegraphics[height=5.1cm]{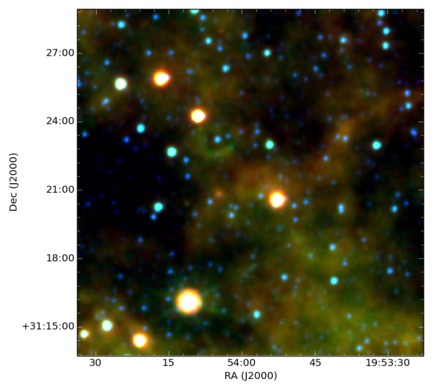}
\includegraphics[height=5.1cm]{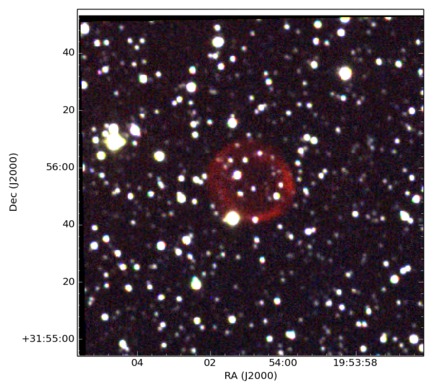}
\includegraphics[height=5.1cm]{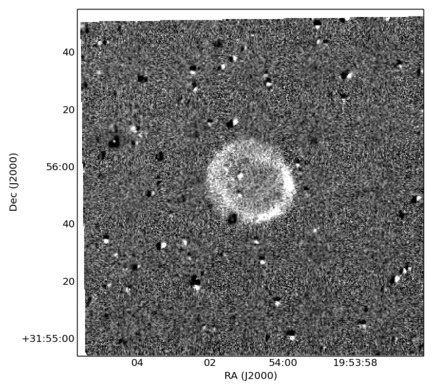}
\includegraphics[height=5.1cm]{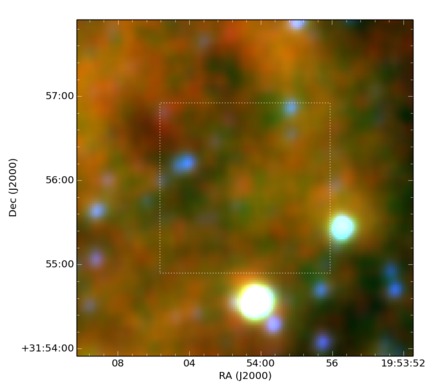}
\includegraphics[height=5.1cm]{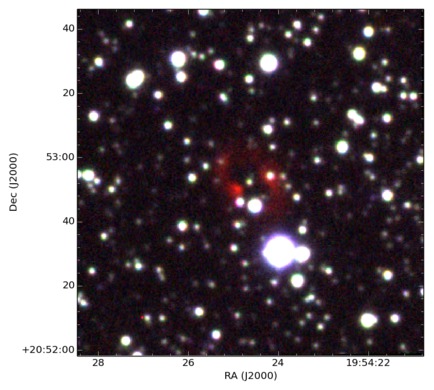}
\includegraphics[height=5.1cm]{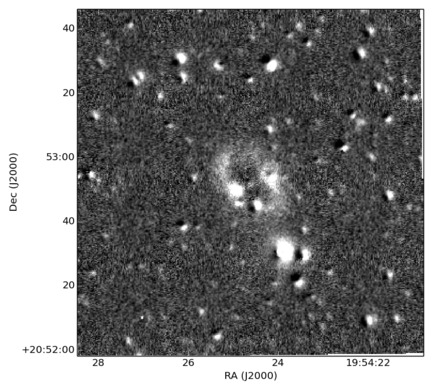}
\includegraphics[height=5.1cm]{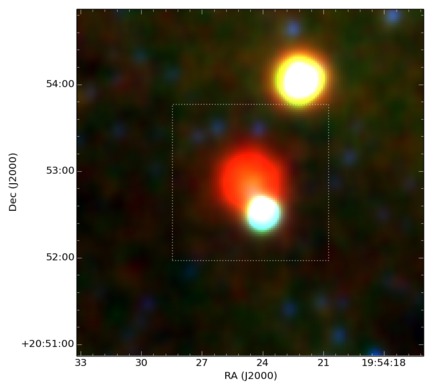}
\includegraphics[height=5.1cm]{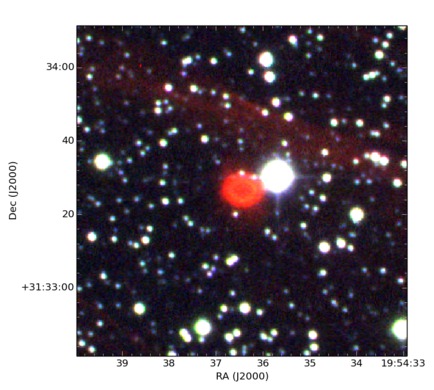}
\includegraphics[height=5.1cm]{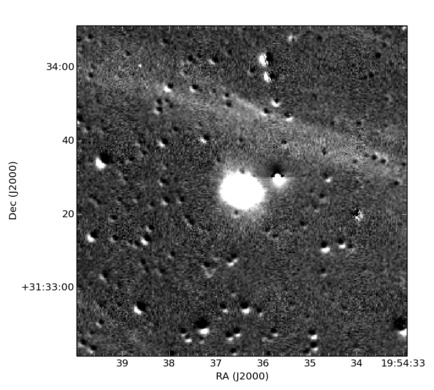}
\includegraphics[height=5.1cm]{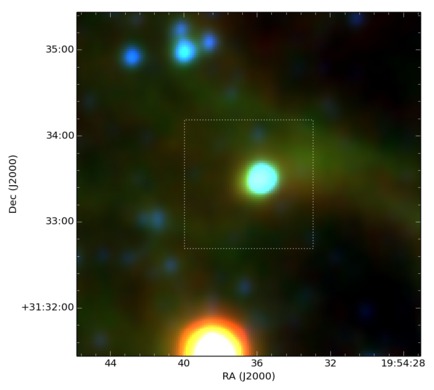}
\caption{\label{imagelabel} Same as in Fig.~\ref{image1}. Objects shown (from top to bottom):  PN G067.5+01.8,PN G068.0+02.1,PN G058.6-03.6,PN G067.8+01.8}
\end{figure*}
\clearpage
\begin{figure*}
\includegraphics[height=5.1cm]{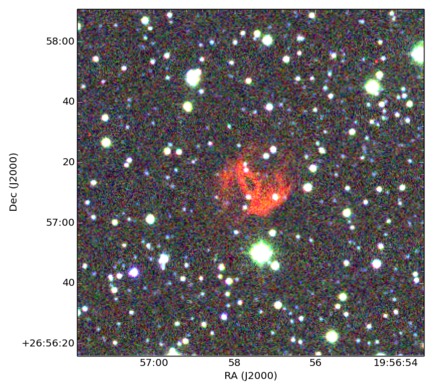}
\includegraphics[height=5.1cm]{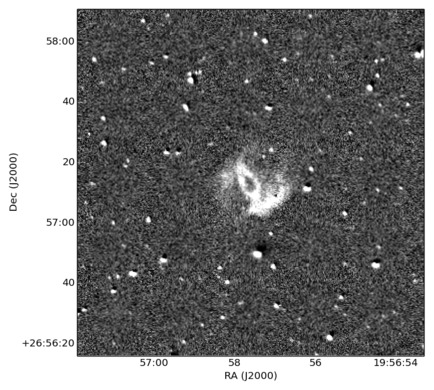}
\includegraphics[height=5.1cm]{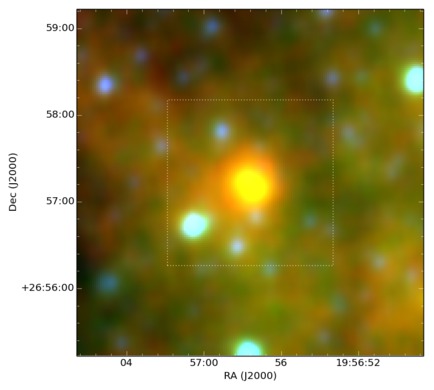}
\includegraphics[height=5.1cm]{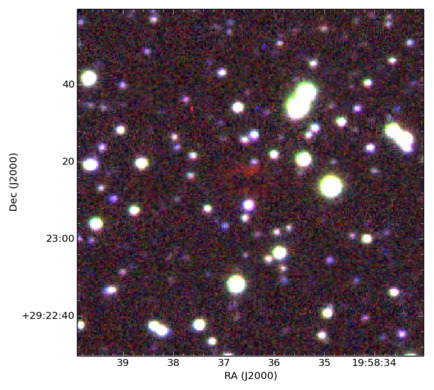}
\includegraphics[height=5.1cm]{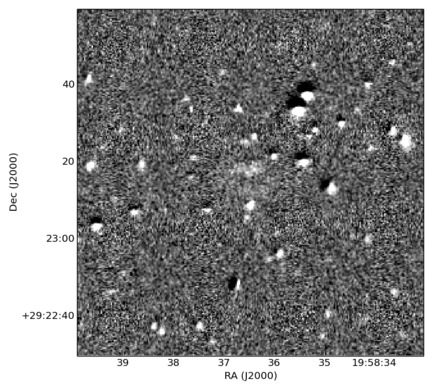}
\includegraphics[height=5.1cm]{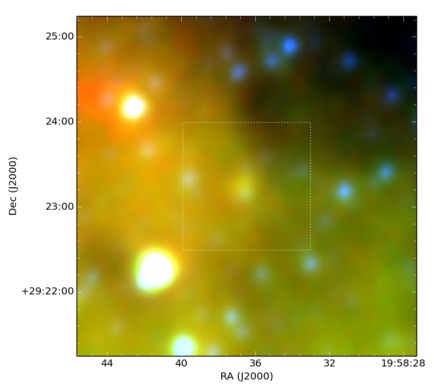}
\includegraphics[height=5.1cm]{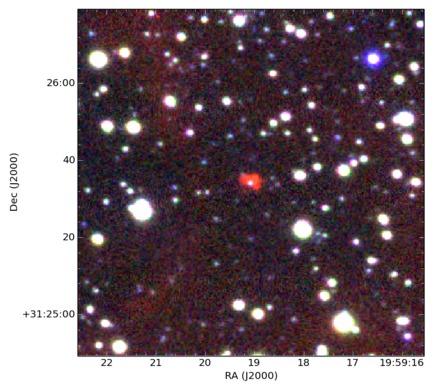}
\includegraphics[height=5.1cm]{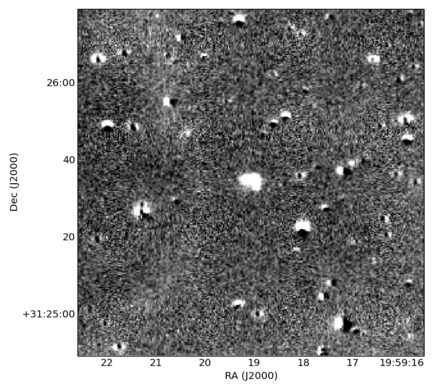}
\includegraphics[height=5.1cm]{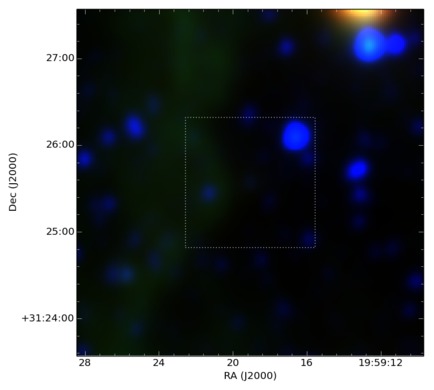}
\includegraphics[height=5.1cm]{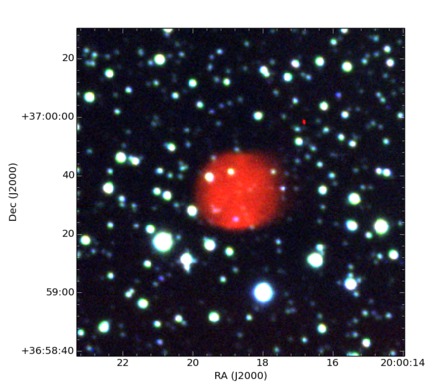}
\includegraphics[height=5.1cm]{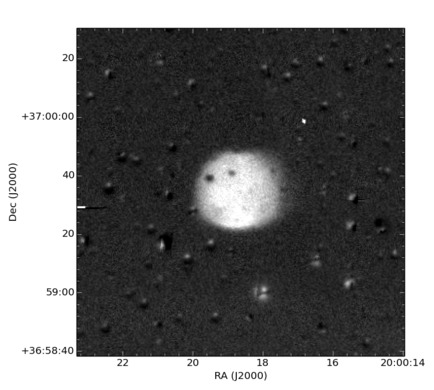}
\includegraphics[height=5.1cm]{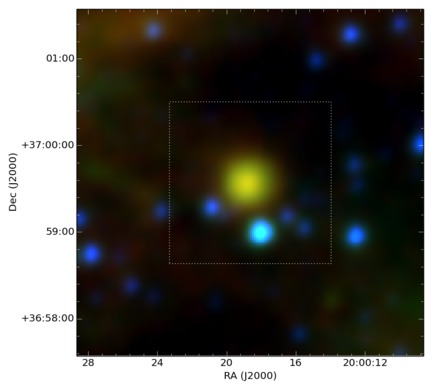}
\caption{\label{imagelabel} Same as in Fig.~\ref{image1}. Objects shown (from top to bottom):  PN G064.1-00.9,PN G066.4-00.0,PN G068.2+00.9,PN G073.0+03.6}
\end{figure*}
\clearpage
\begin{figure*}
\includegraphics[height=5.1cm]{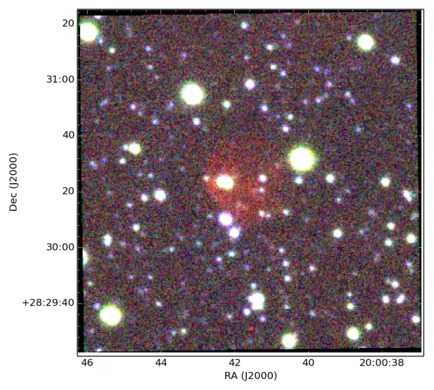}
\includegraphics[height=5.1cm]{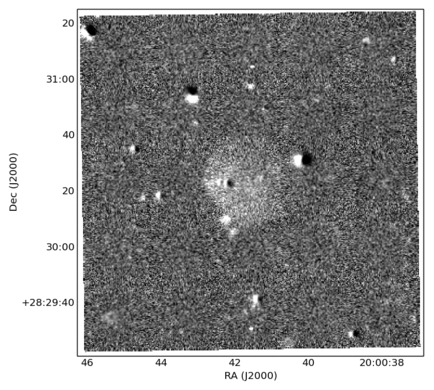}
\includegraphics[height=5.1cm]{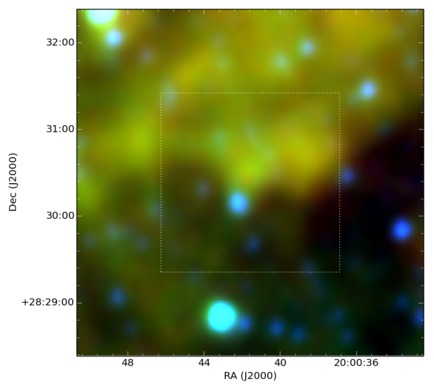}
\includegraphics[height=5.1cm]{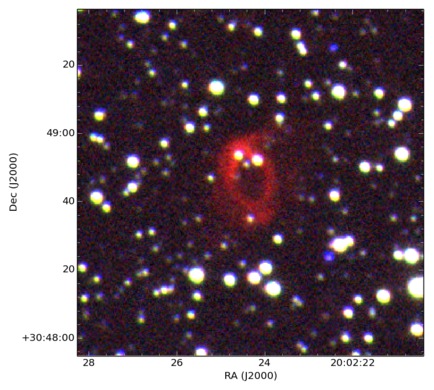}
\includegraphics[height=5.1cm]{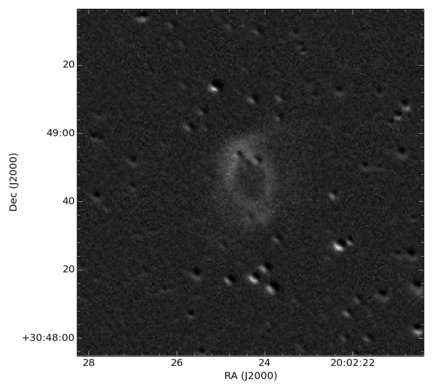}
\includegraphics[height=5.1cm]{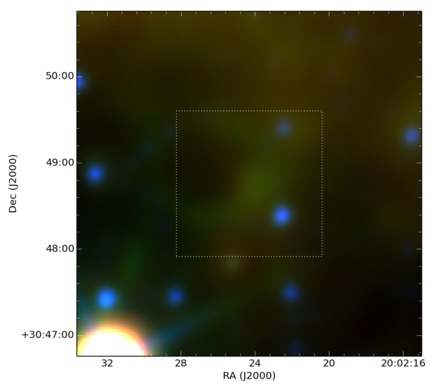}
\includegraphics[height=5.1cm]{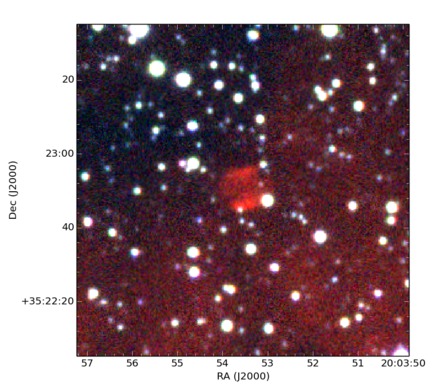}
\includegraphics[height=5.1cm]{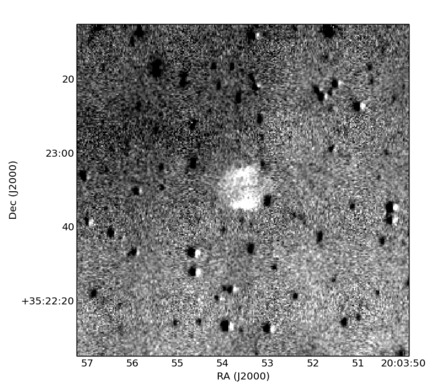}
\includegraphics[height=5.1cm]{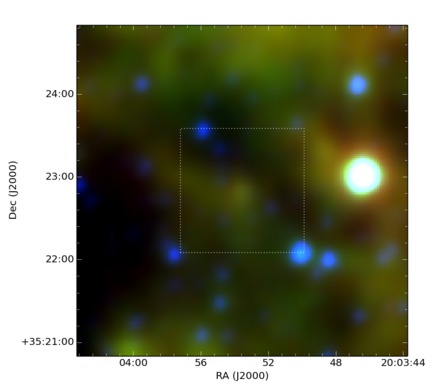}
\includegraphics[height=5.1cm]{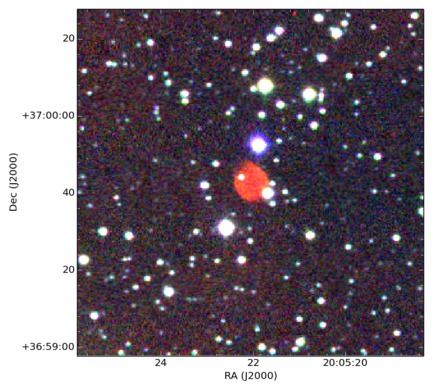}
\includegraphics[height=5.1cm]{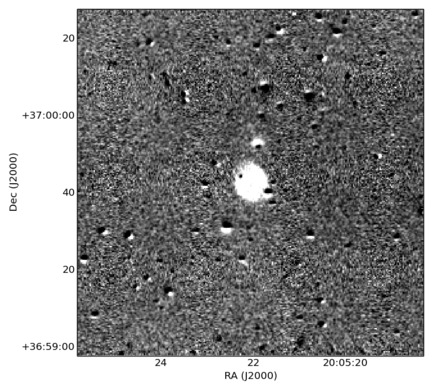}
\includegraphics[height=5.1cm]{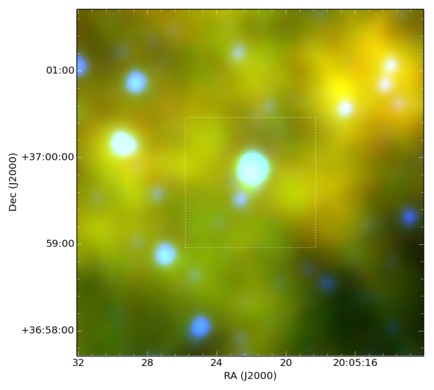}
\caption{\label{imagelabel} Same as in Fig.~\ref{image1}. Objects shown (from top to bottom):  PN G065.8-00.8,PN G068.0+00.0,PN G072.0+02.2,PN G073.6+02.8}
\end{figure*}
\clearpage
\begin{figure*}
\includegraphics[height=5.1cm]{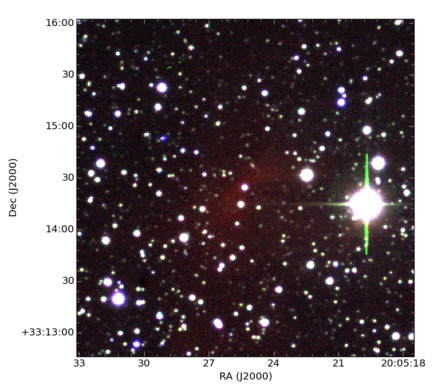}
\includegraphics[height=5.1cm]{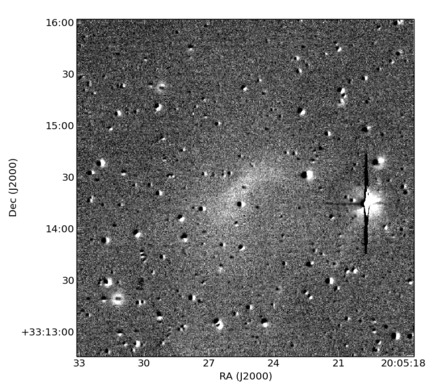}
\includegraphics[height=5.1cm]{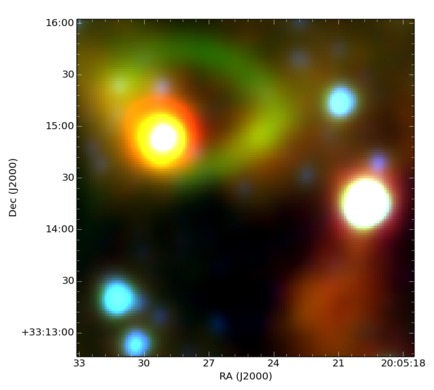}
\includegraphics[height=5.1cm]{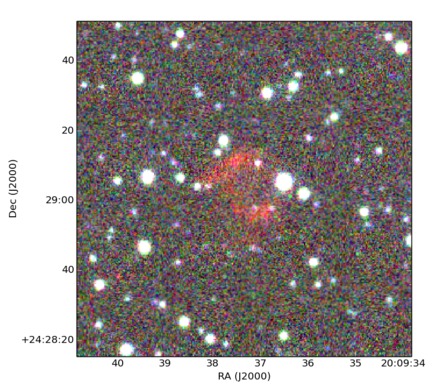}
\includegraphics[height=5.1cm]{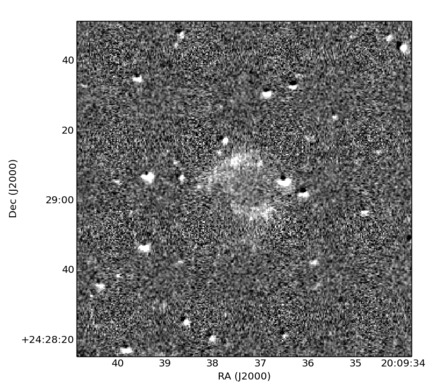}
\includegraphics[height=5.1cm]{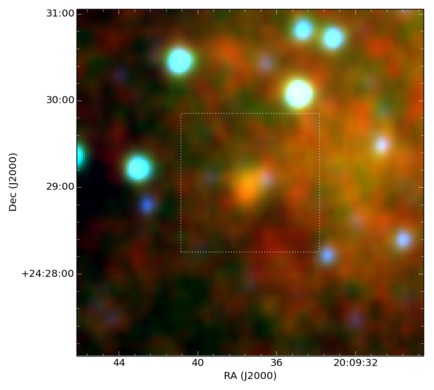}
\includegraphics[height=5.1cm]{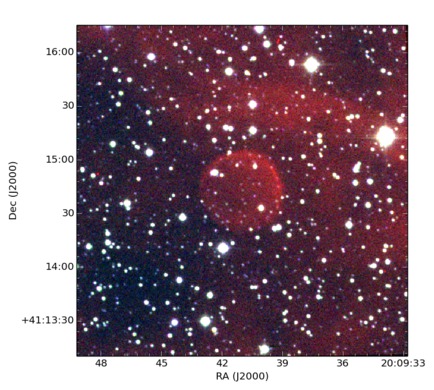}
\includegraphics[height=5.1cm]{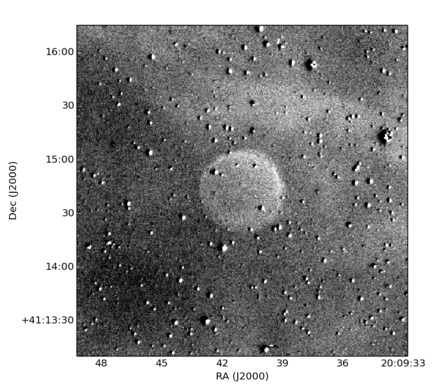}
\includegraphics[height=5.1cm]{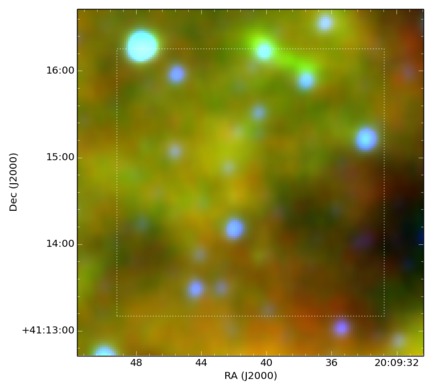}
\includegraphics[height=5.1cm]{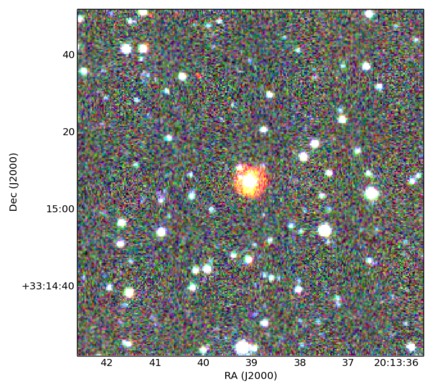}
\includegraphics[height=5.1cm]{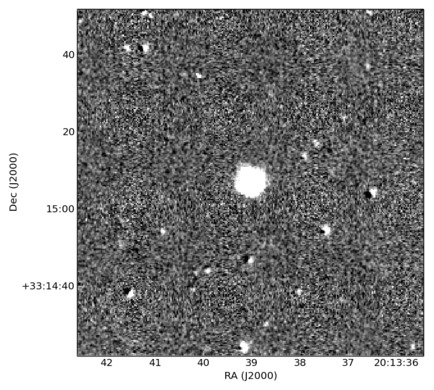}
\includegraphics[height=5.1cm]{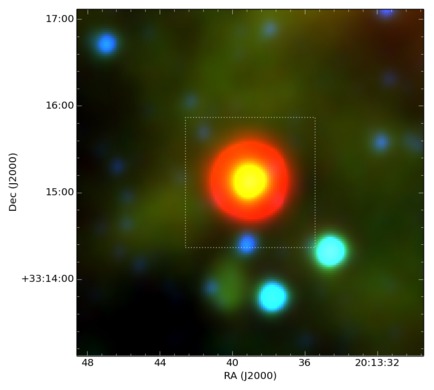}
\caption{\label{imagelabel} Same as in Fig.~\ref{image1}. Objects shown (from top to bottom):  PN G070.4+00.7,PN G063.5-04.7,PN G077.6+04.3,PN G071.3-00.6}
\end{figure*}
\clearpage
\begin{figure*}
\includegraphics[height=5.1cm]{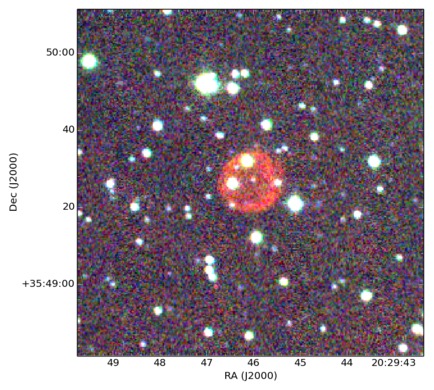}
\includegraphics[height=5.1cm]{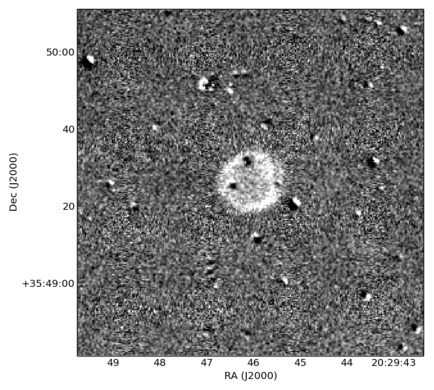}
\includegraphics[height=5.1cm]{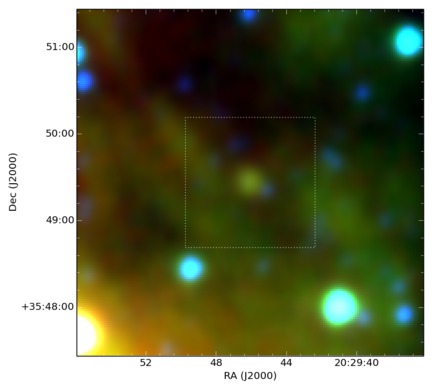}
\includegraphics[height=5.1cm]{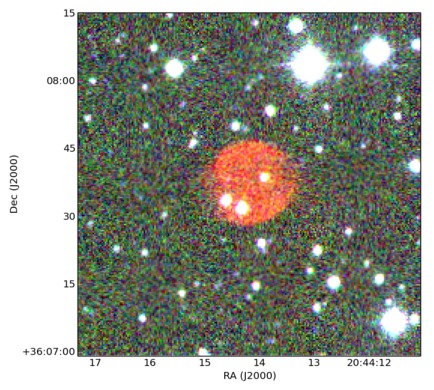}
\includegraphics[height=5.1cm]{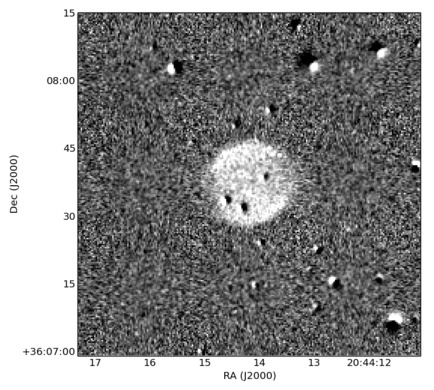}
\includegraphics[height=5.1cm]{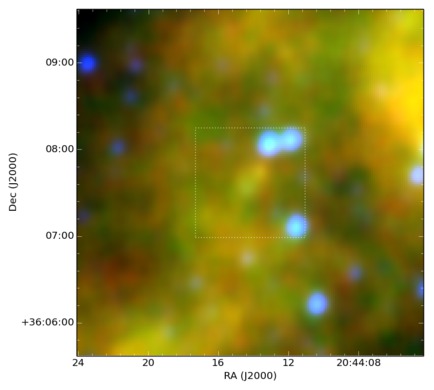}
\includegraphics[height=5.1cm]{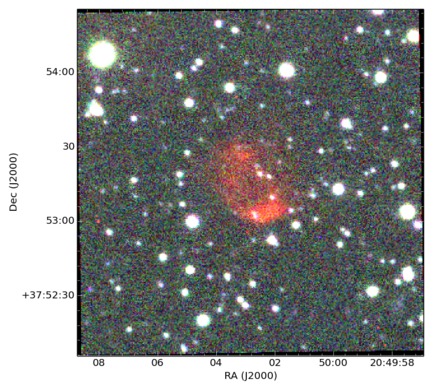}
\includegraphics[height=5.1cm]{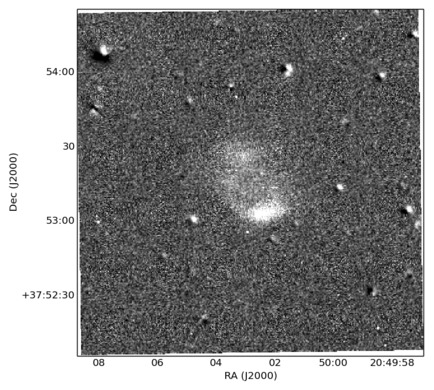}
\includegraphics[height=5.1cm]{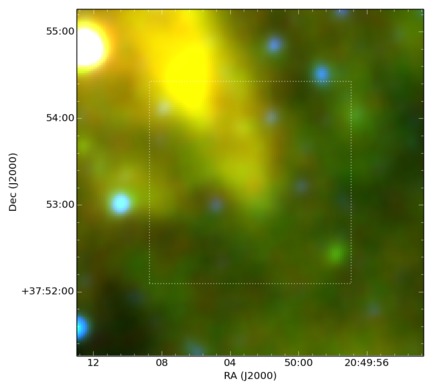}
\includegraphics[height=5.1cm]{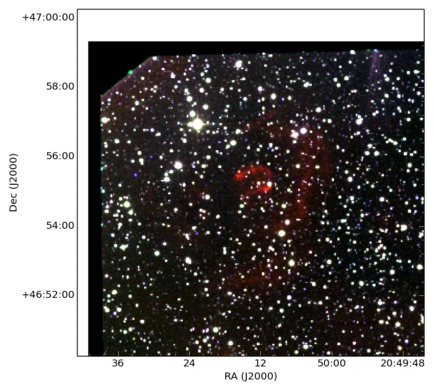}
\includegraphics[height=5.1cm]{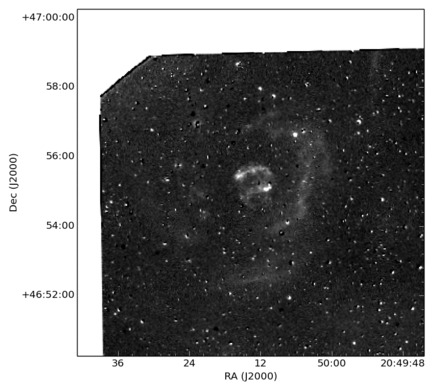}
\includegraphics[height=5.1cm]{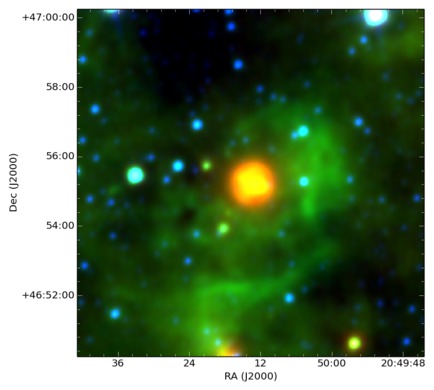}
\caption{\label{imagelabel} Same as in Fig.~\ref{image1}. Objects shown (from top to bottom):  PN G075.3-01.9,PN G077.4-04.0,PN G079.5-03.8,PN G086.5+01.8}
\end{figure*}
\clearpage
\begin{figure*}
\includegraphics[height=5.1cm]{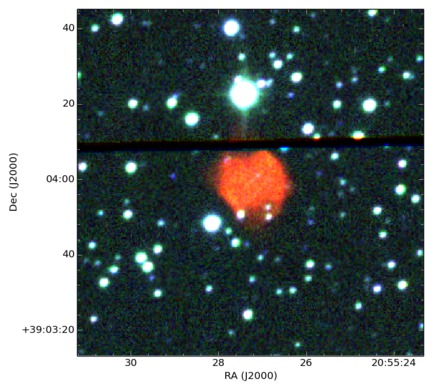}
\includegraphics[height=5.1cm]{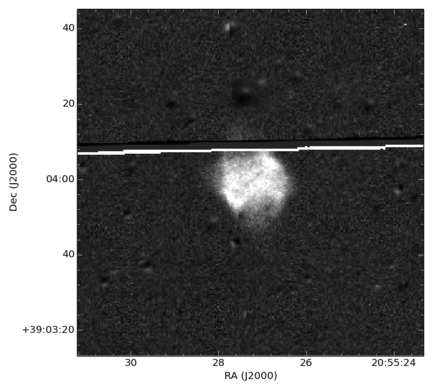}
\includegraphics[height=5.1cm]{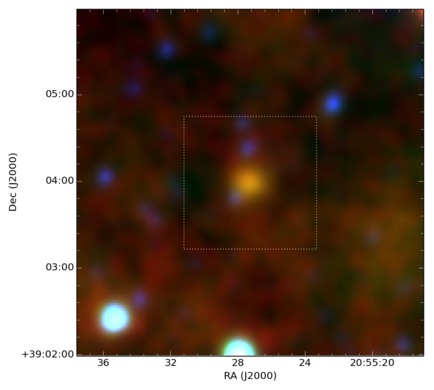}
\includegraphics[height=5.1cm]{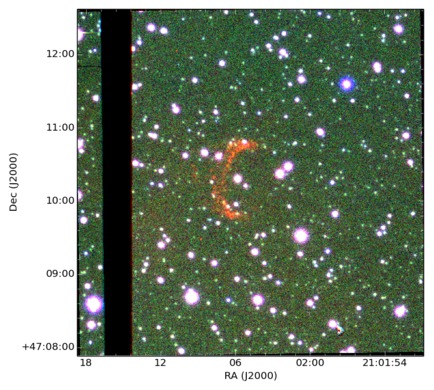}
\includegraphics[height=5.1cm]{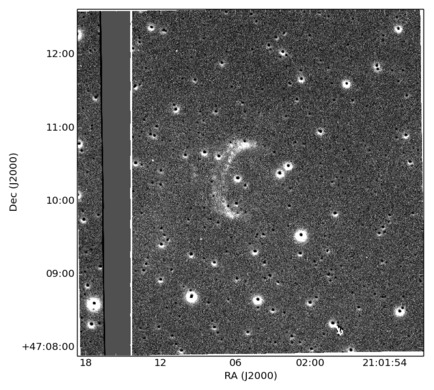}
\includegraphics[height=5.1cm]{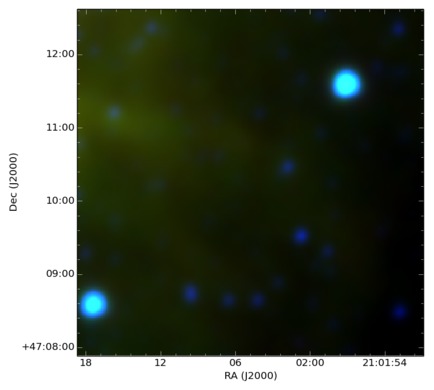}
\includegraphics[height=5.1cm]{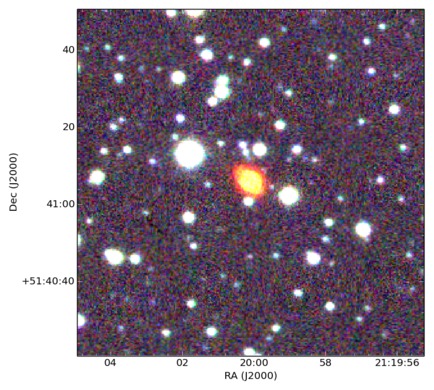}
\includegraphics[height=5.1cm]{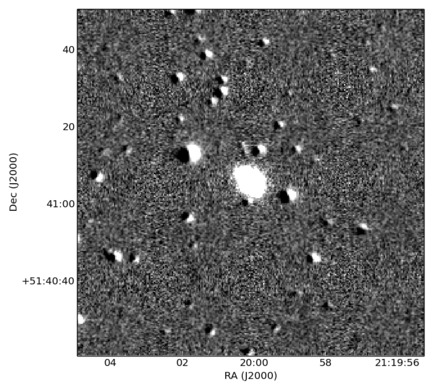}
\includegraphics[height=5.1cm]{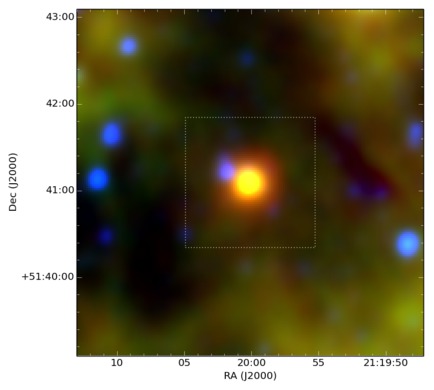}
\includegraphics[height=5.1cm]{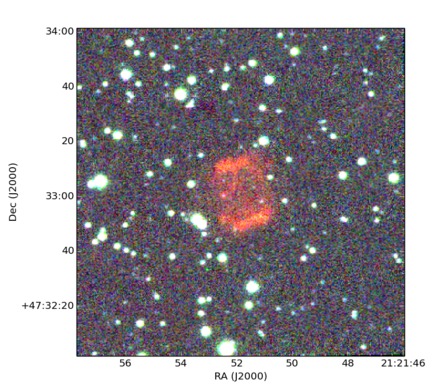}
\includegraphics[height=5.1cm]{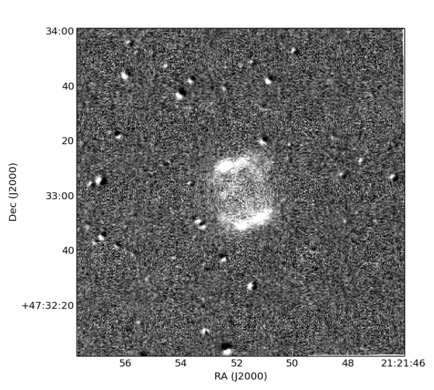}
\includegraphics[height=5.1cm]{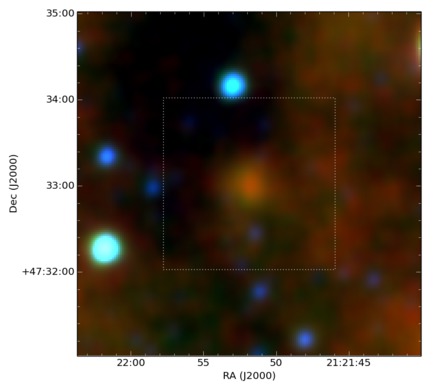}
\caption{\label{imagelabel} Same as in Fig.~\ref{image1}. Objects shown (from top to bottom):  PN G081.0-03.9,PN G088.0+00.4,PN G093.3+01.4,PN G090.5-01.7}
\end{figure*}
\clearpage
\begin{figure*}
\includegraphics[height=5.1cm]{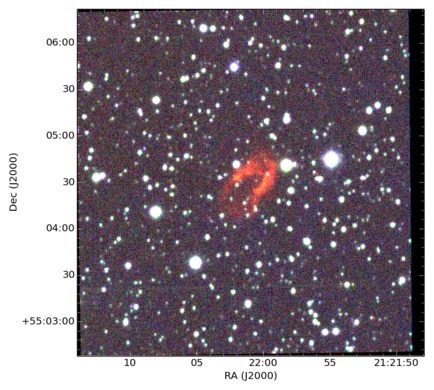}
\includegraphics[height=5.1cm]{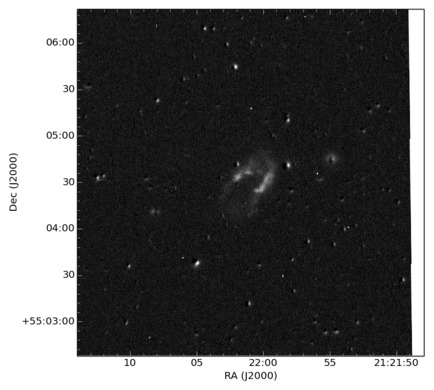}
\includegraphics[height=5.1cm]{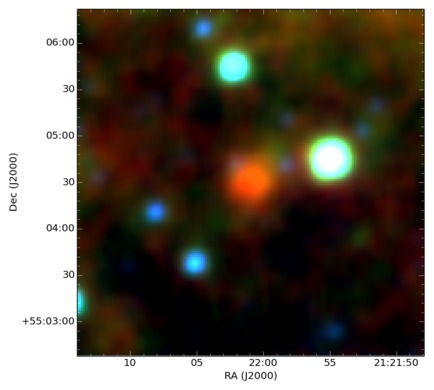}
\includegraphics[height=5.1cm]{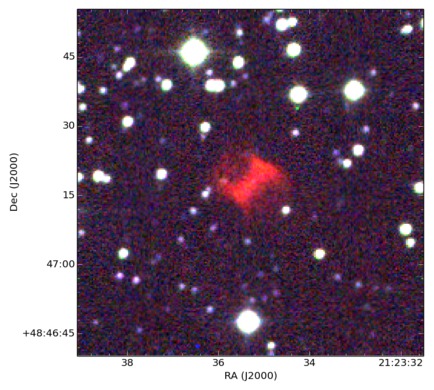}
\includegraphics[height=5.1cm]{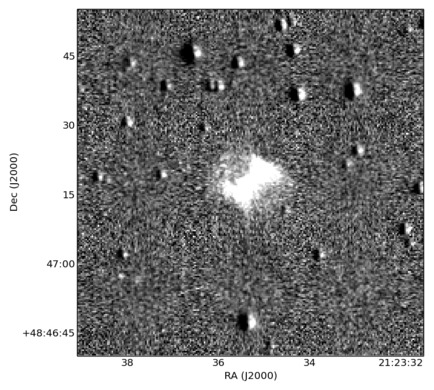}
\includegraphics[height=5.1cm]{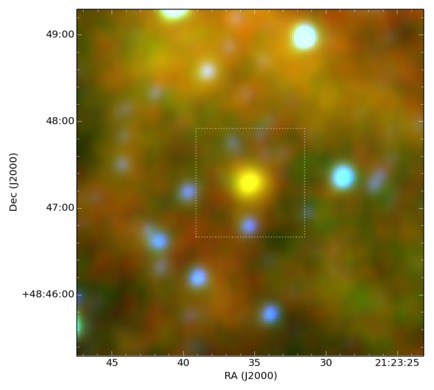}
\includegraphics[height=5.1cm]{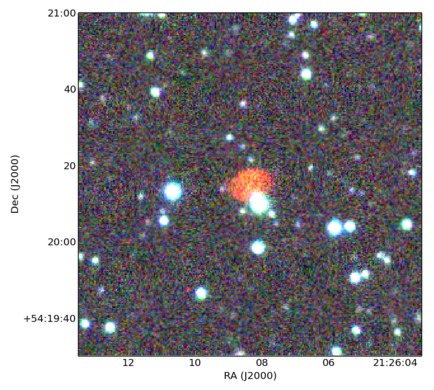}
\includegraphics[height=5.1cm]{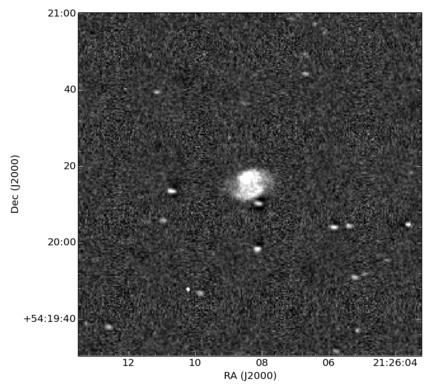}
\includegraphics[height=5.1cm]{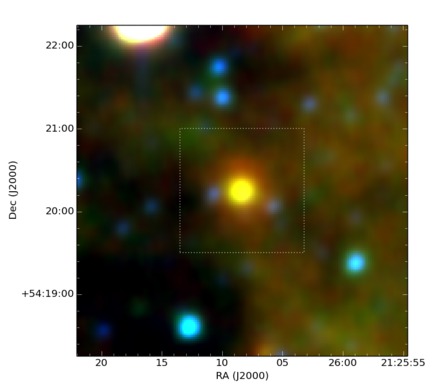}
\includegraphics[height=5.1cm]{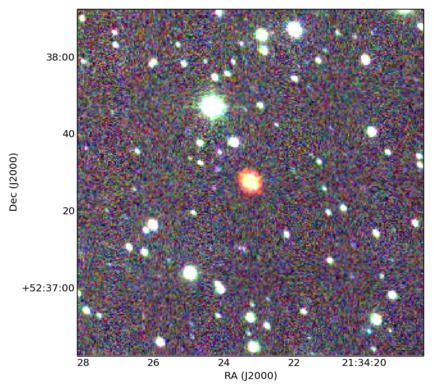}
\includegraphics[height=5.1cm]{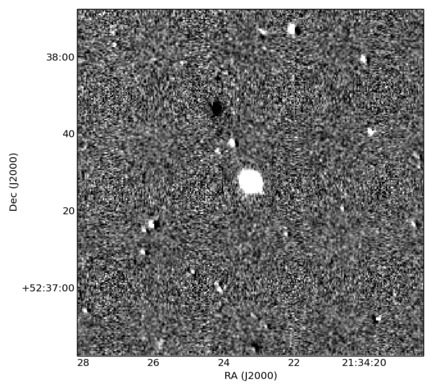}
\includegraphics[height=5.1cm]{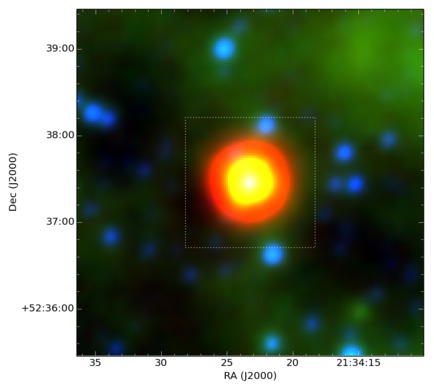}
\caption{\label{imagelabel} Same as in Fig.~\ref{image1}. Objects shown (from top to bottom):  PN G095.9+03.5,PN G091.6-01.0,PN G095.8+02.6,PN G095.5+00.5}
\end{figure*}
\clearpage
\begin{figure*}
\includegraphics[height=5.1cm]{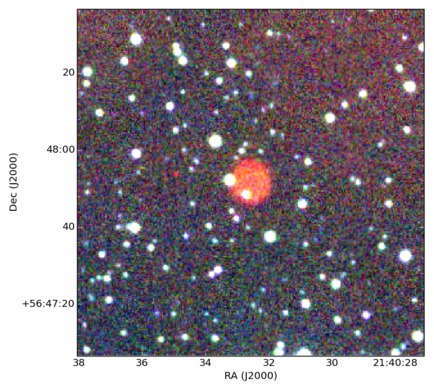}
\includegraphics[height=5.1cm]{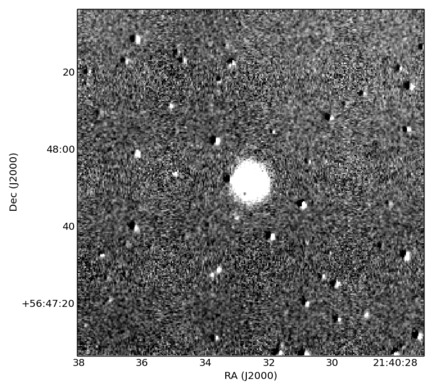}
\includegraphics[height=5.1cm]{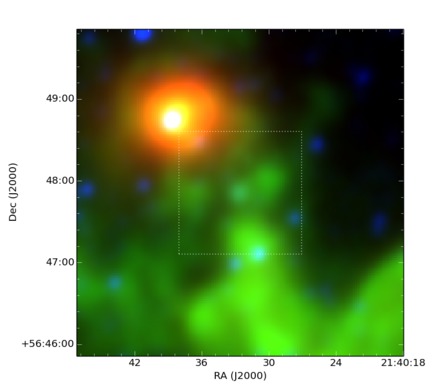}
\includegraphics[height=5.1cm]{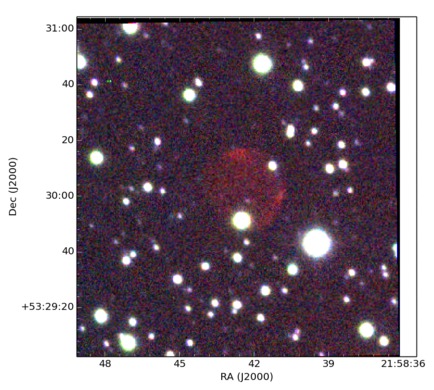}
\includegraphics[height=5.1cm]{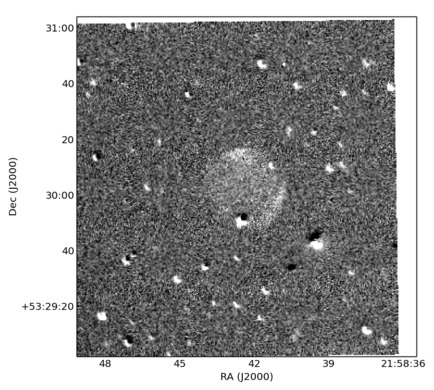}
\includegraphics[height=5.1cm]{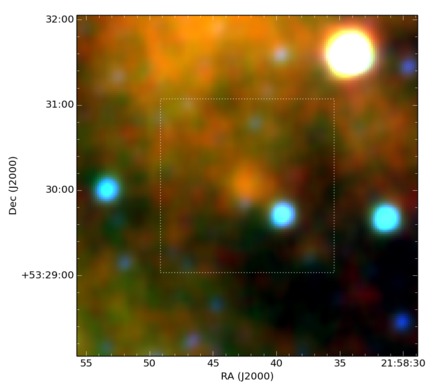}
\includegraphics[height=5.1cm]{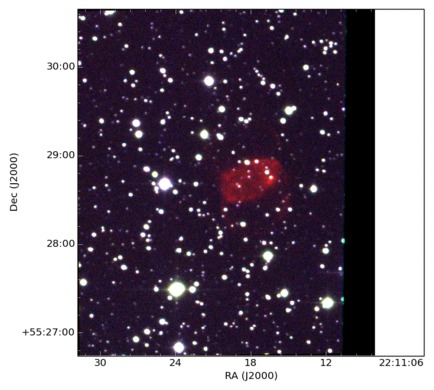}
\includegraphics[height=5.1cm]{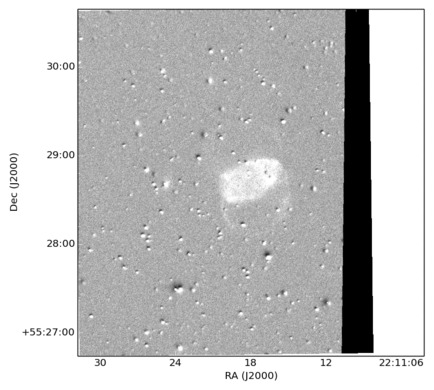}
\includegraphics[height=5.1cm]{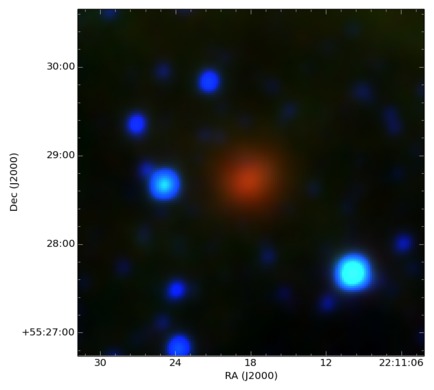}
\includegraphics[height=5.1cm]{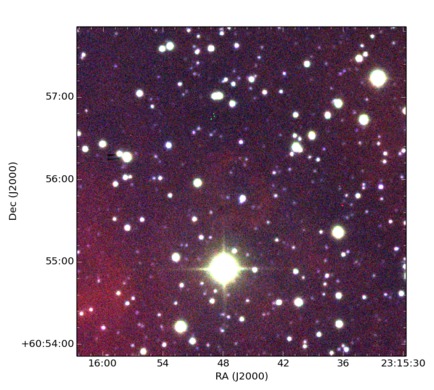}
\includegraphics[height=5.1cm]{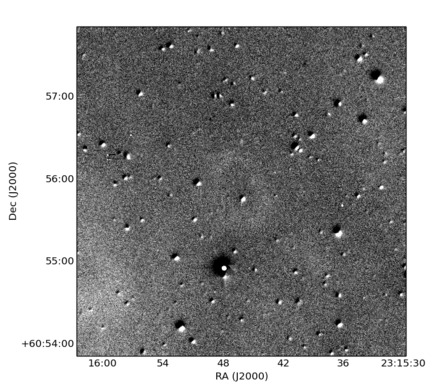}
\includegraphics[height=5.1cm]{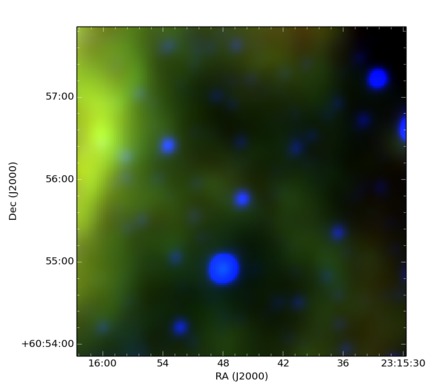}
\caption{\label{imagelabel} Same as in Fig.~\ref{image1}. Objects shown (from top to bottom):  PN G098.9+03.0,PN G098.9-01.1,PN G101.5-00.6,PN G111.5+00.1}
\end{figure*}
\clearpage
\begin{figure*}
\includegraphics[height=5.1cm]{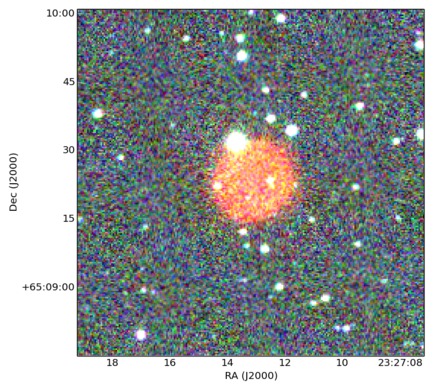}
\includegraphics[height=5.1cm]{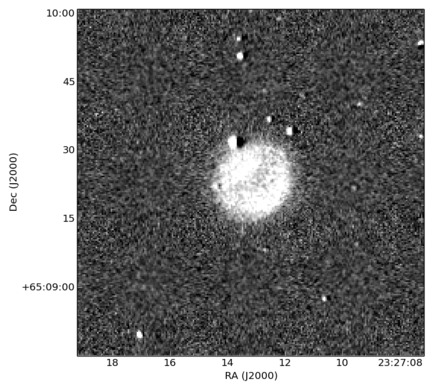}
\includegraphics[height=5.1cm]{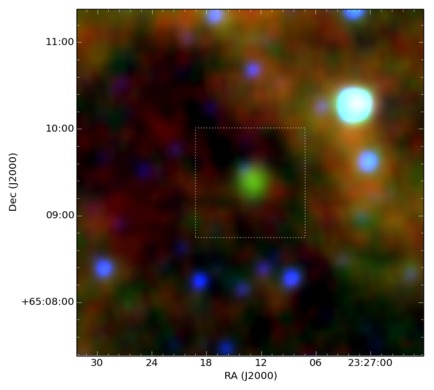}
\includegraphics[height=5.1cm]{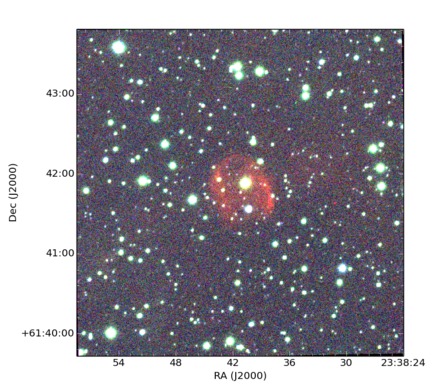}
\includegraphics[height=5.1cm]{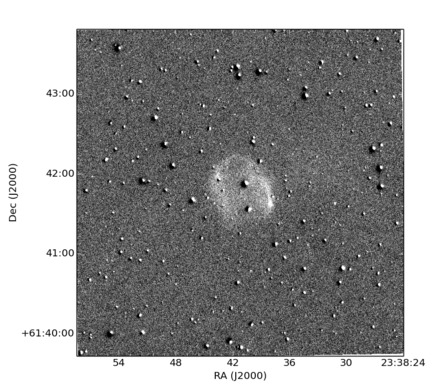}
\includegraphics[height=5.1cm]{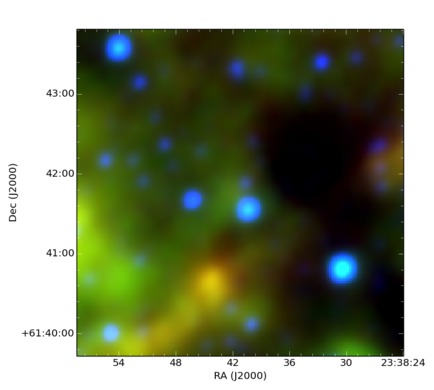}
\includegraphics[height=5.1cm]{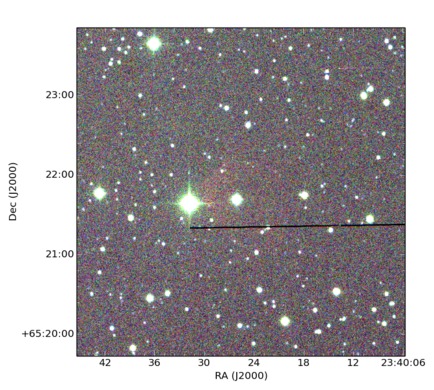}
\includegraphics[height=5.1cm]{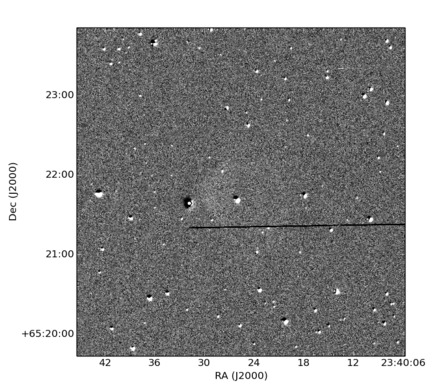}
\includegraphics[height=5.1cm]{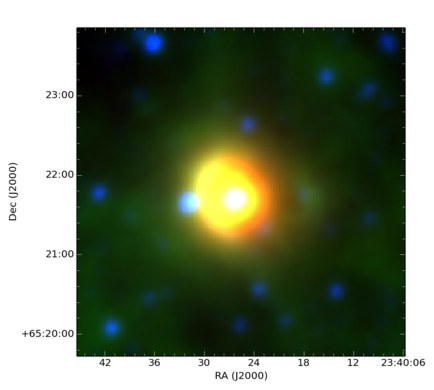}
\includegraphics[height=5.1cm]{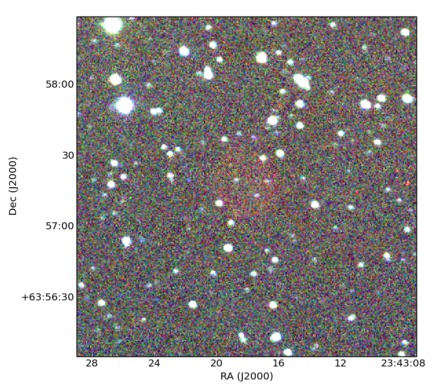}
\includegraphics[height=5.1cm]{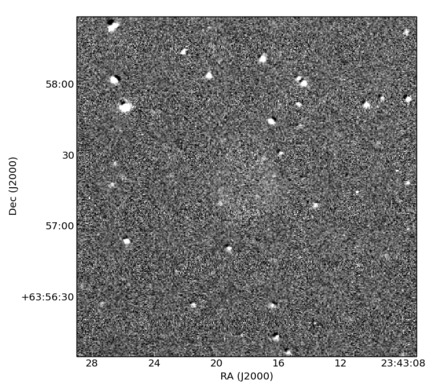}
\includegraphics[height=5.1cm]{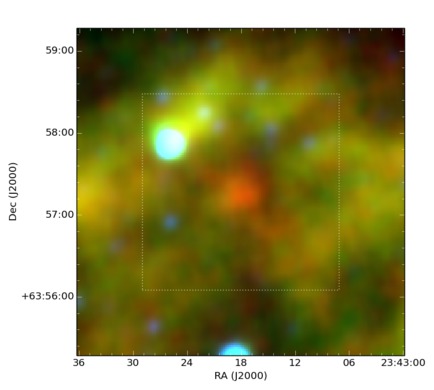}
\caption{\label{imagelabel} Same as in Fig.~\ref{image1}. Objects shown (from top to bottom):  PN G114.2+03.7,PN G114.4+00.0,PN G115.6+03.5,PN G115.5+02.0}
\end{figure*}
\clearpage
\begin{figure*}
\includegraphics[height=5.1cm]{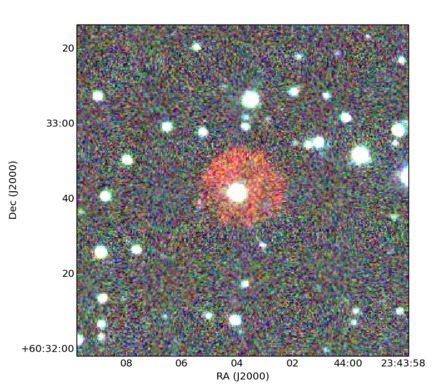}
\includegraphics[height=5.1cm]{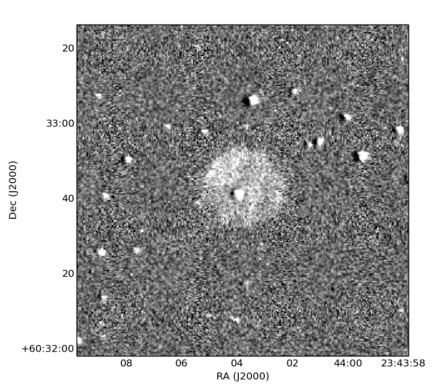}
\includegraphics[height=5.1cm]{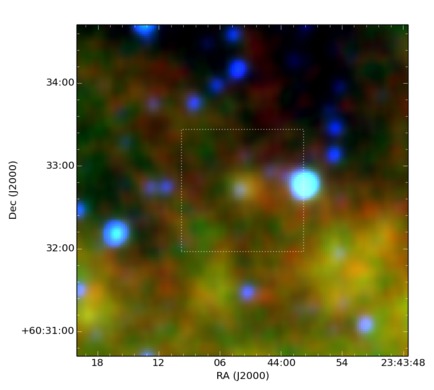}
\includegraphics[height=5.1cm]{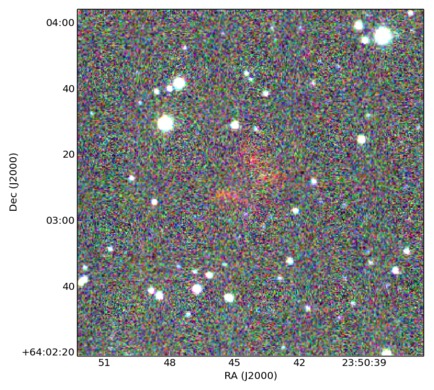}
\includegraphics[height=5.1cm]{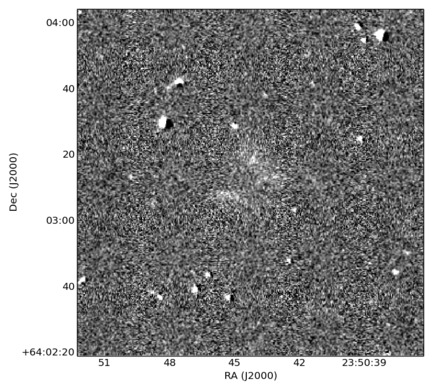}
\includegraphics[height=5.1cm]{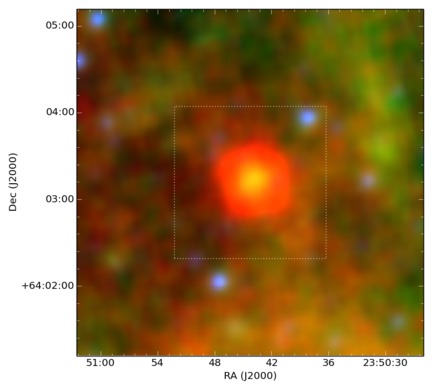}
\includegraphics[height=5.1cm]{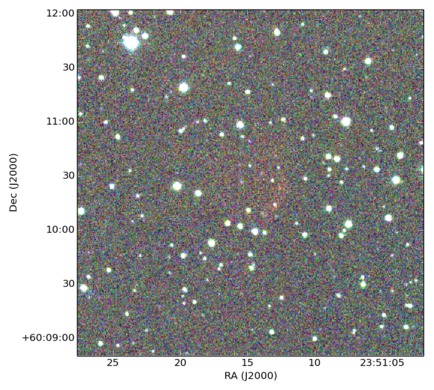}
\includegraphics[height=5.1cm]{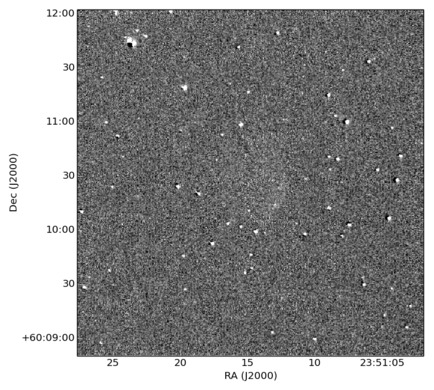}
\includegraphics[height=5.1cm]{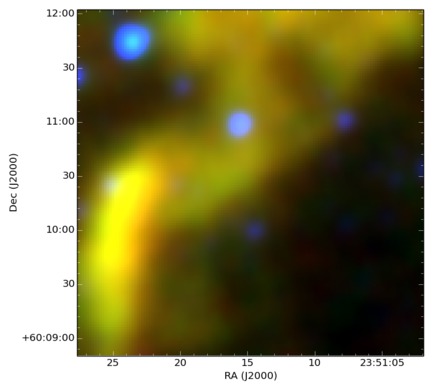}
\caption{\label{imagelabel} Same as in Fig.~\ref{image1}. Objects shown (from top to bottom):  PN G114.7-01.2,PN G116.3+01.9,PN G115.5-01.8}
\end{figure*}
\clearpage

\end{document}